\documentclass[jac]{iucrjournals}

\usepackage{amsmath}
\usepackage{amssymb}
\usepackage{bm}
\usepackage{graphicx}
\graphicspath{{images/}}

\newcommand*\patchAmsMathEnvironmentForLineno[1]{%
  \expandafter\let\csname old#1\expandafter\endcsname\csname #1\endcsname
  \expandafter\let\csname oldend#1\expandafter\endcsname\csname end#1\endcsname
  \renewenvironment{#1}%
    {\linenomath\csname old#1\endcsname}%
    {\csname oldend#1\endcsname\endlinenomath}}
\newcommand*\patchBothAmsMathEnvironmentsForLineno[1]{%
  \patchAmsMathEnvironmentForLineno{#1}%
  \patchAmsMathEnvironmentForLineno{#1*}}
\AtBeginDocument{%
  \patchBothAmsMathEnvironmentsForLineno{equation}%
  \patchBothAmsMathEnvironmentsForLineno{align}}

\newcommand{\tvecd}{\ensuremath{\bm{\mathrm{t}}_d}}
\newcommand{\tvecs}{\ensuremath{\bm{\mathrm{t}}_s}}
\newcommand{\tvecc}{\ensuremath{\bm{\mathrm{t}}_c}}
\newcommand{\rmatd}{\ensuremath{\mathsf{R}_d}}
\newcommand{\rmats}{\ensuremath{\mathsf{R}_s}}
\newcommand{\gvec}{\ensuremath{\bm{\mathrm{G}}}}
\newcommand{\ghat}{\ensuremath{\hat{\bm{\mathrm{G}}}}}
\newcommand{\bhat}{\ensuremath{\hat{\bm{\mathrm{b}}}}}
\newcommand{\dhat}{\ensuremath{\hat{\bm{\mathrm{d}}}}}
\newcommand{\ehat}{\ensuremath{\hat{\bm{\mathrm{e}}}}}
\newcommand{\eye}{\ensuremath{\bm{\mathrm{I}}}}
\newcommand{\detpt}{\ensuremath{\bm{\mathrm{x}}}}
\newcommand{\labcomps}[1]{\left[#1\right]_l}
\newcommand{\detcomps}[1]{\left[#1\right]_d}
\newcommand{\sampcomps}[1]{\left[#1\right]_s}
\newcommand{\xray}{X-ray}
\newcommand{\xrays}{X-rays}
\newcommand{\twotheta}{2\theta}

\newcommand{\capyes}{\ensuremath{\bullet}}
\newcommand{\cappart}{\ensuremath{\circ}}
\newcommand{\capno}{\textendash}

\newcommand{\Pzero}{\ensuremath{\mathrm{P0}}}
\newcommand{\Pone}{\ensuremath{\mathrm{P1}}}
\newcommand{\Ptwo}{\ensuremath{\mathrm{P2}}}
\newcommand{\Pthree}{\ensuremath{\mathrm{P3}}}
\newcommand{\Pfour}{\ensuremath{\mathrm{P4}}}

\newcommand{\Xl}{\ensuremath{\hat{\bm{\mathrm{X}}}_l}}
\newcommand{\Yl}{\ensuremath{\hat{\bm{\mathrm{Y}}}_l}}
\newcommand{\Zl}{\ensuremath{\hat{\bm{\mathrm{Z}}}_l}}
\newcommand{\labframe}{\ensuremath{\{\Xl,\Yl,\Zl\}}}

\newcommand{\Xd}{\ensuremath{\hat{\bm{\mathrm{X}}}_d}}
\newcommand{\Yd}{\ensuremath{\hat{\bm{\mathrm{Y}}}_d}}
\newcommand{\Zd}{\ensuremath{\hat{\bm{\mathrm{Z}}}_d}}
\newcommand{\detframe}{\ensuremath{\{\Xd,\Yd,\Zd\}}}

\newcommand{\Xs}{\ensuremath{\hat{\bm{\mathrm{X}}}_s}}
\newcommand{\Ys}{\ensuremath{\hat{\bm{\mathrm{Y}}}_s}}
\newcommand{\Zs}{\ensuremath{\hat{\bm{\mathrm{Z}}}_s}}
\newcommand{\sampframe}{\ensuremath{\{\Xs,\Ys,\Zs\}}}

\newcommand{\Xc}{\ensuremath{\hat{\bm{\mathrm{X}}}_c}}
\newcommand{\Yc}{\ensuremath{\hat{\bm{\mathrm{Y}}}_c}}
\newcommand{\Zc}{\ensuremath{\hat{\bm{\mathrm{Z}}}_c}}
\newcommand{\xtlframe}{\ensuremath{\{\Xc,\Yc,\Zc\}}}

\title{\textit{HEXRD} and \textit{HEXRDGUI}: a flexible toolkit for
X-ray diffraction analysis in dynamic compression experiments}

\author[a]{Patrick Avery\IUCrCemaillink{patrick.avery@kitware.com}\IUCrOrcidlink{0000-0003-2254-1345}}%
\author[b]{Joel V. Bernier\IUCrOrcidlink{0000-0001-5418-8094}}%
\author[a]{Brianna Major\IUCrOrcidlink{0000-0003-4968-5701}}%
\author[b]{Saransh Singh\IUCrOrcidlink{0000-0002-0286-9549}}

\affil[a]{Kitware, Inc., Clifton Park, New York 12065, USA}
\affil[b]{Lawrence Livermore National Laboratory, Livermore, California 94550, USA}

\begin{document}
\nolinenumbers
\maketitle
\nolinenumbers

\begin{synopsis}
\textit{HEXRD} and its graphical interface \textit{HEXRDGUI} provide a
comprehensive, open-source toolkit for multi-detector \xray\ diffraction
analysis at major laser and synchrotron facilities; this article
describes the toolkit's instrument calibration and powder pattern
fitting capabilities in the context of dynamic compression
experiments.
\end{synopsis}

\begin{abstract}
This article describes the application of \textit{HEXRD}, an open-source Python
library for \xray\ diffraction analysis, and \textit{HEXRDGUI}, its
companion graphical application, to dynamic compression experiments.
Built on a generic multi-detector geometric model, the
toolkit supports arbitrary instrument configurations with independently
positioned planar or cylindrical detector panels, arbitrary beam
directions, and multiple \xray\
sources.  It calibrates instruments from powder rings, Laue spots or
the rings of an unknown phase, fits whole patterns by Le~Bail or
Rietveld refinement, and imports data from the TARDIS, FIDDLE and
PXRDIP platforms at the National Ignition Facility (NIF) and OMEGA~EP.  
Although originally developed for high-energy diffraction
microscopy at synchrotron sources, \textit{HEXRD} has been adopted as an
analysis tool for dynamic compression \xray\ diffraction experiments at
facilities including the NIF, the Matter
in Extreme Conditions end-station at the Linac Coherent Light Source, the Dynamic
Compression Sector (DCS) at the Advanced Photon Source, OMEGA~EP at the
Laboratory for Laser Energetics, and the High Energy Density (HED) instrument at the European
XFEL.  Calibrated from the same ring positions on a single planar
detector, \textit{HEXRD} and the independent code \textit{pyFAI} predict
the same scattering angle $\twotheta$ to numerical precision.  On a
dual-source dataset from the NIF, calibrating
the detectors against both \xray\ sources at once fits the rings of
both to an RMS of 0.18$^{\circ}$ in $\twotheta$, about one
image-plate pixel, whereas a fit to either source alone, with the plate
tilts held at their nominal values, fits its own rings nearly as well
but mispredicts the other's by about 0.6$^{\circ}$.  The software is freely available under a
BSD license.
\end{abstract}


\section{Introduction}

In situ \xray\ diffraction (XRD) is a primary diagnostic for probing the structure of matter under dynamic compression, where materials are subjected to extreme pressures and temperatures using laser or gas-gun drivers \cite{Wark2022review}.  Over the past three decades, XRD measurements during dynamic loading have enabled a series of landmark observations. Early experiments demonstrated that the lattice of a shocked crystal could be probed directly on sub-nanosecond timescales \cite{Wark1989}, and later measurements resolved solid--solid phase transitions in situ, including the $\alpha$--$\varepsilon$ transition in shock-compressed iron \cite{Kalantar2005}.  The advent of \xray\ free-electron lasers enabled femtosecond visualization of lattice dynamics and plasticity in shock-compressed copper \cite{Milathianaki2013} and tantalum \cite{Wehrenberg2017}, and shear-driven phase-transition lowering in silicon \cite{McBride2019}.  More recently, XRD has revealed the structures of planetary and exotic materials at extreme conditions, including a high-pressure phase of magnesium oxide at exoplanet pressures \cite{Coppari2013mgo,Wicks2024}, the formation of diamond and lonsdaleite from shocked graphite \cite{Kraus2016}, superionic water ice \cite{Millot2019ice}, measurement of the high-pressure melt curve of iron \cite{Kraus2022}, and the structure of liquid carbon \cite{Kraus2025carbon}.  These techniques have also extended crystallography into the terapascal regime, measuring solid tin to 1.2~TPa \cite{Lazicki2015sn}, revealing the exotic electride phase in elemental magnesium at $\sim$1~TPa \cite{Gorman2022}, and demonstrating the metastability of diamond ramp-compressed to 2~TPa \cite{Lazicki2021diamond}.

These experiments typically employ non-standard instrument configurations: multiple detector panels at various orientations, pinhole collimation, curved image plates, and polychromatic or pink-beam \xray\ sources.  Facilities such as the National Ignition Facility (NIF) \cite{Rygg2020}, the Matter in Extreme Conditions (MEC) end-station at the Linac Coherent Light Source (LCLS) \cite{Nagler2015mec,Briggs2019sn}, the Dynamic Compression Sector (DCS) at the Advanced Photon Source (APS) \cite{Coleman2022ag}, OMEGA~EP at the Laboratory for Laser Energetics (LLE) \cite{Polsin2025}, and the European X-ray Free Electron Laser (European XFEL) \cite{Gorman2024xfel} each present unique instrumental challenges that general-purpose XRD software packages were not designed to address.

Established tools such as \textit{FIT2D} \cite{Hammersley2016fit2d}, \textit{Dioptas} \cite{Prescher2015dioptas}, and \textit{pyFAI} \cite{Ashiotis2015pyFAI} provide excellent support for standard single and multi-detector setups at synchrotrons and XFEL sources, and \textit{DIALS} \cite{Winter2018dials} with its experiment toolbox \textit{dxtbx} \cite{Parkhurst2014dxtbx} models arbitrary multi-panel detector geometries and beam directions, though its data-reduction pipelines are designed primarily for single-crystal data. Comprehensive Rietveld refinement capabilities are also available in many open-source or commercial packages such as \textit{GSAS-II} \cite{Toby2013gsas2}, \textit{MAUD} \cite{Lutterotti1999MAUDAF}, \textit{FullProf} \cite{rodriguez1993fullprof}, \textit{Profex} \cite{Doebelin2015Profex}, \textit{RIETAN} \cite{izumi1989rietan}, \textit{HighScore Plus} \cite{highscoreplus2014}, and \textit{TOPAS} \cite{Coelho2018topas}, to name a few.  However, none of these packages natively supports the combination of powder and Laue calibration of multi-panel detectors with arbitrary orientations and beam directions, refinable detector distortion models, pinhole and target-assembly corrections, and facility-specific import workflows required for many dynamic compression experiments (Table~\ref{tab:comparison}).  Many beamlines and research laboratories have developed home-grown software for their specific instruments, such as the \textit{IGOR Pro} programs used at Lawrence Livermore National Laboratory (LLNL) for the  Powder X-Ray Diffraction Image Plate (PXRDIP) and Target Diffraction In-Situ (TARDIS) platforms \cite{Millot2026}.  Such programs can be effective, but each new user must learn a different tool for each facility.  The TARDIS and PXRDIP analysis described by \citeasnoun{Rygg2020}, for instance, already optimizes the image-plate geometry jointly with the direction of each \xray\ source and corrects for the target assembly, but only for those platforms.

\begin{table*}
\caption{Capabilities relevant to dynamic compression XRD across
representative open-source and commercial packages.  Columns are:
(1)~multiple detector panels at arbitrary positions and orientations;
(2)~cylindrical detector panels; (3)~multiple \xray\ sources with
independent energies and beam vectors; (4)~pinhole and layered-target-assembly
corrections, which account for the position of the diffracting layer
within the layered target assembly (the ``physics package'',
\S\ref{sec:physpkg}); (5)~Le~Bail fitting; (6)~Rietveld refinement;
(7)~facility-specific import workflows.  \capyes\ denotes native
support, \cappart\ partial or indirect support (for example, achievable
through scripting or an external preprocessing step), and \capno\ no
support.  \textit{HEXRD}'s Le~Bail and Rietveld implementations are
marked partial because they support a narrower range of structural and
microstructural models than dedicated Rietveld packages such as
\textit{GSAS-II} or \textit{TOPAS}; \textit{MAUD}'s facility import is
marked partial for its legacy LCLS import tool.  Entries reflect the
versions released as of October 2026.  Platform-specific in-house
programs, such as the LLNL \textit{IGOR Pro} codes discussed in the
text, are not listed.}
\label{tab:comparison}
\begin{center}
\begin{tabular}{@{}lccccccc@{}}
\hline
\rule{0pt}{2.6ex}Package & (1) & (2) & (3) & (4) & (5) & (6) & (7) \\
\hline
\rule{0pt}{2.6ex}%
\textit{HEXRD}/\textit{HEXRDGUI}               & \capyes  & \capyes  & \capyes  & \capyes & \cappart & \cappart & \capyes \\
\textit{FIT2D} \cite{Hammersley2016fit2d}      & \capno   & \capno   & \capno   & \capno  & \capno  & \capno  & \capno \\
\textit{Dioptas} \cite{Prescher2015dioptas}    & \cappart & \cappart & \capno   & \capno  & \capno  & \capno  & \cappart \\
\textit{pyFAI} \cite{Ashiotis2015pyFAI,Kieffer2020pyFAI} & \capyes  & \capyes  & \cappart & \capno  & \capno  & \capno  & \cappart \\
\textit{DIALS}/\textit{dxtbx} \cite{Winter2018dials,Parkhurst2014dxtbx}
                                               & \capyes  & \cappart & \cappart & \capno  & \capno  & \capno  & \capyes \\
\textit{GSAS-II} \cite{Toby2013gsas2}          & \cappart & \capno   & \capno   & \capno  & \capyes & \capyes & \cappart \\
\textit{MAUD} \cite{Lutterotti1999MAUDAF}      & \cappart & \cappart & \capno   & \capno  & \capyes & \capyes & \cappart \\
\textit{FullProf} \cite{rodriguez1993fullprof} & \capno   & \capno   & \capno   & \capno  & \capyes & \capyes & \capno \\
\textit{TOPAS} \cite{Coelho2018topas}          & \capno   & \capno   & \capno   & \capno  & \capyes & \capyes & \capno \\
\hline
\end{tabular}
\end{center}
\end{table*}

\textit{HEXRD} addresses these limitations through three key capabilities. First, its flexible geometry model (\S\ref{sec:geometry}) supports arbitrary numbers of independently positioned and oriented detector panels with no assumptions about configuration, accommodating non-standard panel shapes, such as curved image plates, and unusual arrangements, such as arbitrarily oriented multi-panel arrays; the subpanels of a segmented detector can be modeled as independent panels so that their misalignment is refined, and detector distortion models are refined alongside the geometry.  Second, corrections for pinhole collimation and layered target assemblies (\S\ref{sec:pinhole}) are integrated directly into the calibration process.  Third, \textit{HEXRD} provides native support for multiple \xray\ sources with independent energies and beam vectors, enabling \textit{simultaneous} calibration of detector position and orientation against diffraction from all beams, essential for instruments like the dual X-ray source (2XRS) configuration at NIF \cite{Rygg2020} where consistency between sources is critical.  Dedicated import workflows for the NIF and OMEGA~EP platforms, and a general one for other detectors (\S\ref{sec:import}), then streamline preprocessing, after which the analysis is the same everywhere.  Joint dual-source calibration and target-assembly corrections already exist in the platform-specific analyses noted above; \textit{HEXRD} provides all three capabilities in one open, scriptable code that is not tied to any single platform.  When a new platform comes online, the software can often be deployed with minimal or no modifications to the source code.

\textit{HEXRD} (Highly Extensible X-Ray Diffraction toolkit) was originally developed at Lawrence Livermore National Laboratory for far-field high-energy diffraction microscopy (ff-HEDM) at synchrotron sources \cite{Boyce2013hexrd, Bernier2011, Bernier2020review}.  Its geometric model, written for multi-panel ff-HEDM detectors, also describes the flat and curved image plates and the multi-panel XFEL detectors used in dynamic compression experiments.  The companion graphical application, \textit{HEXRDGUI}, was subsequently developed in collaboration with Kitware, Inc.\ to provide interactive visualization, calibration, and analysis workflows accessible to experimentalists without programming expertise.  The released software is in routine use by groups other than the developers; analyses led by users outside the development team include shock-compressed liquid tin at the European XFEL \cite{Gorman2024xfel}, shock-ramp-compressed gold at OMEGA~EP \cite{Coleman2025}, shock-compressed nanopolycrystalline diamond \cite{Hari2026} and high-entropy alloys \cite{Lin2026,Huang2026} at the SPring-8 Angstrom Compact Free Electron Laser (SACLA), and the commissioning of the FIDDLE diagnostic at the NIF \cite{Vennari2024}.

In this paper we describe the salient features and capabilities of both \textit{HEXRD} and \textit{HEXRDGUI}, with emphasis on capabilities most relevant to dynamic compression experiments.  In particular, we focus on laser-driven platforms, where most dynamic compression XRD is currently performed.  We note that while \textit{HEXRD} retains its original ff-HEDM capabilities and continues to be widely used for that purpose, including by groups beyond the core developers
\cite{Hurley2018grain,Gustafson2020fatigue,Gordon2021hea,Oh2025lab3dxrd}, the discussion here focuses on the dynamic compression workflows.

The remainder of the article follows the order of a typical analysis.
The instrument geometry is defined (\S\ref{sec:geometry}), together
with the materials and their equations of state
(\S\ref{sec:materials}); the images are imported (\S\ref{sec:import})
and the target assembly is described, with the corrections it implies
(\S\ref{sec:physpkg}); the data are inspected, masked and corrected
(\S\ref{sec:views}); the instrument is calibrated
(\S\ref{sec:calibration}); and the calibrated lineouts are refined by
whole-pattern fitting (\S\ref{sec:wppf}), whose lattice parameters
the equation of state may convert to a pressure at a supplied
temperature.  These two sections also test \textit{HEXRD}
against independent codes and synthetic ground-truth data.
Section~\ref{sec:program} describes the software itself, and
Section~\ref{sec:examples} applies it on six platforms.  For a
laser-driven shot the user supplies the detector images, a starting
instrument configuration (for example, the defaults loaded by the
import tools at the NIF and OMEGA~EP), the energy of each \xray\ source, the layer thicknesses of the target
assembly, and, for calibration, a material of known lattice parameter
in the same exposure, such as ambient diffraction from the pinhole, or,
failing that, the rings alone for the structureless calibration of
\S\ref{sec:structureless}.  Every validation number reported here can be
reproduced from the state files and scripts deposited on Zenodo
\cite{paper_data_zenodo}.

\section{Geometric model}\label{sec:geometry}

The geometric model implemented in \textit{HEXRD} is designed to be
sufficiently general to accommodate any diffraction instrument composed
of planar or cylindrical detector panels, with a common software
interface (an abstract base class) from which other panel shapes can
be derived.  (Throughout this
article, ``geometry'' refers to the position and orientation of the
panels; the surface itself, planar or cylindrical, is the panel
shape.) A collection of individual panels forms an instrument. An initial description was
given by \citeasnoun{Bernier2011}; here we summarize the essential
features.

\subsection{Coordinate systems}

\begin{figure}[!t]
\begin{center}
\begin{minipage}[t]{0.52\columnwidth}
\begin{center}
\includegraphics[width=\linewidth]{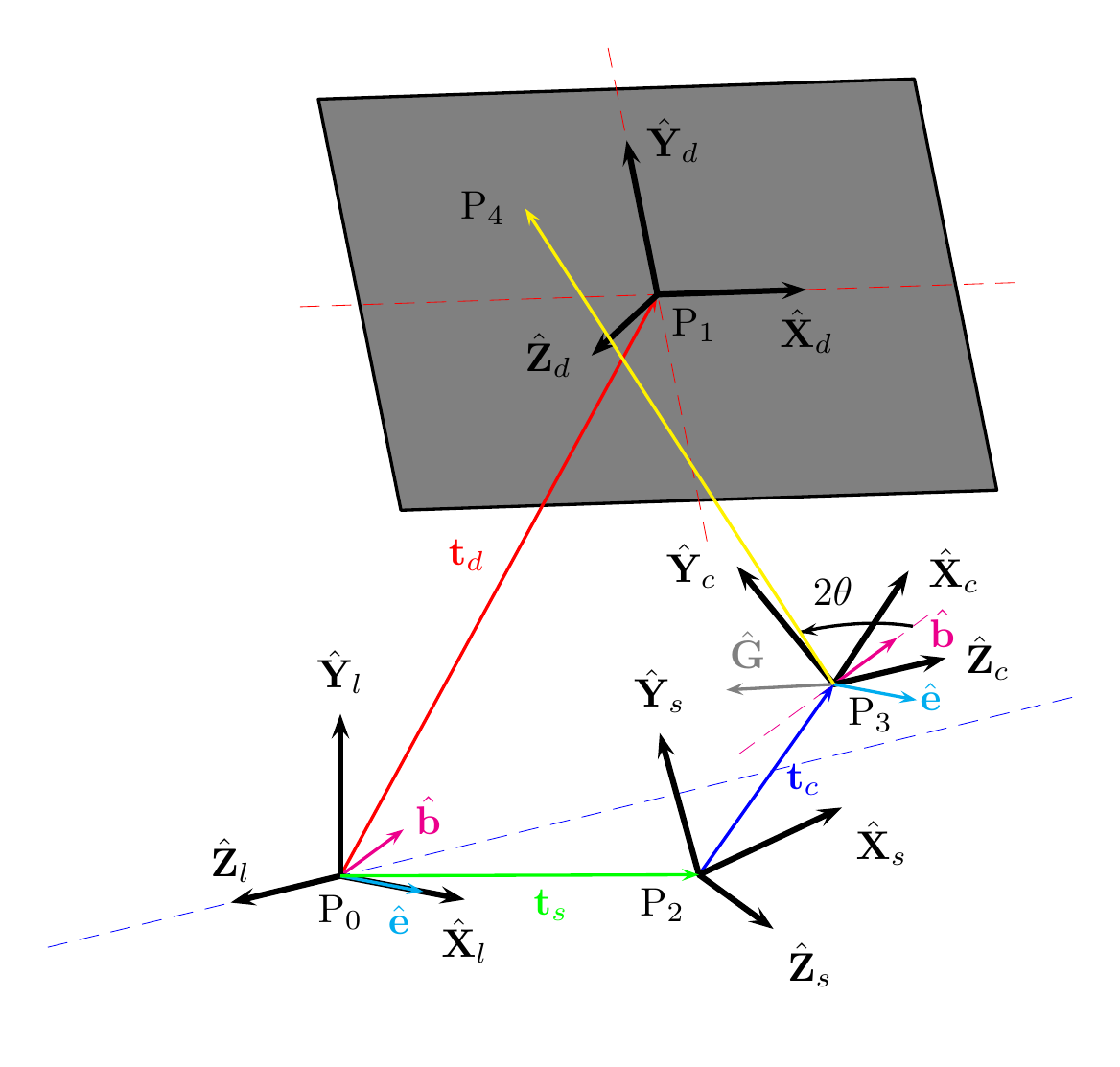}\\
(a)
\end{center}
\end{minipage}\hfill
\begin{minipage}[t]{0.47\columnwidth}
\begin{center}
\includegraphics[width=\linewidth]{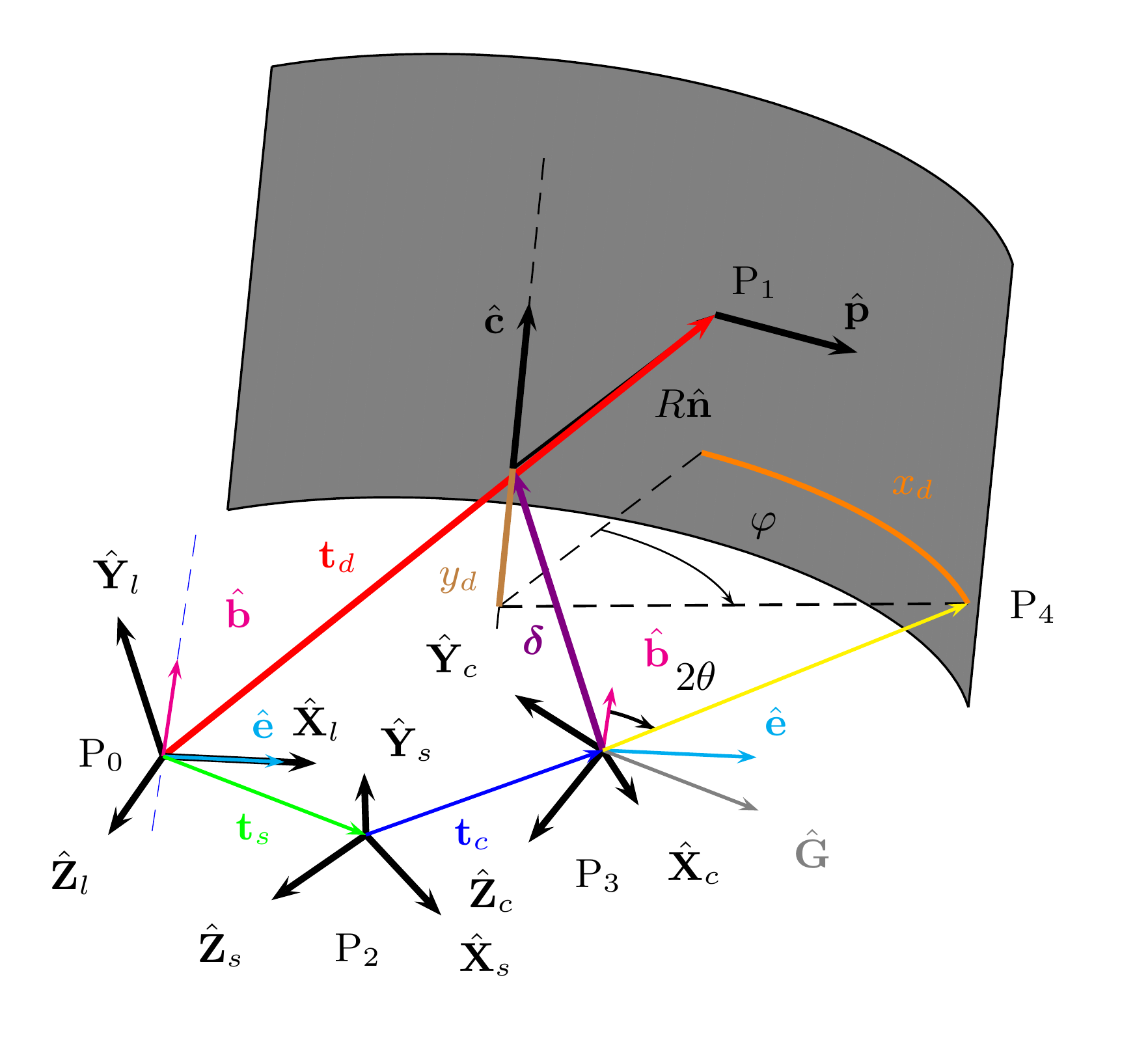}\\
(b)
\end{center}
\end{minipage}
\end{center}
\caption{Diffraction geometry for (a) a planar and (b) a cylindrical
detector panel, illustrating the laboratory ($l$), beam ($b$), detector
($d$), sample ($s$), and crystal ($c$) coordinate frames.  Points
$\Pzero$--$\Pfour$ define the geometric transfer function mapping
reciprocal lattice vectors to detector coordinates; the reciprocal
lattice vector $\ghat$ and scattering angle $\twotheta$ are shown for the
diffracted beam reaching $\Pfour$ (magenta arrow: incident beam $\bhat$;
green: $\tvecs$; blue: $\tvecc$; red: $\tvecd$; yellow: diffracted ray).
In (b), $\bm{\delta}$ (violet) connects the diffracting volume at
$\Pthree$ to the cylinder axis (dashed), $\tvecd$ again points to the
panel center $\Pone$, and the diffracted ray meets the cylinder at
$\Pfour = \Pthree + \beta\dhat$; unwrapping the surface about
$\varphi = 0$ gives its detector coordinates, the arc length
$x_d = R\varphi$ (orange) and the axial coordinate $y_d$ (brown); the
scalar $\beta$ and the cylinder polar angle $\varphi$ are defined in
\S\ref{sec:transfer}.}
\label{fig:geometry}
\end{figure}

Five coordinate frames are used (Fig.~\ref{fig:geometry}):
\begin{itemize}
  \item The \textbf{laboratory frame} $\labframe$, a stationary global
    reference with origin $\Pzero$ at the centroid of the beam
    cross-section, at an arbitrary position along the beam; with
    several \xray\ sources, every beam passes through $\Pzero$.
  \item The \textbf{beam frame}, defined by the incident beam direction
    $\bhat$ and a reference azimuth $\ehat$, providing natural
    representations for the scattering angle $\twotheta$ and azimuth $\eta$,
    which is measured right-handedly about $-\bhat$ from the projection
    of $\ehat$ onto the plane normal to the beam.
    The symbol $\eta$ for the ring azimuth is inherited from
    \textit{HEXRD}'s HEDM origins and is used throughout the interface;
    the powder literature often denotes this angle $\phi$.
  \item The \textbf{detector frame(s)} $\detframe_i$, one per detector
    panel, with origin $\Pone^{(i)}$ at the panel centroid and $\Zd$
    along the panel normal.
  \item The \textbf{sample frame} $\sampframe$, with origin $\Ptwo$
    attached to the specimen, defining the sample coordinate system.
  \item The \textbf{crystal frame} $\xtlframe$, with origin $\Pthree$
    defining the position and orientation of an individual grain relative
    to the sample frame.  This distinction is important for samples
    which are spatially extended such that the diffraction signal
    recorded does not necessarily originate from the same point.
\end{itemize}

In dynamic compression experiments, the points $\Pzero$, $\Ptwo$, and
$\Pthree$ are typically all located at the same physical position (the
sample/target location), but the framework retains the flexibility to
separate them, as required for special circumstances in dynamic compression experiments such as diffraction from different layers offset in space, or significant motion of the sample induced by dynamic compression. The sample frame is placed in the laboratory frame by a translation $\tvecs$ and a rotation $\rmats$,
and a crystal by a translation $\tvecc$ within the sample frame, so
that $\labcomps{\Pthree} = \labcomps{\tvecs} + \rmats\sampcomps{\tvecc}$.

\subsection{Detector parameterization}

Each detector panel $i$ is parameterized by a translation vector
$\tvecd^{(i)}$ and a rotation matrix $\rmatd^{(i)}$, together defining
the affine transformation between detector and laboratory coordinates:
\begin{equation}
  \labcomps{\detpt} = \rmatd \detcomps{\detpt} + \labcomps{\tvecd}.
  \label{eq:det_transform}
\end{equation}
The rotation is stored natively as the three components of an
exponential map, a rotation vector whose direction is the rotation
axis and whose length is the rotation angle; the interface can
display and edit tilts in a choice of Euler conventions, but
configuration files always record the exponential map. The pixel grid
is defined by the number of rows and
columns and the pixel pitch (the center-to-center pixel spacing) in
each direction.

This per-panel parameterization allows \textit{HEXRD} to represent
instruments with any number of detectors at arbitrary positions and
orientations. Detectors may also be grouped into physical units with
known internal geometry (\textit{e.g.}\ a Dexela~2923 detector composed
of four subpanels), allowing subpanel misalignment within a detector to
be taken into account. The grouping may be enforced as a rigid-body
constraint during calibration.  Cylindrical detectors, such as
IMAGE-PLATE-3 on the TARDIS platform at
NIF \cite{Rygg2020}, are also supported.  Each panel may also carry a
distortion model, which maps measured pixel positions to their ideal
ones and whose coefficients can be refined during calibration, such as
the radial-polynomial GE\_41RT model applied to the pincushion
distortion of the GE detector in \S\ref{sec:validation}.  Detector
distortion is distinct from the target-assembly corrections of
\S\ref{sec:pinhole}, which shift the apparent angles of diffraction
from layers that are not at the nominal sample position.

\subsection{Beam parameterization}

The incident beam direction $\bhat$ is specified by spherical polar
angles $(\phi_b, \gamma)$ in the laboratory frame, where the polar
angle $\phi_b$ is measured from $\Yl$ and the azimuth $\gamma$ is
measured from $\Xl$ in the $\Xl$--$\Zl$ plane:
\begin{equation}
  \labcomps{\bhat} = -\begin{pmatrix}
    \sin\phi_b\cos\gamma \\ \cos\phi_b \\ \sin\phi_b\sin\gamma
  \end{pmatrix},
  \label{eq:beam_vec}
\end{equation}
so that the default $(\phi_b, \gamma) = (90^{\circ}, 90^{\circ})$
recovers $\bhat = -\Zl$.  These two angles appear in the interface and
in instrument configuration files as the beam ``polar angle'' and
``azimuth''; note that they parameterize the reversed direction
$-\bhat$, per the leading sign of equation~(\ref{eq:beam_vec}).
Every physical source is, of course, at a finite distance; when that
distance is short enough to matter (\textit{e.g.}\ pinhole geometries
at the NIF), the source--sample distance becomes an additional
parameter that enters into the pinhole distortion corrections described
in \S\ref{sec:pinhole}; it is taken from the facility metadata and is
not among the refinable calibration parameters.

\textit{HEXRD} also supports instruments with multiple \xray\ sources,
each with independent energy, beam vector, and source distance.
Simulated diffraction patterns and calibration are associated with their
respective sources, so that reflections from all beams are satisfied
simultaneously during refinement (\S\ref{sec:composite_cal}); this is the
case, for example, for the dual-source (2XRS) configuration at the NIF
\cite{Rygg2020}.

\subsection{Diffraction and the transfer function}\label{sec:transfer}

Given a reciprocal lattice vector $\gvec$ satisfying the Bragg
condition, the unit diffracted beam vector is
\begin{equation}
  \dhat = (\eye - 2\ghat \otimes \ghat) \bhat.
  \label{eq:dhat}
\end{equation}
The Bragg condition has both a directional and a magnitude part.  The
scattering vector must bisect the incident and diffracted beams,
\begin{equation}
  \ghat \cdot \bhat = -\sin\theta,
  \label{eq:bragg_dir}
\end{equation}
and its length must match the interplanar spacing of the reflection,
\begin{equation}
  \|\gvec\| = \frac{1}{d_{hkl}} = \frac{2\sin\theta}{\lambda},
  \label{eq:bragg_mag}
\end{equation}
for \xray\ wavelength $\lambda$.  Equation~(\ref{eq:bragg_dir})
constrains the orientation of $\ghat$ relative to the beam, from which
equation~(\ref{eq:dhat}) gives the direction of the diffracted ray, and
equation~(\ref{eq:bragg_mag}) fixes the Bragg angle $\theta$ at which
the reflection is excited.
The detector coordinates of the diffraction spot are then obtained by
computing the ray--plane intersection of $\dhat$ with each detector
panel \cite{Bernier2011}:
\begin{equation}
  \detcomps{\Pfour} = \rmatd^T \!\left[ \labcomps{\Pthree - \Pone}
    - \frac{\labcomps{\Zd}^T \labcomps{\Pthree - \Pone}}
           {\labcomps{\Zd}^T \labcomps{\dhat}} \labcomps{\dhat} \right]\!.
  \label{eq:transfer}
\end{equation}
Here $\detcomps{\Pfour}$ denotes the in-plane coordinates of the
intersection point measured relative to the panel origin $\Pone$ and
expressed in the detector frame, which is the form in which pixel
coordinates are stored.
This transfer function is the core computational kernel of
\textit{HEXRD}. It maps reciprocal lattice vectors to detector
coordinates for any instrument configuration, and its inverse provides
the angular coordinates $(\twotheta, \eta)$ from measured detector
positions.  Both directions are implemented as compiled C and C++
kernels that operate on arrays of vectors; the cylindrical case is
vectorized NumPy.

The same construction applies to cylindrical detectors, where the
diffracted ray is intersected with a cylindrical surface rather than a
plane. The cylinder can be unwrapped into a rectangle and this intersection point can be represented using planar Cartesian coordinates. A diffracted ray along $\dhat$, emanating from the diffracting volume at $\Pthree$, meets a cylinder of radius $R$ at the point
$\Pthree + \beta\dhat$. With reference to Fig.~\ref{fig:geometry}(\textit{b}), we define
\begin{align}
\labcomps{\hat{\bm{\mathrm{c}}}} &= \rmatd \detcomps{\Yd}, \nonumber \\
\labcomps{\hat{\bm{\mathrm{p}}}} &= \rmatd \detcomps{\Xd}, \nonumber \\
\labcomps{\hat{\bm{\mathrm{n}}}} &= \labcomps{\hat{\bm{\mathrm{c}}}} \times \labcomps{\hat{\bm{\mathrm{p}}}} = -\rmatd \detcomps{\Zd}, \nonumber \\
\labcomps{\bm{\mathrm{\delta}}} &= \labcomps{\tvecd} - \left(R\labcomps{\hat{\bm{\mathrm{n}}}} + \labcomps{\tvecs} + \rmats\sampcomps{\tvecc}\right).
\end{align}
$\hat{\bm{\mathrm{c}}}$ is the cylinder axis, $\hat{\bm{\mathrm{p}}}$ the tangential direction and $\hat{\bm{\mathrm{n}}}$ the outward radial direction at the center of the panel; they are the detector-frame axes $\Yd$, $\Xd$ and $-\Zd$ expressed in the laboratory frame. $\bm{\delta}$ is the vector connecting the diffracting volume to the axis of the cylindrical detector. Note that the translation vector for the detector, $\labcomps{\tvecd}$, is defined to point to the center of the cylindrical panel, consistent with the planar detector. Since the diffracting volume lies inside the cylinder ($W < 0$ below), the two roots of the quadratic
$U\beta^2 - 2V\beta + W = 0$ have opposite signs, and the physical intersection is the positive root:
\begin{equation}
  \beta = \frac{V + \sqrt{V^2 - U\,W}}{U},
  \label{eq:cyl_transfer}
\end{equation}
with coefficients
\begin{equation}
  \begin{aligned}
    U &= 1 - \left(\labcomps{\dhat}^{T}\labcomps{\hat{\bm{\mathrm{c}}}}\right)^2, \\
    V &= \left(\labcomps{\dhat}^{T} \labcomps{\bm{\delta}}\right)
         - \left(\labcomps{\dhat}^{T} \labcomps{\hat{\bm{\mathrm{c}}}}\right)\,\left(\labcomps{\hat{\bm{\mathrm{c}}}}^{T} \labcomps{\bm{\delta}}\right), \\
    W &= \labcomps{\bm{\delta}}^{T} \labcomps{\bm{\delta}} - \left(\labcomps{\hat{\bm{\mathrm{c}}}}^{T}\labcomps{\bm{\delta}}\right)^2 - R^2.
  \end{aligned}
  \label{eq:cyl_coeffs}
\end{equation}
Next we define a new vector, $\labcomps{\bm{d}^{\prime}}$ (not a unit
vector), which points from the cylinder axis to the intersection point,
\begin{equation}
    \labcomps{\bm{d}^{\prime}} = \beta\labcomps{\dhat} - \labcomps{\bm{\delta}}.
\end{equation}
The intersection point is then described by the cylindrical coordinates
$(R, \varphi, y_d)$, where the polar angle $\varphi$ is measured from the
center of the panel about $\hat{\bm{\mathrm{c}}}$ and takes its sign from
$\hat{\bm{\mathrm{p}}}$:
\begin{equation}
  \varphi = \mathrm{sgn}\!\left(\labcomps{\bm{d}^{\prime}}^{T}
            \labcomps{\hat{\bm{\mathrm{p}}}}\right)
            \arccos{\left(\frac{1}{R}\labcomps{\bm{d}^{\prime}}^{T}
            \labcomps{\hat{\bm{\mathrm{n}}}} \right)}.
  \label{eq:cyl_phi}
\end{equation}
Unwrapping the cylinder about $\varphi = 0$ finally gives the Cartesian
coordinates of the intersection point on the (flattened) detector,
expressed in the detector frame,
\begin{equation}
    \detcomps{\Pfour} = \left(R\varphi,\;
      \labcomps{\bm{d}^{\prime}}^{T}\labcomps{\hat{\bm{\mathrm{c}}}} \right),
  \label{eq:cyl_dewarp}
\end{equation}
so that cylindrical and planar panels are treated identically by the
rest of the code.  (For the planar case, intersections behind the diffracting volume are
rejected in the projection itself, and rays parallel to the panel yield
unbounded coordinates that the subsequent on-panel clipping discards.)  The symbol $\varphi$ is used here for the cylinder
polar angle to avoid confusion with the Bragg angle $\theta$ and the
beam polar angle $\phi_b$.

Because equation~(\ref{eq:cyl_phi}) recovers $\varphi$ from an
inverse cosine, it returns a principal value in $[0, \pi]$ to which the
sign of $\labcomps{\bm{d}^{\prime}}^{T}\labcomps{\hat{\bm{\mathrm{p}}}}$
is then applied, so $\varphi \in (-\pi, \pi]$ and the unwrapped
coordinate $x_d = R\varphi$ is single valued.  A panel is therefore
limited to a total subtended angle of less than $2\pi$, with the seam
placed diametrically opposite the panel center at $\varphi = \pm\pi$;
the image plates considered here subtend well under a full revolution.  (The
degenerate case $U \to 0$, a ray parallel to the cylinder axis,
cannot occur for a ray that intersects the panel.)  As a numerical
check of equations~(\ref{eq:cyl_transfer})--(\ref{eq:cyl_dewarp}), forward
projecting $1.2\times10^{5}$ random diffraction directions onto a
tilted cylindrical panel and inverting the resulting coordinates
recovers the input angles to better than $10^{-7}$ degrees in double
precision.  Checked against the independent cylindrical-detector model
of \textit{pyFAI} \cite{Ashiotis2015pyFAI}, with the sample on and off
the cylinder axis and with the cylinder tilted, the two codes assign
the same pixels the same $\twotheta$ to 0.01~mdeg and $\eta$ to
0.11~mdeg, and the forward projection places \textit{pyFAI}'s angles on
the panel to 0.01~\textmu m.  The same comparison for planar panels,
over 400 random geometries with panels 50--300~mm from the sample,
tilted by up to 53$^{\circ}$ and reaching $\twotheta = 123^{\circ}$,
half of them with the beam along a non-default direction, agrees to
better than $10^{-8}$~mdeg in $\twotheta$ and $10^{-6}$~mdeg in $\eta$.

\section{Crystallography and materials}\label{sec:materials}

\textit{HEXRD} includes a crystallographic materials model for
simulated pattern generation, calibration, and whole-pattern fitting.

\subsection{Materials model}

The materials model stores space group, lattice parameters, atomic
site positions with fractional coordinates, site occupancies, and
thermal displacement parameters (isotropic $U$ or $B$ factors, or
anisotropic $U_{ij}$ tensors).  Space groups are selected by lattice
type and identified by number, Hall symbol, or Hermann--Mauguin
notation; lattice parameters are automatically constrained according to
the lattice symmetry.  Only the standard descriptions of the space
groups are supported: those that admit two origin choices are
represented in the first origin choice tabulated in International
Tables (a CIF recorded in any other description must be transformed to
the standard one before import), trigonal crystals are always represented in
their hexagonal basis, and monoclinic groups in the unique-axis-$b$,
cell-choice-1 setting.  The
material model can also store the stiffness matrix, as well as other
parameters related to the third-order Birch--Murnaghan
thermal equation of state described later in the text.

\begin{figure}[htbp]
\begin{center}
\includegraphics[width=0.6\columnwidth]{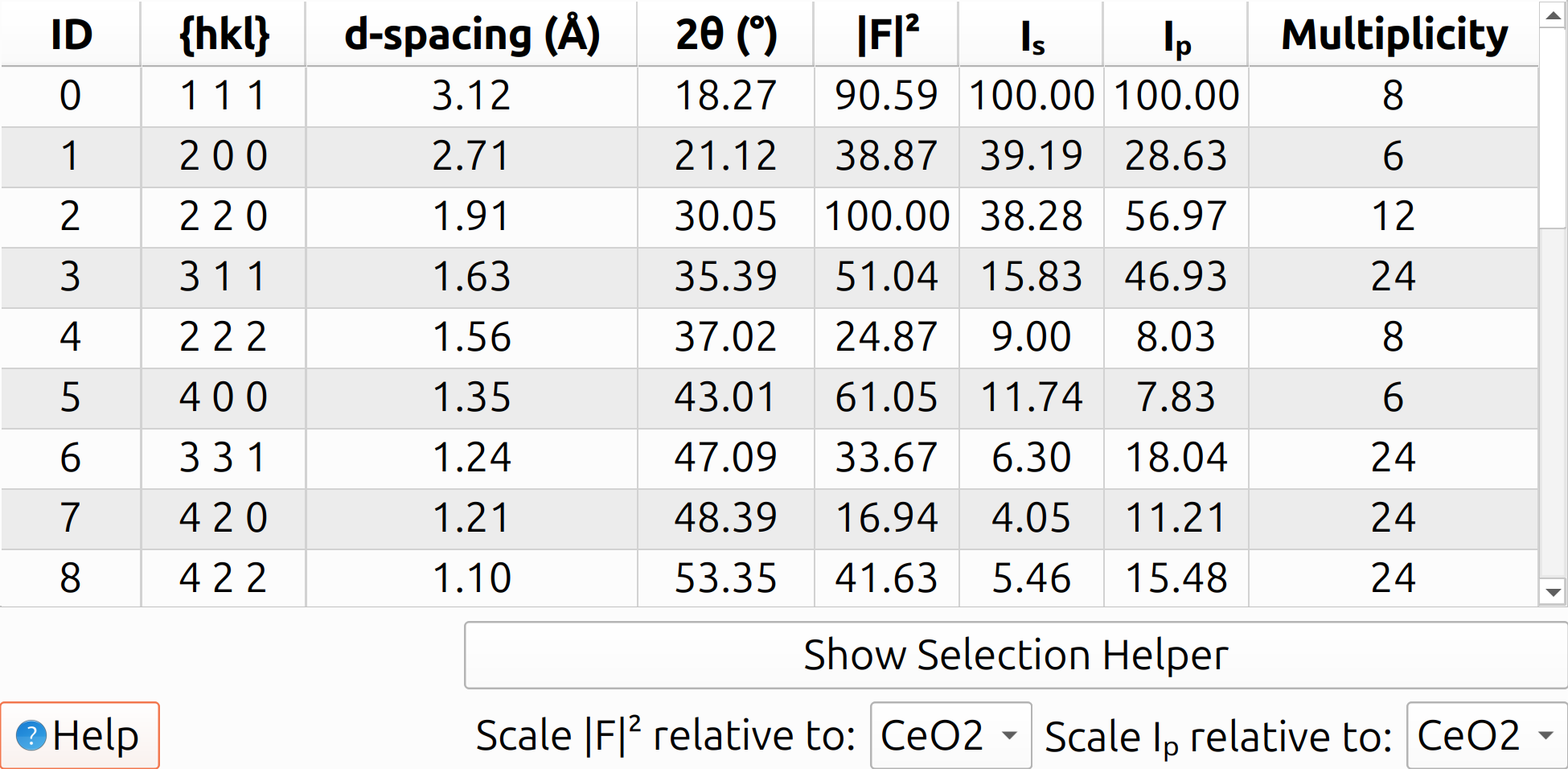}
\end{center}
\caption{Reflections table for CeO$_2$, showing $\{hkl\}$, d-spacing,
$\twotheta$ (at a beam energy of 12.5~keV), structure factor $|F|^2$,
intensity $I_s$ without the multiplicity (computed with the powder
Lorentz factor and the polarization factor for an unpolarized beam), powder intensity $I_p$, and multiplicity for each
reflection.  The $|F|^2$ and $I_p$ columns are rescaled linearly so that the
weakest reflection reads 0 and the strongest 100.}
\label{fig:reflections_table}
\end{figure}

Structure factors are computed using tabulated atomic scattering
factors \cite{Waasmaier1995} with dispersion corrections
\cite{Chantler1995,Chantler2000} and nuclear Thomson scattering
\cite{Berant1977}.  A reflections table
(Fig.~\ref{fig:reflections_table}) allows users to inspect d-spacings,
$\twotheta$ values, structure factors, and intensities for each
reflection, and to select which reflections are displayed as simulated
patterns.

Materials can be imported from CIF files or entered manually through the
GUI.  A default library of materials commonly encountered in dynamic
compression experiments (\textit{e.g.}\ CeO$_2$, LaB$_6$, Pt, Ta, MgO, LiF, diamond)
is bundled with the application.

\subsection{Equation of state}

For high-pressure experiments, \textit{HEXRD} implements the thermal
third-order Birch--Murnaghan equation of state (EOS)
\cite{BirchMurnaghan1947} of the JCPDS format, as used by
\textit{Dioptas} \cite{McHardy2023}, so JCPDS parameter files are used
directly.  With $\Delta T = T - T_0$ and $T_0 = 298$~K, the thermal
expansivity and the moduli at temperature $T$ are
\begin{equation}
  \begin{aligned}
    \alpha_T &= \alpha_0 + \alpha_1 \Delta T, \\
    K_T &= K_0 + \left(\partial K_0 / \partial T\right) \Delta T, \\
    K_T' &= K_0' + \left(\partial K_0' / \partial T\right) \Delta T,
  \end{aligned}
  \label{eq:eos_thermal}
\end{equation}
and the pressure is the third-order Birch--Murnaghan isotherm about the
room-temperature volume $V_0$, evaluated with these moduli, plus the
thermal pressure $\alpha_T K_T \Delta T$,
\begin{equation}
  P(V,T) = 3 K_T f \left(1 + 2f\right)^{5/2}
           \left[1 - \tfrac{3}{2}\left(4 - K_T'\right) f\right]
           + \alpha_T K_T \Delta T,
  \qquad
  f = \tfrac{1}{2}\left[\left(\frac{V_0}{V}\right)^{2/3} - 1\right],
  \label{eq:eos_bm3}
\end{equation}
with $f$ the Eulerian finite strain.  The EOS is therefore
parameterized by $V_0$, $K_0$, $K_0'$, their temperature derivatives, and the
thermal expansion coefficients (an empirical form); parameters may be loaded from JCPDS
files or entered manually.  The EOS serves two roles.  First, the
interactive pressure--temperature slider: for the pressure and
temperature it is set to, it solves equation~(\ref{eq:eos_bm3})
numerically for the volume and moves the simulated diffraction lines
accordingly, so adjusting it until the lines match the data gives a
quick estimate of the pressure.  Second, the Python API works the
other way round: the lattice parameters refined by whole-pattern
fitting (\S\ref{sec:wppf}) give the unit-cell volume $V$, and
equation~(\ref{eq:eos_bm3}) then gives the pressure directly.

Because the temperature enters equation~(\ref{eq:eos_bm3}) only
through these parameters and the thermal pressure, it is an input
rather than a fitted quantity: it must be supplied from an independent measurement or from
hydrodynamic simulation of the drive.  In dynamic compression this is
often a major source of systematic uncertainty in a reported 
pressure, and it can exceed the uncertainty propagated
from the refined lattice parameters themselves; the same machinery
applied to static compression data, where the temperature is well
known, is correspondingly more accurate.  Reported pressures
should therefore be quoted together with the assumed temperature and
the EOS parameter set used.  It also assumes hydrostatic strain: under
uniaxial loading the rings are not circular, and fitting narrow
azimuthal slices of the polar view (\S\ref{sec:views}) instead resolves
the strain along and across the load, as \citeasnoun{Hari2026} did to
measure the flow strength of diamond.

\section{Data import and preprocessing}\label{sec:import}

\textit{HEXRDGUI} supports loading diffraction images in common
formats (HDF5, TIFF, and \textit{HEXRD}'s NumPy-based \texttt{.npz}
frame caches) and additional formats via the FabIO
library \cite{Knudsen2013fabio}, with single- or multi-frame import for
time-resolved experiments.  Dark-frame subtraction, geometric
transforms, and frame selection are configured during import.

\subsection{Facility-specific import tools}

\begin{figure*}[htbp]
\begin{center}
\begin{minipage}[c]{0.38\textwidth}\centering(a)\\[2pt]
\includegraphics[width=\textwidth]{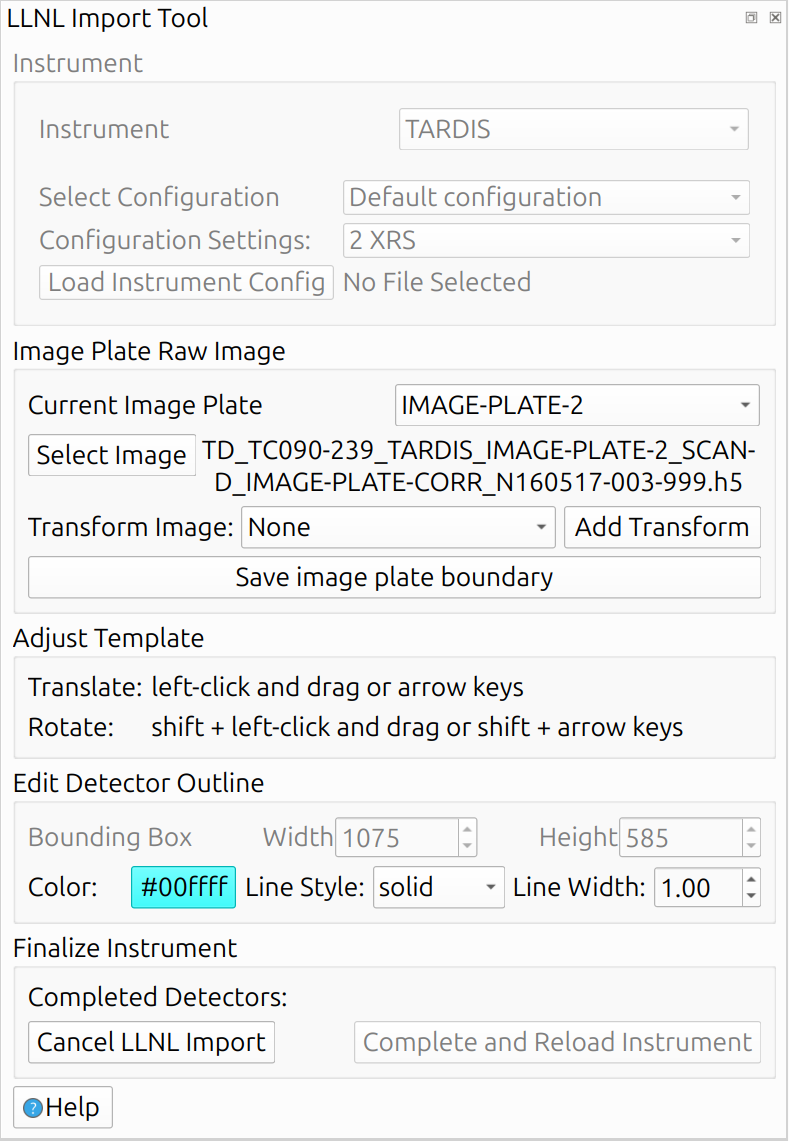}\end{minipage}\hfill
\begin{minipage}[c]{0.58\textwidth}\centering(b)\\[2pt]
\includegraphics[width=\textwidth]{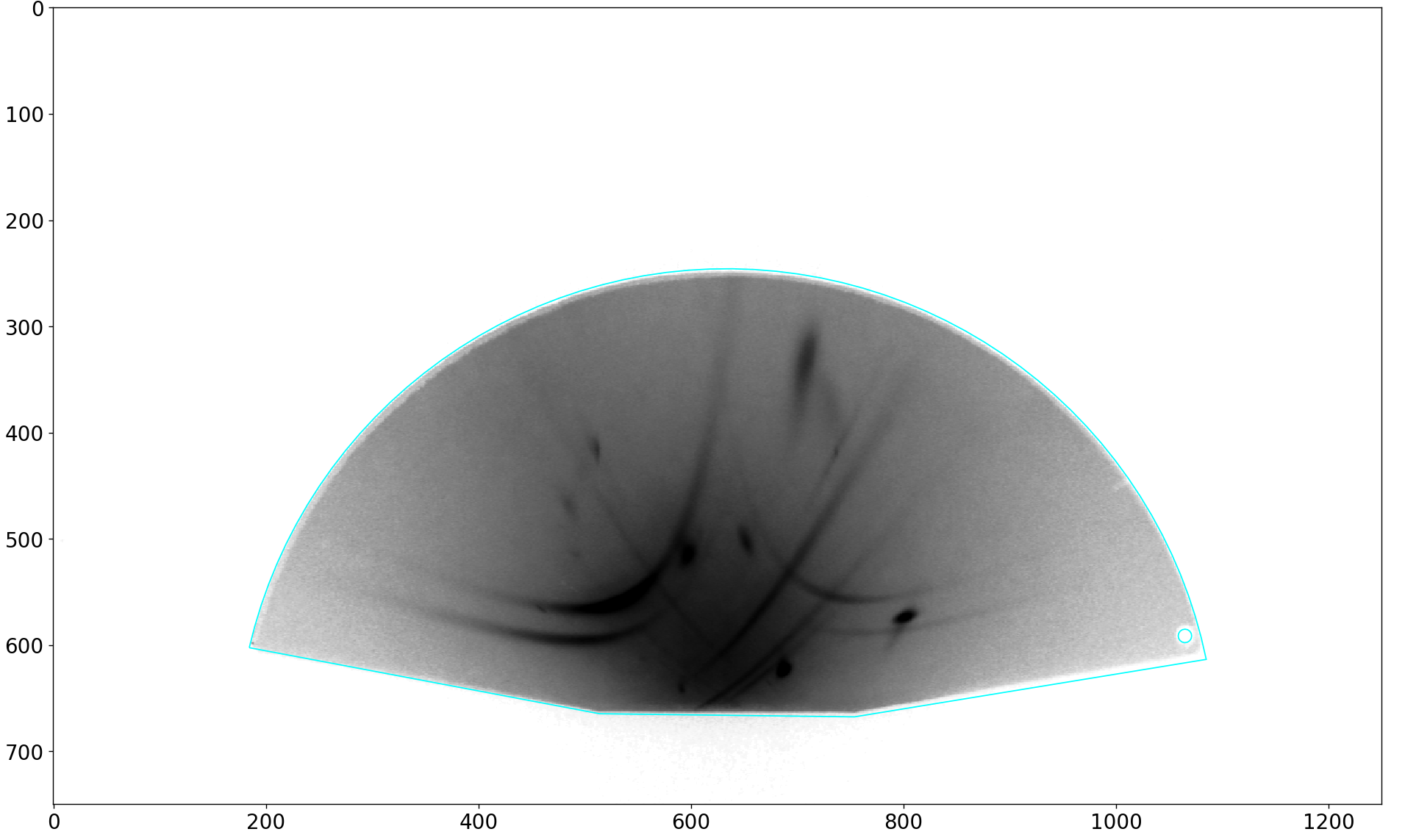}\\[8pt]
(c)\\[2pt]
\includegraphics[width=\textwidth]{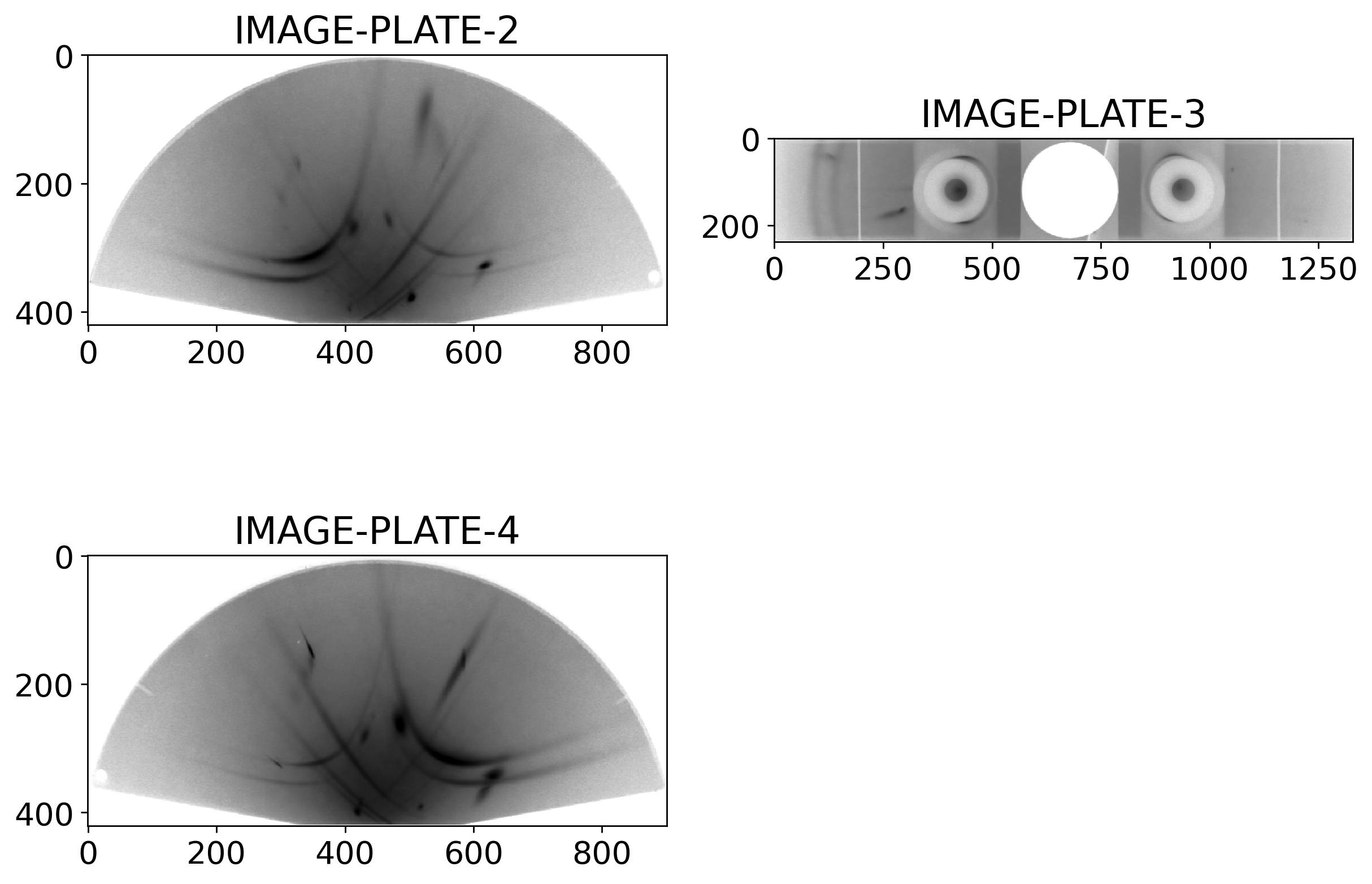}\end{minipage}
\end{center}
\caption{Importing TARDIS image-plate scans with the LLNL import tool.
(a)~The tool's panel, with the instrument, configuration and current
image plate selected.  (b)~The raw scan of that plate with the
interactive detector boundary template (cyan outline), which is
translated and rotated into alignment before cropping.  (c)~Raw view
of the imported data with all three image plates cropped.}
\label{fig:import}
\end{figure*}

Facility-specific import tools provide streamlined workflows for the
TARDIS and FIDDLE (Flexible Imaging Diffraction Diagnostic for Laser
Experiments) platforms at the NIF \cite{Vennari2024,Werellapatha2024}
and the PXRDIP platform at OMEGA~EP \cite{Rygg2012} (Fig.~\ref{fig:import}).  These tools load default instrument configurations, display interactive detector
templates for boundary adjustment on image-plate scans, and handle
facility-specific features such as instrument location and tilt
calibration using the Advanced Tracking Laser Alignment System (ATLAS)
on the NIF \cite{Brunton2018}.  Data from other instruments,
including synchrotron and XFEL detectors, go through a general import
workflow that supports per-panel file assignment for multi-detector arrays (ePix10k, Rayonix,
GE, Pilatus, Varex, etc.).

\section{Physics package and target-assembly corrections}\label{sec:physpkg}

\begin{figure}[htbp]
\begin{center}
\includegraphics[width=\columnwidth]{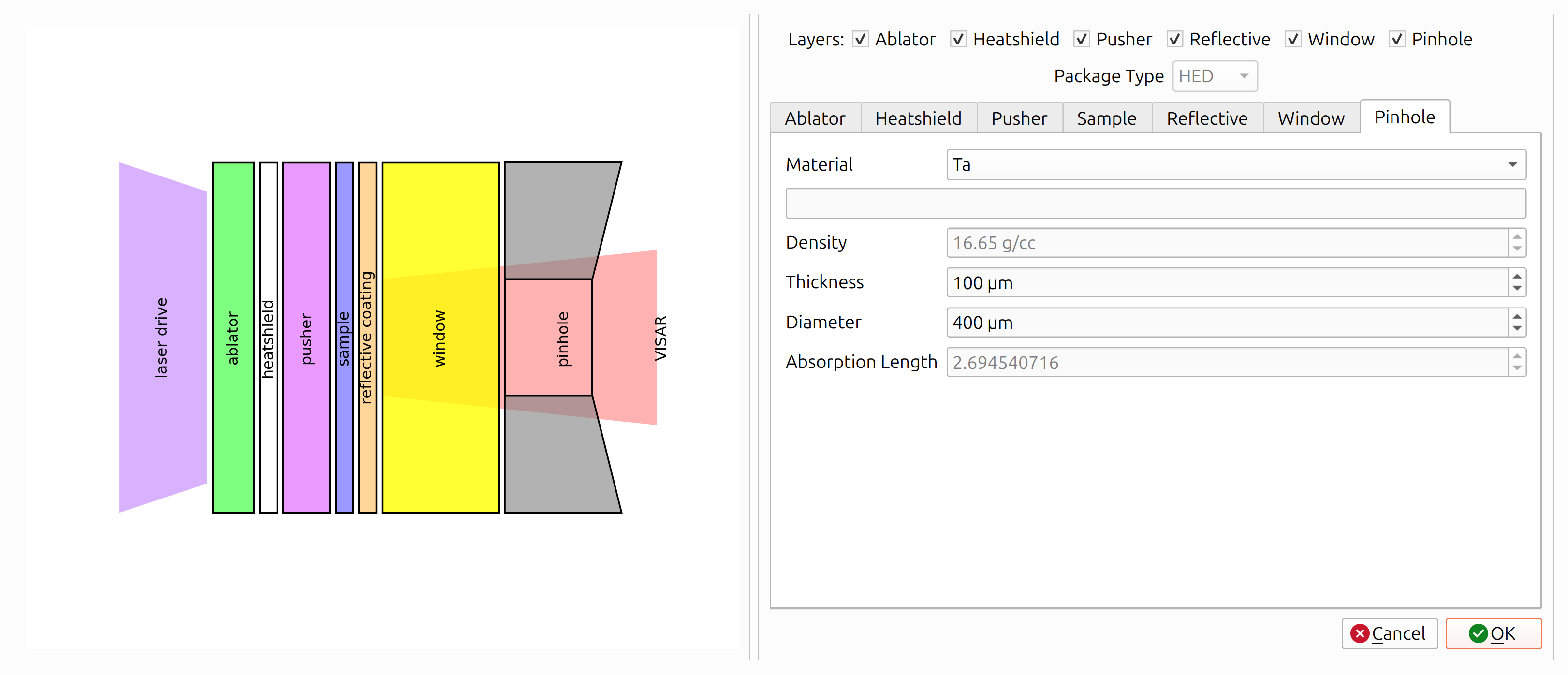}
\end{center}
\caption{The physics-package editor showing an illustrative target
assembly for a NIF experiment with all layers defined (ablator, heat shield, pusher,
sample, reflective, window, and pinhole).  The optional layers are
enabled via the checked boxes (the sample layer is always present), and
the assembly is illustrated by the corresponding diagram on the left.}
\label{fig:physpkg}
\end{figure}

In dynamic compression experiments, the \xray\ beam passes through
multiple layers of the target assembly (the ``physics package'') before
and after diffracting from the sample.  \textit{HEXRDGUI} provides a
physics-package editor (Fig.~\ref{fig:physpkg}) that allows users to
define the target assembly geometry, including the following layers (in
order of \xray\ propagation): ablator, heat shield, pusher, sample,
reflective coating, window, and pinhole.  Each
layer is specified by its material, density, and thickness.  The pinhole
layer additionally requires a diameter.  Sample tilt can be specified
using the globally configured Euler angle convention.

The physics-package definition serves multiple purposes within
\textit{HEXRD}: it provides the geometric parameters needed for pinhole
distortion corrections (\S\ref{sec:pinhole}), enables correction for
absorption of the scattered \xrays\ through the target stack, and is used for
$\twotheta$ corrections that account for the offset between the
different layers.  The editor includes a visual diagram of
the layer stack and automatically detects the instrument type (TARDIS,
PXRDIP, or general) to apply appropriate defaults.

\subsection{Layer and pinhole corrections}\label{sec:pinhole}

In many dynamic compression experiments at NIF and similar facilities,
the diffracted \xrays\ leave the target assembly through a pinhole
collimator, and the diffracting layer sits tens to hundreds of
micrometers from the pinhole rather than at the point assumed by the
instrument geometry.  The apparent $\twotheta$ of its rings is then
shifted, by up to about half a degree in the geometries tested below.
\textit{HEXRD} implements two correction models:
\begin{itemize}
  \item \textbf{Layer distortion}: the first-order correction of
    equation~(8) of \citeasnoun{Rygg2020},
    $\Delta\twotheta = \arctan\!\left[\sin\twotheta_{\mathrm{nom}} /
    \left((r_x/z_s)\cos\beta_{\mathrm{p}} - \cos\twotheta_{\mathrm{nom}}\right)\right]$,
    where $r_x$ is the source-to-pinhole-center distance,
    $\beta_{\mathrm{p}}$ the angle of the diffracted ray from the pinhole
    axis ($\beta$ in the notation of Rygg \textit{et al.}), and $z_s$ the
    offset of the layer's midplane (\textit{e.g.}\ sample, window,
    ablator) from the pinhole center, measured along the pinhole axis
    and computed from the physics package as the layer's standoff from
    the pinhole plus half the thicknesses of the layer and the pinhole.
    This is the point-pinhole approximation: the neglected second-order
    term, of order $(d_p/r_x)^2$ for a pinhole of diameter $d_p$, is
    about 0.01$^{\circ}$ for the standard TARDIS geometry
    \cite{Rygg2020}.  For a driven sample, $z_s$ and the
    layer thickness at the time of the exposure can differ from their
    initial values and must then be supplied by the user, for example from
    the velocity-interferometer (VISAR) record \cite{Rygg2020}.
  \item \textbf{Pinhole distortion}: corrects for the geometric effect
    of the pinhole collimator, as described by \citeasnoun{Rygg2020}.
    The implementation is a vectorized port of the reference
    implementation contributed to the \textit{HEXRD} repository by the
    first author of \citeasnoun{Rygg2020}; a test distributed with the
    validation scripts of \S\ref{sec:validation} drives the port and
    the reference on identical geometry and reproduces the reference to
    better than $10^{-2}$~mdeg across 102 geometries with the \xray\
    source on the pinhole axis and 2520 with it 22.5--45$^{\circ}$ off
    the axis, at source and detector azimuths of both signs, for
    corrections as large as half a degree.
\end{itemize}
These corrections can be applied per overlay (allowing different
corrections for different materials at different positions within the
target assembly), and can optionally be applied directly to the polar
view so that rings associated with the selected layer appear as straight
vertical lines.
The pinhole dimensions are taken from the physics-package definition
above.  Beyond the critical exit angle of the pinhole
the corrections are undefined, and picks there are excluded from the
calibration residual.

\section{Visualization and data preparation}\label{sec:views}

Once an instrument and its data are loaded, \textit{HEXRDGUI} provides
the interactive views, simulated overlays, and masking and correction
tools used to inspect a dataset and prepare it for calibration and
whole-pattern fitting.

\subsection{View modes}

\begin{figure}[htbp]
\begin{center}
(a)\\[2pt]
\includegraphics[width=0.7\columnwidth]{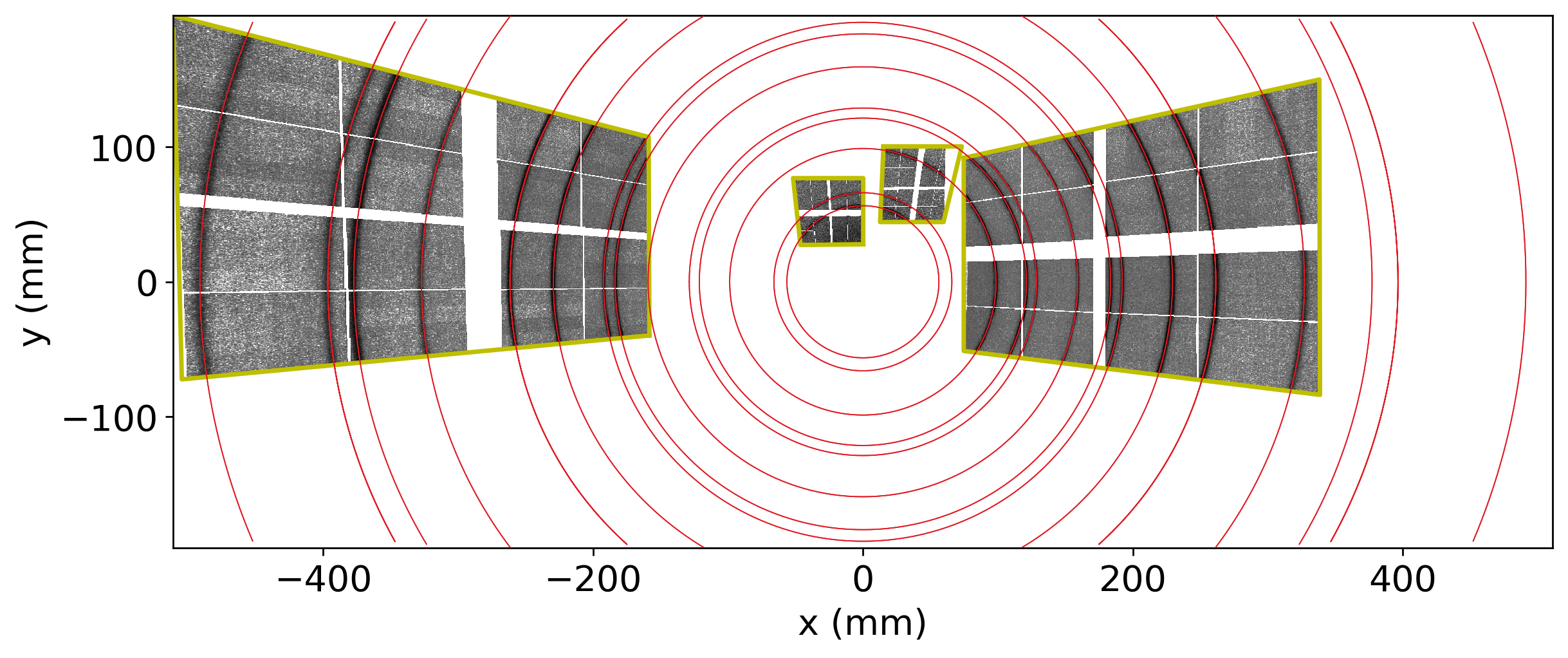}\\[6pt]
(b)\\[2pt]
\includegraphics[width=0.7\columnwidth]{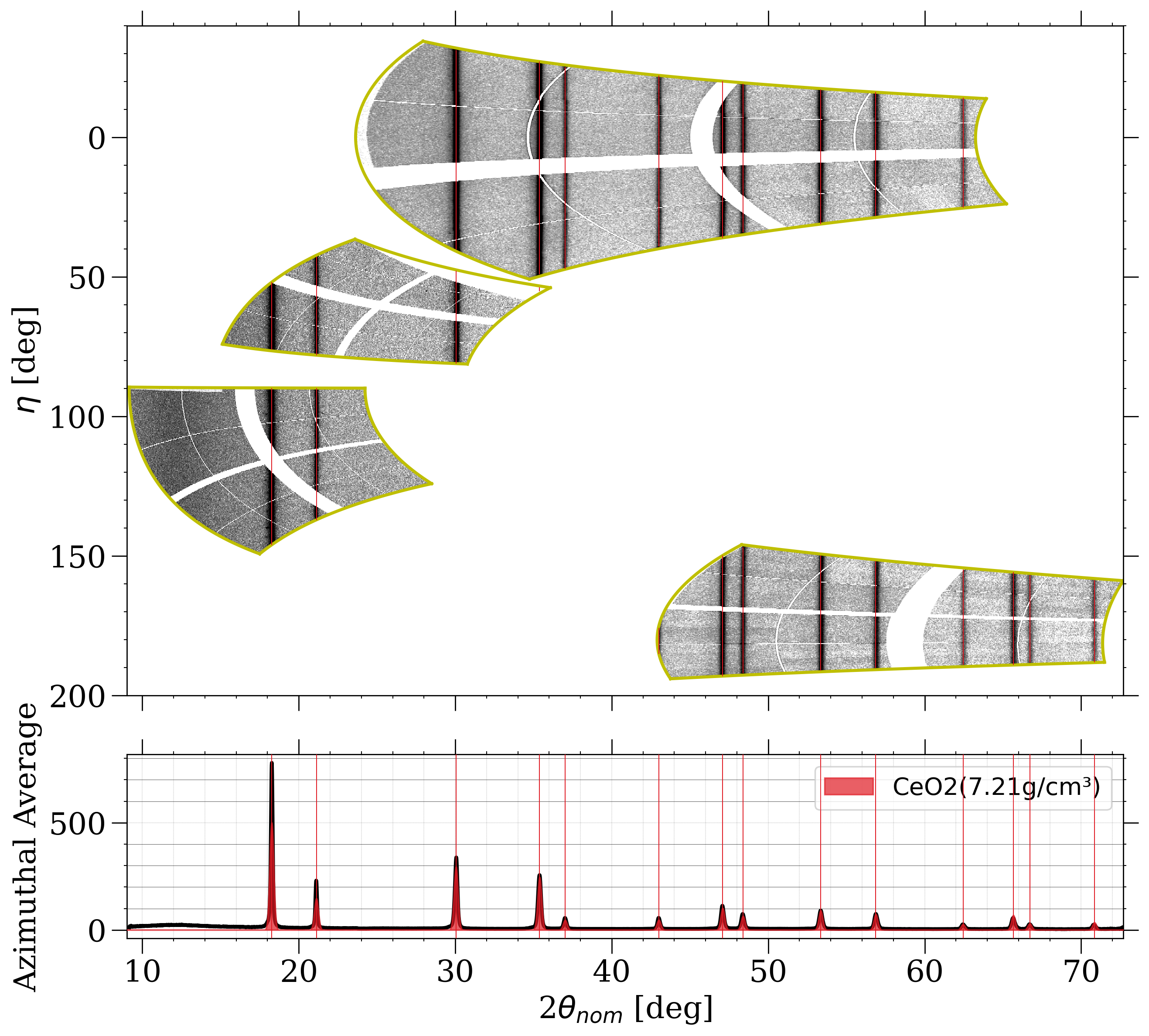}
\end{center}
\caption{MEC data from four ePix10k quad detectors (LCLS experiment
LV7318), with a simulated
CeO$_2$ powder overlay (red) shown in both views.  (a)~Cartesian view
projecting all panels onto a common virtual plane, where the simulated
Debye--Scherrer rings appear as concentric arcs.  (b)~Polar view
($\twotheta$ vs.\ $\eta$) of the same data, in which those rings become
vertical straight lines, with the azimuthally averaged lineout below
(the $\eta$ axis spans only the range the panels cover; intensities
are in arbitrary units here and in the later lineouts).
The lineout shows the observed average (black) together with a
one-dimensional azimuthal overlay (red) of the same phase, drawn as
filled peak profiles on the lineout rather than as vertical lines, so
that it simulates the relative peak intensities and profiles rather
than only the line positions.  The \textit{nom} subscript on angular
axes, here and in later figures, marks nominal angles, computed
without the target-assembly corrections of \S\ref{sec:pinhole}.}
\label{fig:views}
\end{figure}

\textit{HEXRDGUI} provides four interactive view modes for inspecting
diffraction data:
\begin{itemize}
  \item \textbf{Raw view}: displays detector images as acquired, with
    multi-detector data shown side-by-side or in tabs
    (Fig.~\ref{fig:import}(\textit{c})).
  \item \textbf{Cartesian view}: projects all detector panels onto a
    common virtual plane as seen from the sample, revealing gaps and
    overlaps between panels (Fig.~\ref{fig:views}(\textit{a})); cylindrical panels
    are drawn flattened, so distances in this view are not physical
    for them and the view is intended for planar arrangements.
  \item \textbf{Polar view}: resamples detector data onto a regular grid of $\twotheta$
    (diffraction angle) vs.\ $\eta$ (azimuthal angle) with a user-set pixel size,
    where Debye--Scherrer rings appear as vertical straight lines
    (Fig.~\ref{fig:views}(\textit{b})).  An azimuthally averaged lineout is shown
    below the image.
  \item \textbf{Stereographic view}: projects all panels onto a
    stereographic circle, giving a single view of the full angular
    coverage of a multi-panel instrument
    (see the TARDIS example in \S\ref{sec:examples}).
\end{itemize}
The polar view is the primary working mode for powder calibration and
whole-pattern fitting.  It supports configurable ranges in $\twotheta$
and $\eta$ for display and azimuthal averaging, selectable $x$-axis
units ($\twotheta$ or momentum transfer,
$Q = 4\pi\sin\theta/\lambda = 2\pi\|\gvec\|$), and optional layer or pinhole
distortion corrections (\S\ref{sec:pinhole}) so that rings from the
selected layer appear as straight vertical lines.  Lineouts and
whole-pattern fits are computed from this resampled grid, so its pixel
size also sets how many polar pixels contribute to each lineout point
(\S\ref{sec:wppf}).

\subsection{Overlays}\label{sec:overlays}

Simulated diffraction patterns are displayed as overlays on the data in
all view modes.  Three overlay types used in dynamic compression are
managed through the Overlay Manager, a dockable list with per-overlay
controls, and drawn on the main canvas (the manager also offers
rotation-series overlays for high-energy diffraction microscopy):
\begin{itemize}
  \item \textbf{Powder overlays}: Debye--Scherrer rings from
    polycrystalline phases, with selectable reflections, tolerance
    widths, and optional layer distortion corrections such as pinhole.
  \item \textbf{Laue overlays}: single-crystal spots computed from the
    crystal orientation, position, and strain (stretch matrix), together
    with the lattice parameters, for a given energy range and minimum
    d-spacing.  Spot ranges may be displayed as
    ellipses or rectangles, and spots can be optionally labeled by the diffracting plane
    or energy.
  \item \textbf{Constant-$\chi$ overlays}: rings of constant $\chi$
    angle (the angle between the diffraction plane normal and the sample
    normal; unrelated to the $\chi^2$ statistic), useful for identifying preferred orientation, fiber
    textures in particular.
\end{itemize}

Each of these overlay types supports interactive parameter editing and
can be associated with different materials and \xray\ sources.  Overlay
positions update in real time as parameters are adjusted.  Powder
overlays may use layer and pinhole distortion corrections from the
physics-package definition, and Laue overlay parameters (crystal orientation, position, and stretch
matrix) can be loaded from a file or adjusted interactively using
sliders.

Additionally, one-dimensional azimuthally averaged overlays can be
displayed on the azimuthal lineout (Fig.~\ref{fig:views}(\textit{b})), with
configurable FWHM and scale factor, for comparing simulated and observed
relative peak intensities and profiles.  These are managed separately
through the polar view settings rather than the Overlay Manager.

\subsection{Masking and intensity corrections}\label{sec:corrections}

\subsubsection{Masking}

Regions of detector images may be masked using rectangular, elliptical,
or hand-drawn polygonal selections in the raw or polar views, or by
threshold-based intensity filtering.  Masks can also be generated from
powder or Laue overlays to exclude specific diffraction features, and
from the pinhole geometry to mask regions that fall outside the pinhole aperture.
All masks are managed through a mask manager that supports export/import
in HDF5 format.

\subsubsection{Intensity corrections}

Intensity corrections available in \textit{HEXRD} include:
\begin{itemize}
  \item \textbf{Solid-angle corrections}: accounting for the varying
    solid angle subtended by each pixel as a function of detector
    position and orientation.
  \item \textbf{Lorentz correction}: dividing out the Lorentz factor
    $L(\theta) = 1/(\sin^{2}\theta\cos\theta)$, the geometric weighting
    of powder intensities measured on a stationary sample with an area
    detector.
  \item \textbf{Polarization correction}: dividing out the
    polarization factor of the scattering, either
    $\tfrac{1}{2}(1 + \cos^{2}2\theta)$ for an unpolarized source such
    as a laser-plasma backlighter, or its $\eta$-dependent form for a
    linearly polarized synchrotron or XFEL beam with configurable
    horizontal and vertical fractions.
  \item \textbf{Dark-frame subtraction}: applied per-detector during
    import.
  \item \textbf{Absorption corrections}: accounting for \xray\
    attenuation in the sample and window layers of the target assembly
    as defined in the physics package (\S\ref{sec:physpkg}), as well as
    through the detector-side filter and coating (each specified by
    material, density, and thickness) and the angle-dependent response
    of the image-plate phosphor \cite{Rygg2020}, as corrected in its
    erratum \cite{Rygg2026erratum}.
  \item \textbf{Median filtering}: for noise reduction in individual
    frames or frame sequences.
  \item \textbf{SNIP background subtraction}: the Statistics-sensitive
    Non-linear Iterative Peak-clipping algorithm
    \cite{Ryan1988snip} (two-dimensional form after
    \citeasnoun{Morhac1997}) applied to the rebinned polar image for
    background estimation in the polar view.  Three variants are
    available: fast SNIP~1D and SNIP~1D, which operate on each row of
    constant $\eta$ independently, and SNIP~2D, which operates on the
    full two-dimensional image.
\end{itemize}

\section{Calibration}\label{sec:calibration}

Accurate calibration of detector geometry is essential for quantitative
diffraction analysis, particularly in dynamic compression experiments
where small changes in lattice parameters correspond to large pressure
variations.  \textit{HEXRD} provides several calibration methods suited
to different experimental modalities.  All methods minimize a residual
between observed and simulated diffraction features using bounded
nonlinear least-squares optimization via the \textit{lmfit} library
\cite{Newville2014lmfit}, with the Trust Region Reflective algorithm
\cite{Branch1999trf} as the default solver.

Refinable parameters include detector tilts (three components using the
globally configured Euler angle convention) and translations (three
components) for each panel,
beam energy, beam vector orientation, material lattice parameters,
the coefficients of a detector distortion model, and,
for Laue overlays, the crystal orientation, position, and stretch.
Each parameter can be individually toggled for refinement and bounded.

Parameter uncertainties, reported in the whole-pattern fitting dialog
and available from the calibration API, are the standard errors
returned by \textit{lmfit}, obtained from the
covariance matrix estimated from the Jacobian at the solution and scaled
by the reduced $\chi^2$ of the fit.  These
are local, linearized estimates and should be interpreted with care in
two common situations: when a parameter has converged onto one of its
bounds, where the quadratic approximation underlying the covariance
estimate no longer holds, and when parameters are strongly correlated,
as detector distance, beam energy and lattice parameters inevitably are
in a powder calibration.  \textit{HEXRDGUI} highlights parameters that
have converged onto one of their bounds in the parameter tree
(in red; an example appears in \S\ref{sec:composite_cal}) so that these
cases are visible to the user.

\subsection{Relative and engineering constraints}

Two types of constraints help regularize calibration for instruments
with known internal geometry:
\begin{itemize}
  \item \textbf{Rigid-body constraints} enforce that a group of
    detectors (either within a single physical unit or across the entire
    instrument) moves as a rigid body during refinement, preserving
    their known relative positions and orientations.  For example, on the
    FIDDLE instrument at NIF \cite{Vennari2024}, the relative positions
    and orientations of the individual Icarus cameras (gated hybrid-CMOS
    detectors) are measured by
    ATLAS, but the overall position and orientation of the assembly must
    be determined through calibration (\S\ref{sec:fiddle}).
  \item \textbf{Engineering constraints} impose instrument-specific
    physical bounds (\textit{e.g.}\ the distance between the top and
    bottom TARDIS image plates is constrained to a physically reasonable
    range given the known constraints of the NIF target positioner).
\end{itemize}
Both constraint types are set in the calibration dialog; in the TARDIS
example of \S\ref{sec:composite_cal}, relative constraints are not
used, but the engineering constraint on the plate separation is active.

\subsection{Powder calibration}\label{sec:powder_cal}

\begin{figure}[htbp]
\begin{center}
\begin{minipage}[b]{0.416\columnwidth}\centering(a)\\[2pt]
\includegraphics[width=\linewidth]{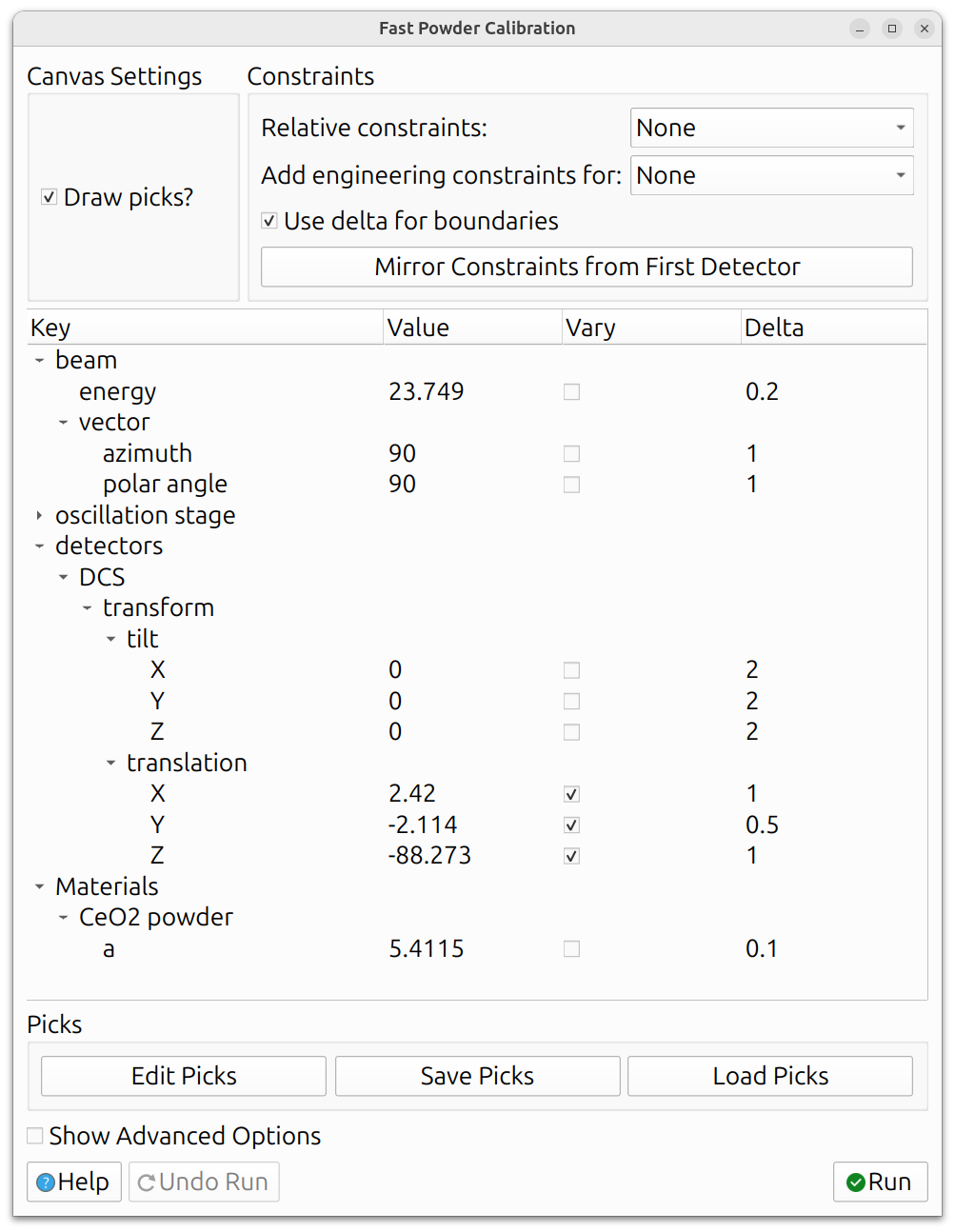}\end{minipage}\hfill
\begin{minipage}[b]{0.544\columnwidth}\centering(b)\\[2pt]
\includegraphics[width=\linewidth]{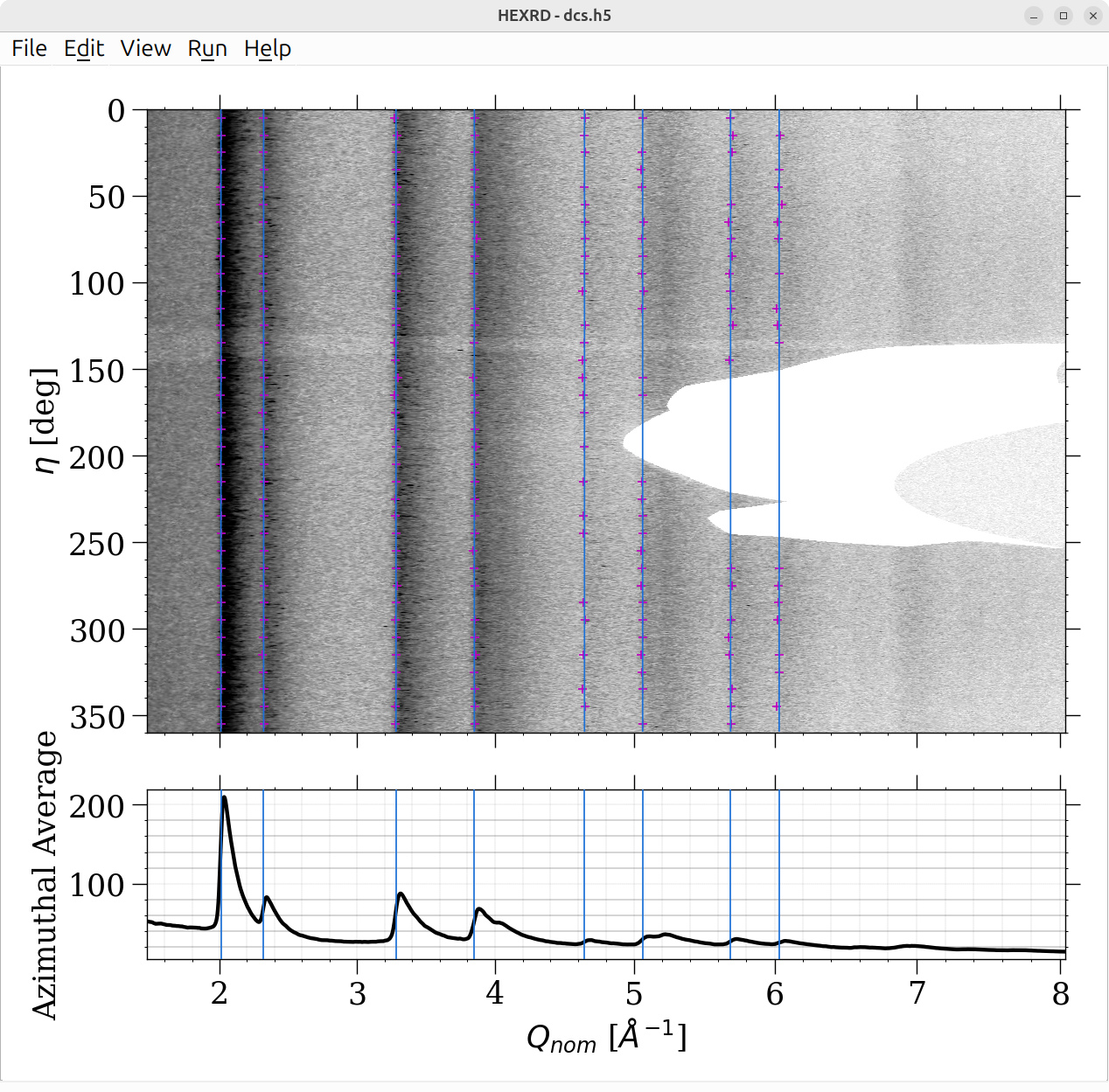}\end{minipage}
\end{center}
\caption{``Fast powder'' calibration of a DCS pink-beam dataset using a
CeO$_2$ standard.  (\textit{a})~Calibration dialog with refinable
parameters and constraints; with the ``Use delta for boundaries''
option checked, the Delta column bounds each parameter to $\pm\Delta$
about its current value.  (\textit{b})~Polar view showing
auto-picked peak positions (magenta plus signs) overlaid on the powder
rings, with simulated ring positions (vertical blue lines), and the
azimuthal lineout below.  In the configuration shown only the three
detector translations are refined.}
\label{fig:fastpowder}
\end{figure}

The powder calibration method operates against a known powder standard
(Fig.~\ref{fig:fastpowder}).  Calibration points may be picked manually
or auto-picked; auto-picking requires that simulated Debye--Scherrer
lines are already close to the observed rings, which can be arranged by
first adjusting the instrument parameters by hand with the calibration
sliders.

For auto-picking, the
image data are projected into polar ($\twotheta$ vs.\ $\eta$)
coordinates, where ring positions are identified automatically by
fitting user-specified peak profiles along radial slices at regular
$\eta$ intervals; Gaussian, pseudo-Voigt, and pink-beam
\cite{VonDreele2021pink} peak profiles are available.
User-configurable parameters include the annular search width in
$\twotheta$ and $\eta$, the minimum peak amplitude, and a maximum
fit-error threshold for rejecting spurious picks.  The objective
function is the unweighted sum of squared differences between picked
and predicted ring positions, measured in $\twotheta$.  A ``Fast powder'' calibration mode provides a
streamlined workflow that operates on a single powder overlay and uses
auto-picking exclusively, bypassing the manual point-picking step
entirely.

\subsection{Laue calibration}

For experiments that produce single-crystal Laue diffraction, for
example from a diamond or LiF window in the target assembly, the
Laue calibration method can refine detector geometry and all crystal
parameters (position, orientation, and stretch matrix) simultaneously.
Laue spots may be picked manually or auto-picked; as with powder
calibration, auto-picking requires that the simulated spots are already
close to the observed positions, and both the crystal and the
instrument sliders can be used to roughly align the simulation first.  The refinement minimizes the
discrepancy between picked and predicted spot positions on the detector.

\subsection{Composite calibration}\label{sec:composite_cal}

\begin{figure}[htbp]
\begin{center}
\begin{minipage}[b]{0.545\columnwidth}\centering(a)\\[2pt]
\includegraphics[width=\linewidth]{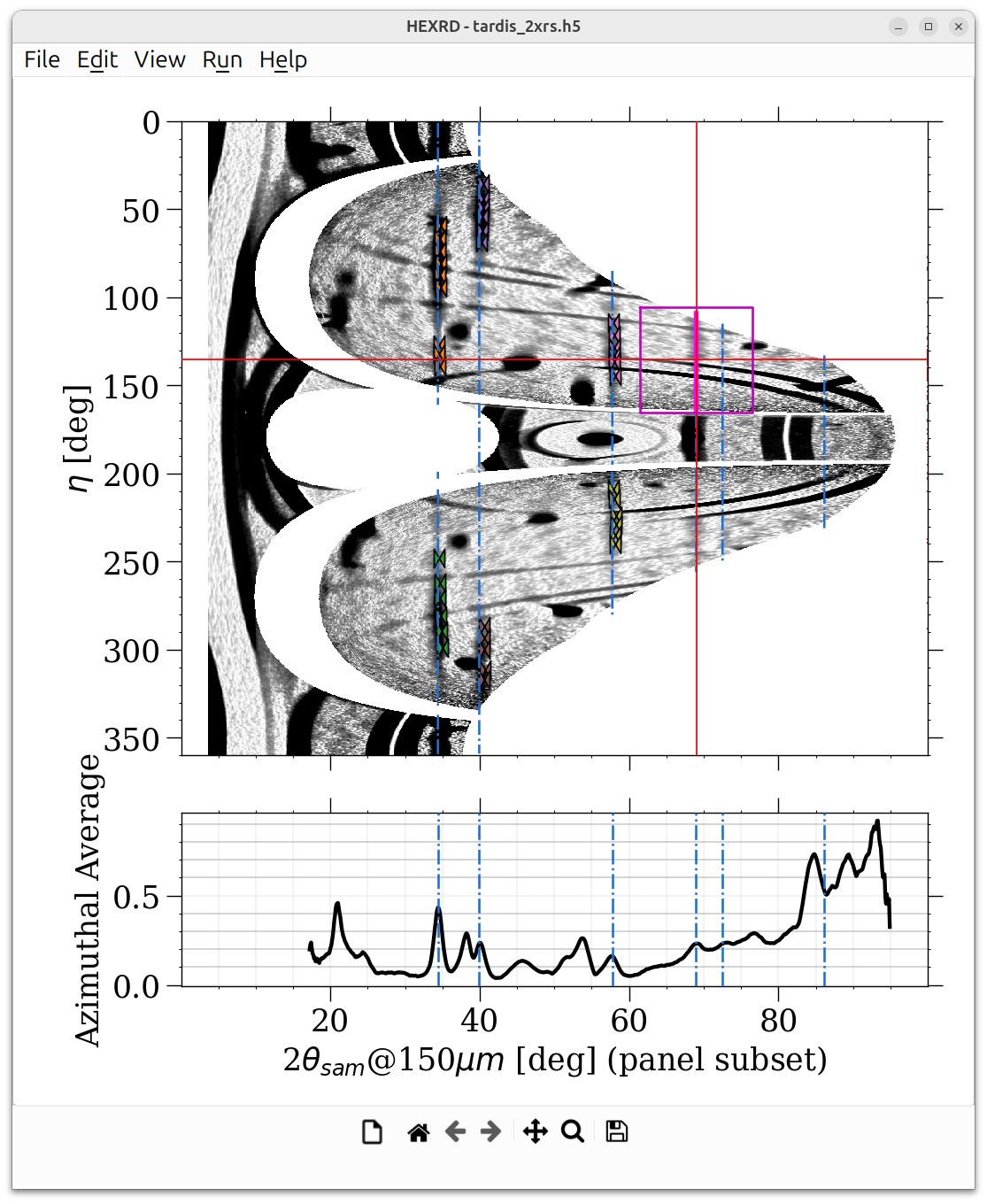}\end{minipage}\hfill
\begin{minipage}[b]{0.415\columnwidth}\centering(b)\\[2pt]
\includegraphics[width=\linewidth]{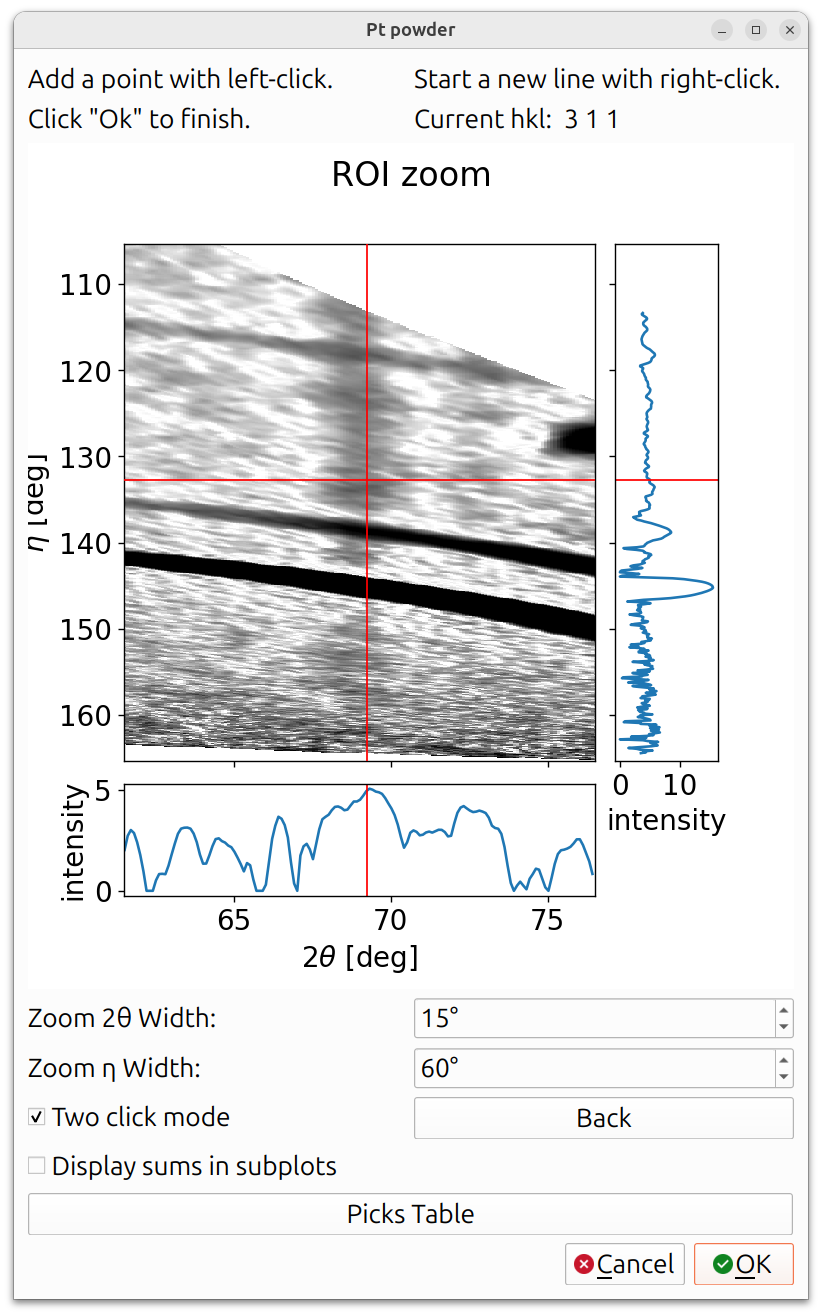}\end{minipage}
\end{center}
\caption{Hand-picking points for composite calibration of TARDIS data.
(\textit{a})~Polar view across multiple image plates, with picked
points shown as colored markers and simulated rings as blue dash-dot
lines; the bold magenta line marks the ring currently being picked and
the magenta rectangle the region shown magnified.  The azimuthal
lineout appears below; the axis label $\twotheta_{sam}$@150\textmu m
indicates that the layer correction of \S\ref{sec:pinhole} is
applied, mapping angles to the sample layer at its 150~\textmu m
standoff.  (\textit{b})~Zoom dialog, which provides the
magnified view, corresponding to the magenta rectangle in
(\textit{a}), used to place each point precisely in two-click mode.}
\label{fig:handpicking}
\end{figure}

\begin{figure}[htbp]
\begin{center}
\begin{minipage}[b]{0.476\columnwidth}\centering(a)\\[2pt]
\includegraphics[width=\linewidth]{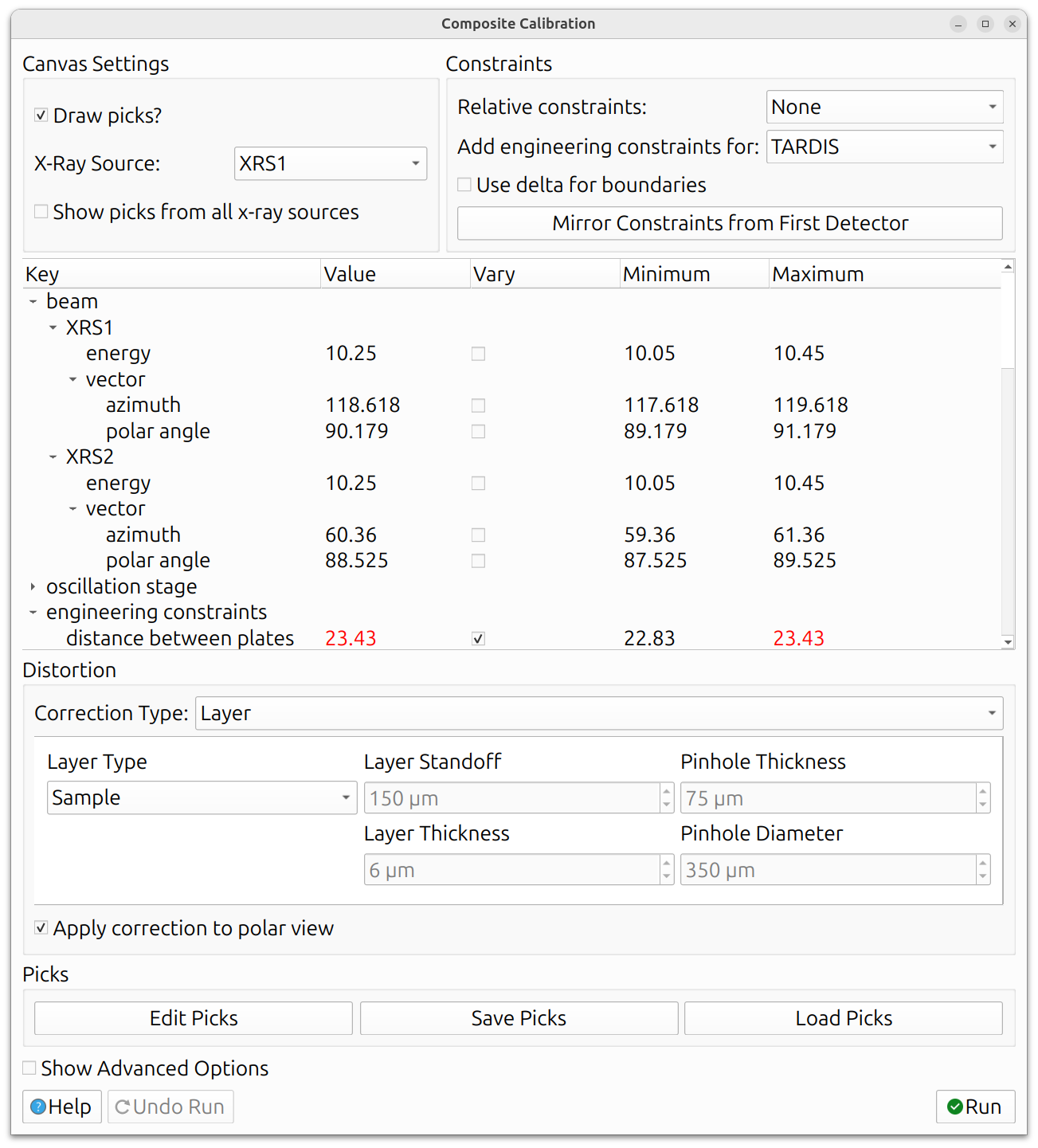}\end{minipage}\hfill
\begin{minipage}[b]{0.484\columnwidth}\centering(b)\\[2pt]
\includegraphics[width=\linewidth]{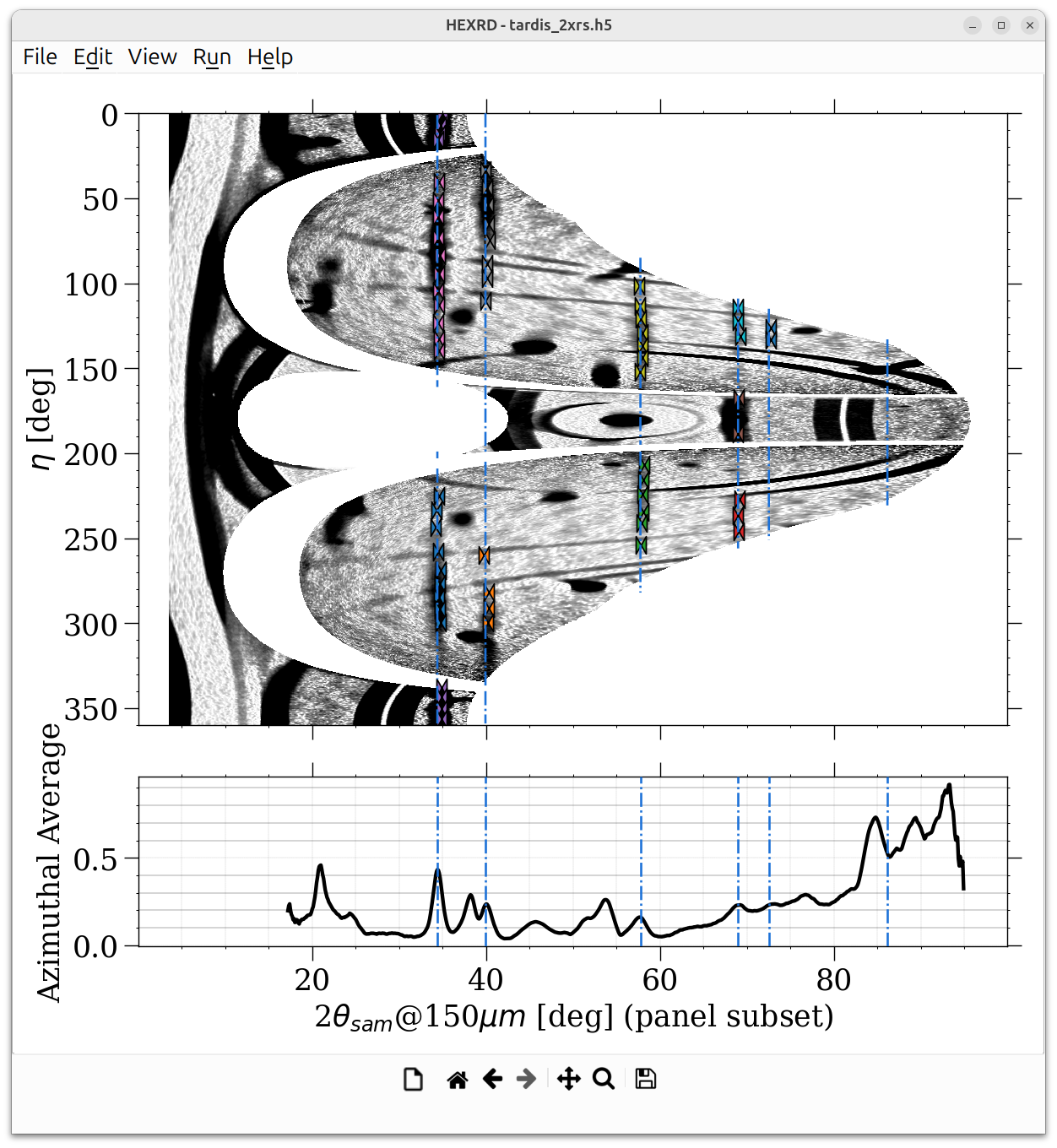}\end{minipage}
\end{center}
\caption{Composite calibration of a TARDIS dual-source (2XRS) dataset.
The same material is assigned to two separate overlays, one per \xray\
source, so that picks from both sources contribute to the overall
residual and the instrument is calibrated against both beams
simultaneously.  (\textit{a})~Calibration dialog showing refinable
parameters for both \xray\ sources (XRS1, XRS2) and the TARDIS
engineering constraints.  The ``distance between plates'' parameter
constrains the Y-axis separation of IMAGE-PLATE-4 and IMAGE-PLATE-2 to
a physically reasonable range; a parameter shown in red has reached its
bound, signaling that the bound is active and should be reviewed.
The ``X-Ray Source'' selector switches the polar view between
sources, and the ``Distortion'' section applies a sample-layer
correction.  (\textit{b})~Polar view with calibration picks (colored
markers) and simulated rings (blue dash-dot lines), above the
azimuthal lineout.}
\label{fig:composite}
\end{figure}

Composite calibration refines the instrument against any number of
powder and Laue overlays at once, each tied to one material, one \xray\
source and one layer of the target assembly.  One mechanism thus covers
both situations common in dynamic compression: diffraction from several
layers of the physics package (for example a powder sample and a
single-crystal window), each with its own correction
(\S\ref{sec:pinhole}), and diffraction from several \xray\ sources,
for which the same material is simply given one overlay per source.
The residuals of all
overlays are joined into one unweighted least-squares objective: powder
picks contribute their $\twotheta$ differences (in radians) and Laue
picks their detector-plane position differences (in millimeters), with
no relative weighting between the two; none of the calibrations in this
paper mixes the two kinds of pick.  As in the
individual methods, points may be auto-picked
or picked by hand; hand picking uses a two-click mode in which the
user left-clicks on the main canvas to freeze the zoom dialog at that
location, then left-clicks on the zoom dialog to place the point
precisely (Fig.~\ref{fig:handpicking}).  For instruments with multiple
\xray\ sources, the composite calibrator handles picks from each source
independently, automatically switching the polar-view projection for
picks associated with different beams (Fig.~\ref{fig:composite}).

Simultaneous refinement against all sources is essential in practice:
calibrating against a single source leaves the geometry
underconstrained.  Rotating a panel about the incident beam through the
sample leaves every Debye--Scherrer cone of that beam unchanged, so one
source cannot fix that rotation, while a second beam from another
direction can.  A single-source fit that appears excellent will
therefore generally degrade the agreement for the other source.  Only a joint refinement, in which
picks from every beam contribute to a common residual, yields a
geometry consistent with all sources at once.

\paragraph{Synthetic test.}
A ground-truth test, distributed with the validation scripts
of \S\ref{sec:validation} \cite{paper_data_zenodo}, quantifies this.  Powder picks were
simulated for a two-source, two-panel instrument of known geometry;
each panel was then displaced by 0.7~mm and its normal tilted by about
0.13$^{\circ}$, and the composite calibrator was run on the picks
alone, refining six degrees of freedom per panel with the beam
parameters and lattice parameter held fixed.  With noise-free picks, refining against both sources recovers the
true geometry to numerical precision (below 1~nm and $10^{-4}$~mdeg);
with Gaussian noise of half a pixel on every pick, it recovers it to
better than 7~\textmu m in translation and 13~mdeg in panel-normal
orientation.  Refining
against the first source alone reproduces that source's noise-free
picks exactly, yet settles 0.34 and 0.66~mm away from the true panel
positions and mispredicts the second source's rings by 88~mdeg in
$\twotheta$: the fit drifts without penalty along the rotation that a
single source cannot fix (the same freedom that forces one tilt to be
fixed in \S\ref{sec:validation}), and how far it drifts depends on
where it starts.  Alternating
single-source refinements, the only option in \textit{HEXRD} before
native multi-source support, merely shift the error from one source to
the other: recalibrating to the second source zeroes its own residual but
displaces the first source's rings by 97~mdeg, returning to the first
displaces the second's by 54~mdeg, and after three passes the geometry
is still 0.25--0.38~mm from the truth.

\paragraph{NIF dual-source shot.}
The real dataset is NIF shot N160517-003 of \citeasnoun{Rygg2020}
(Figs.~\ref{fig:handpicking} and \ref{fig:composite}): a 6~\textmu m Pt
sample ramp-compressed to 180~GPa ($a = 3.542$~\AA) between a diamond
pusher and a 150~\textmu m LiF window, exposed by two co-timed Ge
He-$\alpha$ \xray\ sources (10.25~keV) and recorded on three image
plates through a Ta pinhole 75~\textmu m thick and 350~\textmu m in
diameter.  A simultaneous refinement of both sources converges on this
dataset (116 hand-picked
points across three image plates and both sources, with the
sample-layer correction of \S\ref{sec:pinhole} applied for the
sample's 150~\textmu m standoff from the pinhole, set by the LiF window
between them) with an RMS
residual of 0.18$^{\circ}$ in $\twotheta$ for each source, comparable
with the typical $\twotheta$ uncertainty of 0.2$^{\circ}$ that
\citeasnoun{Rygg2020} report for such data.  This is about one 100~\textmu m
pixel at the $\approx$27~mm plate distances, the precision with which
positions can be picked by hand on the broad, noisy diffraction lines
of these exposures (Fig.~\ref{fig:composite}).  The
refinement varies twelve parameters: the three translations and three
tilts of IMAGE-PLATE-2, the same for IMAGE-PLATE-4 except its vertical
position, and the plate separation, which sets that position through
the TARDIS engineering constraint; IMAGE-PLATE-3 and the beam energies
and directions are fixed.  The plate separation converges onto the
upper bound of its engineering range (23.43~mm, shown in red in
Fig.~\ref{fig:composite}(\textit{a})); releasing the bound moves it by
only 6~\textmu m, to $23.44 \pm 0.07$~mm, and leaves the residuals
unchanged.  Refined against either source alone, even starting from this optimum, the
same picks fit that source more closely (155 and 165~mdeg RMS, against
180 and 183~mdeg jointly) but mispredict the other source's rings by
1.6 and 1.4$^{\circ}$ RMS, with the plates moved by up to 0.7~mm and
tilted by 2.1--2.6$^{\circ}$.  One source leaves each plate's rotation 
about its beam undetermined, so, as in the synthetic test, how far such 
a fit drifts depends on where it starts.  Started instead from the
nominal geometry loaded by the import tool, whose plate tilts differ
from the optimum by about 0.5$^{\circ}$, the joint refinement returns
to the same optimum.  A fit to either source alone must hold the tilts
at assumed values; with them at their nominal values, it fits its own
rings to 195 and 197~mdeg but mispredicts the other source's by 0.63
and 0.58$^{\circ}$, about three pixels.  
A script distributed with the validation scripts reruns
this calibration from the geometry and picks stored in the distributed
state file and confirms that the interactively obtained optimum is
stationary.

The Pt lattice parameter is held throughout at 3.542~\AA, the value
\citeasnoun{Rygg2020} obtained from their own analysis of this shot, so
the calibration reproduces that analysis; where an absolute
$\twotheta$ scale is needed, features of known angle, such as ambient
diffraction from the pinhole substrate, provide it.  The assumed standoff $z_s$ of
the layer correction (\S\ref{sec:pinhole}) enters just as directly:
refitting the lattice parameter alone at the converged geometry shows
that each micrometer of change in $z_s$ moves it by 39~ppm, so $z_s$
must be known to within a few micrometers for the lattice parameter to
be accurate to about $10^{-4}$.  The calibration uses the as-built
standoff.  By the time of the exposure the compression wave has reached
the LiF window and changed its thickness by a few micrometers, which is
not accounted for here, since computing it needs hydrodynamic
simulations of the drive.  The error this introduces is far smaller than the
effects of the spread of densities from the pressure distribution in
the sample, the instrumental broadening and the high background of
these exposures.

\subsection{Structureless calibration}\label{sec:structureless}

When the crystal structure of the sample is unknown at the
high-pressure and temperature conditions achieved during dynamic
compression, and there are no other diffraction features from a crystal
with known structure and density, a structureless calibration can be
performed using only the positions of observed Debye--Scherrer lines
without reference to a structural model.  The user picks points along each ring on every panel with the
two-click zoom tool of Fig.~\ref{fig:handpicking}.  The objective is
simply that all points on one ring map to the same $\twotheta$,
whichever panel or azimuth they came from: in the polar view, each ring
must become a straight vertical line.

In addition to the instrument geometry, the $\twotheta$ positions of the
ideal rings themselves are included as refinable parameters
(\texttt{DS\_ring} parameters), allowing the calibration to
simultaneously determine both the detector geometry and the ring
positions.  Refining both simultaneously is degenerate in at least one
direction (an overall detector-distance/$\twotheta$-scale
trade-off), so in practice the degeneracy is broken by keeping a subset
of the geometric parameters fixed or tightly bounded (\textit{e.g.}\
through the engineering constraints described above), with the ring
positions started near the picked positions (by default at their
mean $\twotheta$) or at values computed from the expected density of
the compressed sample.  When a ring of known
angle is present, such as ambient diffraction from the pinhole, the
powder or composite calibration of \S\ref{sec:powder_cal} or
\S\ref{sec:composite_cal} is used instead.  \citeasnoun{Rygg2020} use the
same equal-$\twotheta$ constraint when no reference lines are
available.

This method is particularly valuable for dynamic compression experiments
where the compressed sample may be in an unknown or novel phase, and
only the requirement that rings be self-consistent is imposed.  It
implicitly assumes that all selected Debye--Scherrer lines originate
from the same layer in the physics package.

\subsection{Validation against independent implementations}
\label{sec:validation}

To test the geometric model and the calibration against independent
implementations, one calibration dataset was analyzed three times over: with \textit{HEXRD}, with \textit{pyFAI}~2026.9.0
\cite{Ashiotis2015pyFAI}, and with \textit{Dioptas}~0.11.0
\cite{Prescher2015dioptas}.  The dataset is NIST standard reference
material SRM~674b (CeO$_2$, certified
$a = 5.41153 \pm 0.00030$~\AA, expanded uncertainty at $k = 2$),
measured at beamline 1-ID-E of the Advanced Photon Source with the
monochromator set to the Au $K$-absorption edge (80.725~keV), on a
single planar GE panel
(2048$\times$2048 pixels of 200~\textmu m) about 1051~mm from the
sample.  A single planar panel is a configuration that all three
codes, including \textit{Dioptas}, describe, so it allows a direct
three-way comparison of the per-panel transfer function from which
every \textit{HEXRD} configuration is built.  (The multi-source and
target-assembly machinery, which those codes do not provide natively,
is checked by the synthetic test of \S\ref{sec:composite_cal} and the
reference comparison of \S\ref{sec:pinhole}; the planar transfer
function at large tilts and other beam directions, and the cylindrical
one, are checked against \textit{pyFAI} in \S\ref{sec:transfer}.)  The panel also has a pincushion
distortion, described by \textit{HEXRD}'s GE\_41RT model (with its
inverse evaluated by Newton iteration, the GE\_41RT\_newton variant,
because the closed-form inverse handles distortion of only one sign
and is exact only when all three radial exponents are 2),
so the
effect of modeling a detector distortion can be measured.  That model maps the radius $r$ of each measured point about
the panel center, at fixed azimuth $\eta_d$, to the corrected radius
$r' = r\,[1 + p_1 (r/r_0)^{n_1}\cos 2\eta_d + p_2 (r/r_0)^{n_2}\cos 4\eta_d
+ p_3 (r/r_0)^{n_3}]$ with $r_0 = 204.8$~mm; here the exponents are
$(n_1, n_2, n_3) = (2, 4, 2)$ and the three coefficients $p_i$ are the
refinable parameters.

For the main comparison, a neutral extraction, independent of all
three codes (a routine in the validation scripts that samples the image
along radial lines across each ring, subtracts a linear background and
locates the peak by a least-squares parabola within one pixel of its
maximum),
fitted 26,318 sub-pixel positions on 24 CeO$_2$ rings directly from
the image (its search windows are centered on the rings predicted by
the stored calibration, but the position fitted within each window
depends only on the image); these agree with \textit{HEXRD}'s own pseudo-Voigt ring
fits to a median of 0.027~pixel, with no systematic offset (median
signed difference 0.002~pixel).  Each package then refined the
sample--detector distance, the beam center, and two detector tilts
against those shared positions, with the same least-squares
minimizer.  Everything else was fixed: the beam
energy at the monochromator setting, the lattice parameter at the
certified value, and the third tilt, which axially symmetric rings
cannot constrain.  \textit{Dioptas} was additionally run end to end
with its own automatic ring search and refinement (2426 points), as a
fully independent analysis.  The two leftmost columns of 
Table~\ref{tab:validation} treat the
distortion equivalently: \textit{HEXRD} applies its model internally,
and for the \textit{pyFAI} ``pre-corrected'' control the same fixed
polynomial was applied to the positions beforehand, so \textit{pyFAI}
refined its geometry without needing to know about the distortion.  The
distortion coefficients themselves were refined on this same image by
\textit{HEXRD}'s powder-calibration workflow, from its own auto-picked
ring fits and with the beam energy and lattice parameter at the fixed
values above, and were then held fixed everywhere; the deposited 
state file stores them.

\begin{table}
\caption{Detector geometry for the GE CeO$_2$ dataset refined
independently by three packages.  \textit{HEXRD} and \textit{pyFAI}
were refined against the same 26,318 independently extracted ring
positions; \textit{Dioptas} used its own automatic ring search.  The
last two columns have no distortion correction.  Beam-center coordinates are
fractional (column, row) pixel indices with the center of the first
pixel at index 0 (0.5 was subtracted from \textit{pyFAI}'s native
values, which place that center at 0.5); the distance is measured along the panel normal, 
and the detector tilt is the angle between
the panel normal and the beam.  The
beam energy and CeO$_2$ lattice parameter were fixed at their physical
values in every refinement; the
quoted standard errors are conditional on those fixed values
(see text), and are given for the
\textit{HEXRD} column.}
\label{tab:validation}
\begin{center}
\begin{tabular}{@{}lcccc@{}}
\hline
\rule{0pt}{2.6ex} & \textit{HEXRD} & \textit{pyFAI} & \textit{pyFAI} & \textit{Dioptas} \\
\hline
\rule{0pt}{2.4ex}ring positions & shared & shared & shared & own search \\
panel distortion & modeled & pre-corrected & not applied & not applied \\
\hline
\rule{0pt}{2.6ex}%
distance (mm) & $1050.7356 \pm 0.0004$ & 1050.7356 & 1051.1617 & 1051.2654 \\
beam center, column (pixel) & 1030.770 & 1030.770 & 1030.773 & 1030.770 \\
beam center, row (pixel) & 1007.334 & 1007.334 & 1007.326 & 1007.309 \\
detector tilt ($^{\circ}$) & $0.18955 \pm 0.00034$ & 0.18955 & 0.18819 & 0.18856 \\
RMS $\twotheta$ residual ($^{\circ}$) & 0.00055 & 0.00055 & 0.00144 & 0.00170 \\
\hline
\end{tabular}
\end{center}
\end{table}

Table~\ref{tab:validation} makes three separate comparisons:
\begin{itemize}
  \item \textbf{Same positions, distortion on both sides}
    (columns 1--2): \textit{HEXRD} and \textit{pyFAI} converge to the
    same geometry.  Evaluated at all 26,318 positions, the two
    geometric models predict $\twotheta$ values agreeing to an RMS of
    $4\times10^{-7}$~mdeg, the level of numerical round-off, a direct
    check of the transfer function of equation~(\ref{eq:transfer})
    against an independent implementation.
  \item \textbf{Withholding the distortion} (column 3) moves the
    refined distance by 0.43~mm, the cost of having no distortion
    model, with everything else held constant.
  \item \textbf{An independent ring search} (column 4):
    \textit{Dioptas} shares \textit{pyFAI}'s geometry engine, so the
    further 0.10~mm difference from column 3 measures how sensitive
    the poorly conditioned distance is to which peaks are picked, and
    to \textit{Dioptas}'s own refinement.  (Its
    search indexed 30 rings; the shared set has 24 because the six
    weakest, (444), (800), (733), (662), (840) and (842), fall below the
    amplitude threshold of the neutral extraction.)
\end{itemize}
In addition, the calibration stored in the state file, which a
different user performed through the \textit{HEXRDGUI} interface using
\textit{HEXRD}'s own auto-picked ring fits instead of the shared
positions, agrees with the refined geometry to 1.7~\textmu m in
distance, 0.02~pixel in beam center and 0.002$^{\circ}$ in tilt; that
state file is the one deposited on Zenodo \cite{paper_data_zenodo}.

\begin{figure}[htbp]
\begin{center}
\includegraphics[width=\columnwidth]{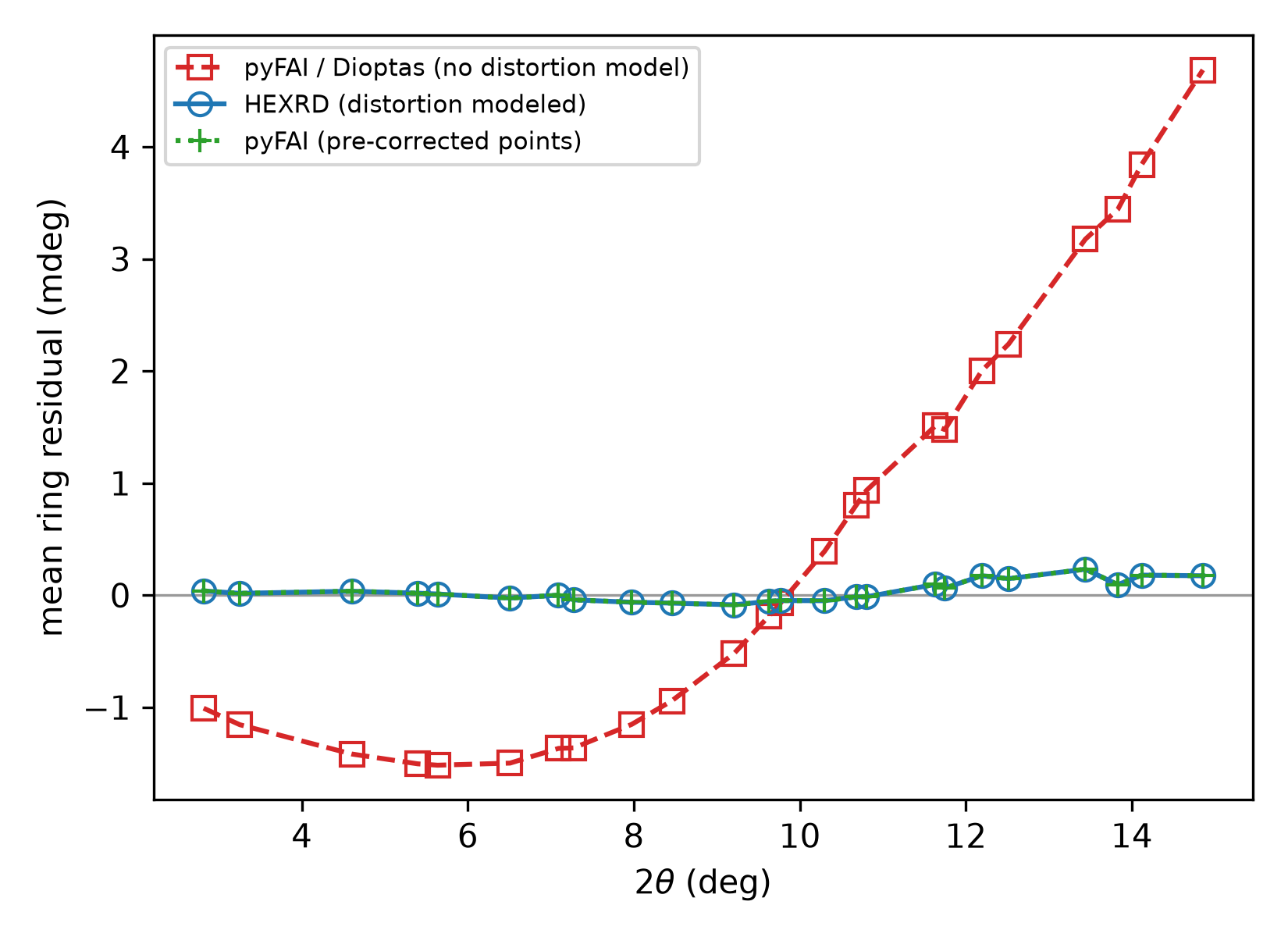}
\end{center}
\caption{Mean $\twotheta$ residual of each ring of the GE CeO$_2$
dataset with respect to the certified CeO$_2$ lattice parameter.  With
the panel distortion modeled, \textit{HEXRD} is flat to within
0.23~mdeg across the whole range; supplying the same pre-corrected
positions to \textit{pyFAI} reproduces that curve exactly.  Without a
distortion model the residual acquires a systematic trend of $-1.5$ to
$+4.7$~mdeg that no adjustment of the geometry can remove (one curve is
plotted for \textit{pyFAI} and \textit{Dioptas}, which share the same
geometry engine; it is the \textit{pyFAI} refinement of the shared
positions).}
\label{fig:validation}
\end{figure}

The distortion of this panel is a pincushion of about 0.5~pixel
(100~\textmu m) at a radius of 200~mm, rising to 1.1~pixels
(220~\textmu m) in the corners where the high-angle rings fall.  \textit{pyFAI} and \textit{Dioptas}
apply a detector distortion as a fixed, separately measured correction,
such as a spline file, and none was used here.
Without a distortion model, their refinement absorbs what it can into
the geometry (the azimuthally symmetric part into the distance,
the 0.43~mm above) and
leaves the rest as the ring-by-ring trend of
Fig.~\ref{fig:validation}.  Supplying the same codes
with pre-corrected positions removes the trend entirely and reproduces
the \textit{HEXRD} result, so the discrepancy comes from the missing
distortion model and not from the underlying geometry.  On this
detector, the three-coefficient distortion model removes a
ring-to-ring systematic error of up to 4.7~mdeg, eight times the
0.55~mdeg scatter of individual positions and far larger than the uncertainty of
each ring mean.  Its coefficients were determined on the same image,
so this measures how much of the residual the model accounts for; the
comparison of the geometric models themselves (columns 1 and 2 of
Table~\ref{tab:validation}) is independent of it.

Two further points set the scope of the comparison.  First, the
lattice parameter, the beam energy and the distance are mutually
degenerate for a monochromatic powder measurement (refining $a$ and
the distance together gives a correlation coefficient of $+1.00$), so
two of the three must be fixed externally.  The comparison therefore
tests the consistency of the geometric models, with the absolute
$\twotheta$ scale set by the fixed values, and the standard errors in
Table~\ref{tab:validation} are conditional on the fixed values (and
somewhat optimistic, since neighboring positions along a ring are not
independent); this is the same degeneracy discussed in \S\ref{sec:structureless}.
Second, the residual floor of 0.00055$^{\circ}$ is set by the scatter
of the extracted peak positions; the two geometric models agree to
numerical round-off, far below it.  The
extraction and comparison scripts, the converters between
\textit{HEXRD} instrument files and \textit{pyFAI} \texttt{poni} files,
and the \textit{Dioptas} calibration with its picked positions are
deposited on Zenodo \cite{paper_data_zenodo} so the comparison can be
repeated.

\section{Whole-pattern peak fitting}\label{sec:wppf}

\begin{figure}[htbp]
\begin{center}
\begin{minipage}[b]{0.417\columnwidth}\centering(a)\\[2pt]
\includegraphics[width=\linewidth]{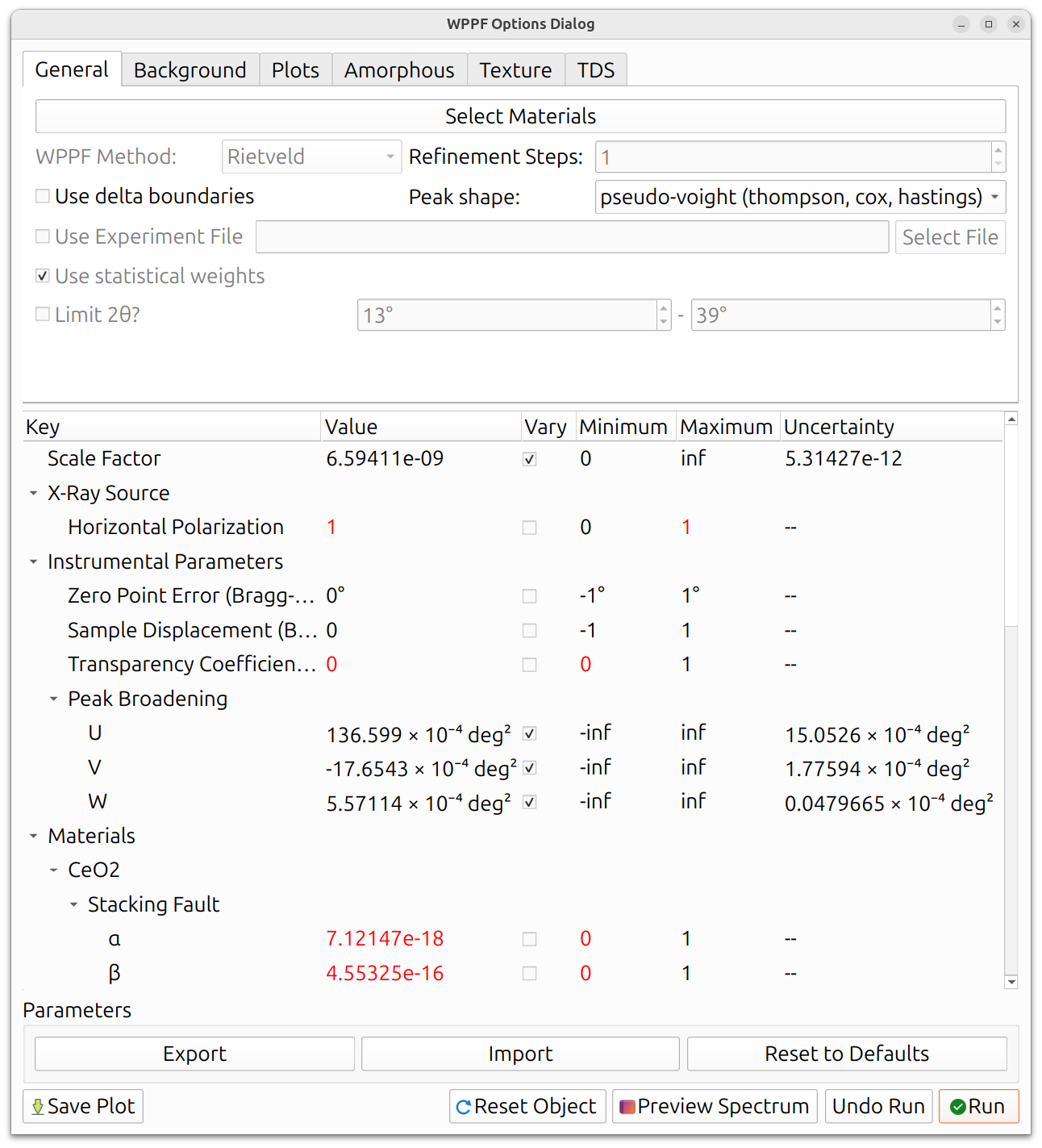}\end{minipage}\hfill
\begin{minipage}[b]{0.543\columnwidth}\centering(b)\\[2pt]
\includegraphics[width=\linewidth]{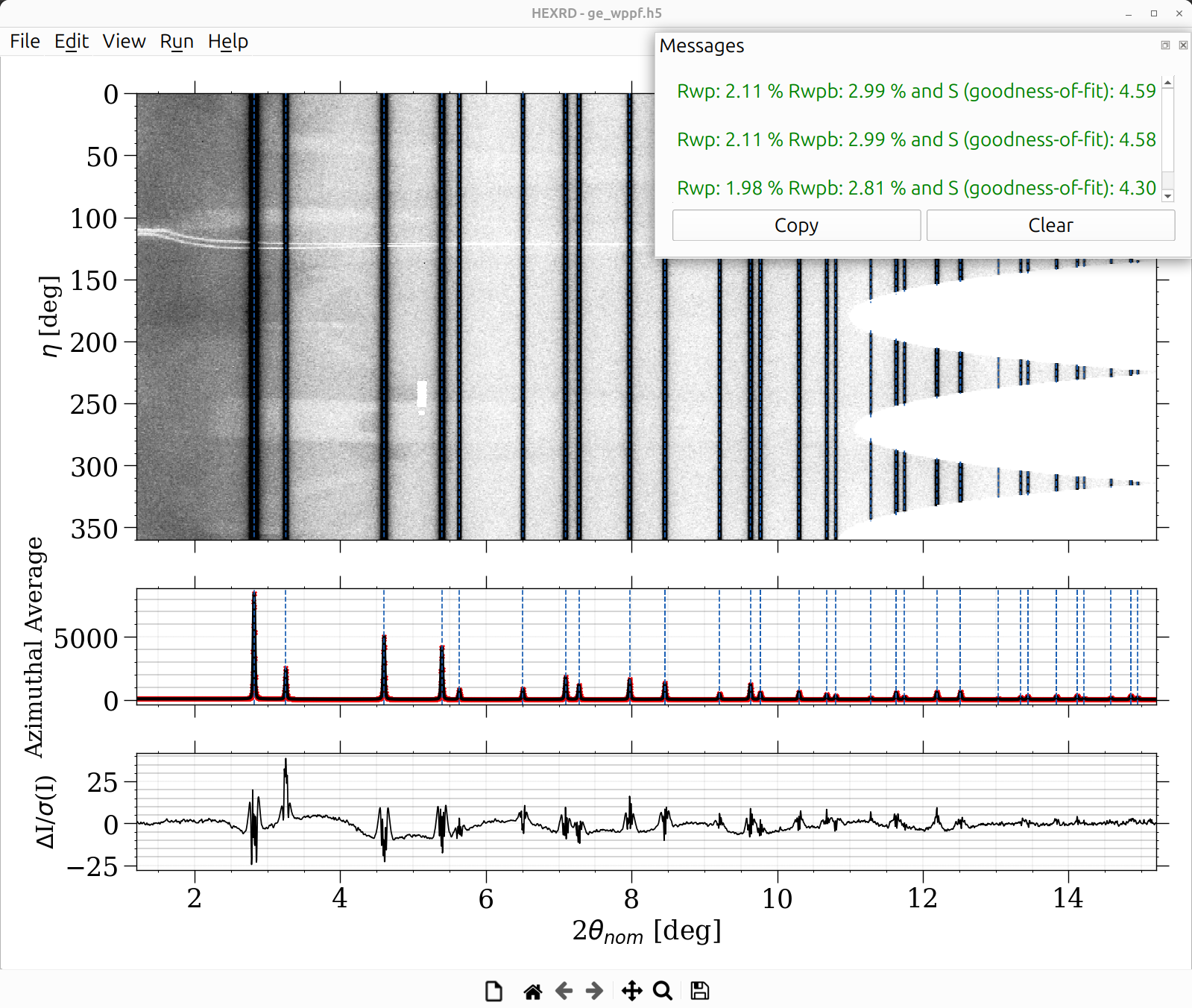}\end{minipage}
\end{center}
\caption{Whole-pattern peak fitting of the GE CeO$_2$ dataset.
(\textit{a})~WPPF options dialog, with tabs for general settings,
background, plots, amorphous contributions, texture and thermal
diffuse scattering.  The parameter tree lists the refinable
parameters, whose number varies with the features enabled, with columns
for value, refinement toggle, bounds and uncertainty (standard error).
(\textit{b})~Polar view with simulated line positions (vertical blue
lines) and a message log reporting $R_{wp}$, $R_{wpb}$ and $S$ for
each iteration of a Rietveld fit of this lineout (the closure test and
the \textit{GSAS-II} comparison in the text are Le~Bail fits of the same
data).  The azimuthal lineout
below it shows the measured intensities (black) overlaid with the
fitted pattern (red), with the difference plot underneath.}
\label{fig:wppf}
\end{figure}

After instrument calibration, \textit{HEXRD} provides whole-pattern
peak fitting (WPPF) for material characterization from the azimuthally
averaged $\twotheta$ pattern (Fig.~\ref{fig:wppf}).  Two methods are
implemented: Le~Bail fitting \cite{LeBail1988}, which extracts peak
intensities without a crystal-structure model, and Rietveld refinement
\cite{Rietveld1969}, which uses a full structural model for phase
quantification and structure refinement.  The Rietveld intensities
include the Lorentz--polarization factor, so the Lorentz and
polarization corrections of \S\ref{sec:corrections} should be left
off a pattern that is to be refined this way.

Five pseudo-Voigt peak-shape models are available: the
Thompson--Cox--Hastings (TCH) profile \cite{Thompson1987}, the
Finger--Cox--Jephcoat (FCJ) profile with axial-divergence correction
\cite{Finger1994}, a pink-beam pseudo-Voigt peak shape
\cite{VonDreele2021pink}, and two one-sided profiles in which the
pseudo-Voigt is convolved with a single exponential or with a half
Gaussian; the latter approximates the spread of lattice parameters in a
sample with a temperature gradient, as when it is heated by the
\xray\ source.  In all models the Gaussian width
follows the Caglioti parameterization \cite{Caglioti1958} (plus a
Gaussian size term) and the Lorentzian width carries Scherrer size and
microstrain terms.  Anisotropic peak broadening may be modeled using
the phenomenological approach of \citeasnoun{Stephens1999}.

Refinement minimizes the weighted sum of squared residuals over the
points of the azimuthally averaged lineout. The basic weighting
scheme uses $w_i = 1/I_{\mathrm{obs},i}$. Points with zero intensity
are assigned $w_i = 0$ and excluded from the number of observations,
$N$. A lineout containing negative intensities, for example after
background subtraction, is first shifted by a constant offset to
make its intensities non-negative. This shift is a numerical
convention: it does not restore Poisson counting statistics and
changes the intensity-based weights without changing the underlying
noise variance. 

The weighted profile $R$-factor, expected $R$-factor, reduced
chi-squared, and goodness-of-fit are defined as
\begin{equation}
  \begin{aligned}
    R_{wp} &=
    \left[
      \frac{\sum_i w_i
        \left(I_{\mathrm{calc},i}-I_{\mathrm{obs},i}\right)^2}
           {\sum_i w_i I_{\mathrm{obs},i}^2}
    \right]^{1/2},
    &
    R_{\mathrm{exp}} &=
    \left[
      \frac{N-P}{\sum_i w_i I_{\mathrm{obs},i}^2}
    \right]^{1/2},
    \\
    \chi^2 &=
    \left(\frac{R_{wp}}{R_{\mathrm{exp}}}\right)^2
    =
    \frac{\sum_i w_i
      \left(I_{\mathrm{calc},i}-I_{\mathrm{obs},i}\right)^2}
         {N-P},
    &
    S &= \sqrt{\chi^2}
       = \frac{R_{wp}}{R_{\mathrm{exp}}},
  \end{aligned}
  \label{eq:rwp}
\end{equation}
where $P$ is the number of independently refined parameters and the
sums run over the included observations.

In high-energy-density experiments, diffraction lineouts frequently
contain a large background from the target assembly and scattered
drive radiation. This background increases the denominator of
$R_{wp}$ and can therefore obscure discrepancies between the
calculated and observed Bragg peaks. To assess profile agreement
relative to the intensity above background, \textit{HEXRD} also
reports the background-adjusted statistic
\begin{equation}
  R_{wpb} = \left[
    \frac{\sum_i w_i \left(I_{\mathrm{calc},i}-I_{\mathrm{obs},i}\right)^2}
         {\sum_i w_i \left(I_{\mathrm{obs},i}-I_{\mathrm{bkg},i}\right)^2}
  \right]^{1/2},
  \label{eq:rwpb}
\end{equation}
where $I_{\mathrm{bkg},i}$ is the refined or estimated background
and the weights are the same as in equation~(\ref{eq:rwp}).
The squared denominator terms accommodate locally negative
background-subtracted intensities; $R_{wpb}$ is non-negative and
finite provided that the denominator is nonzero. Its numerator
retains the residuals over the full fitted profile. For patterns
dominated by a positive background, $R_{wpb}$ is generally larger
than $R_{wp}$ and provides a more informative measure of the
residual relative to the scattering above background.
Available background representations include cubic spline
interpolation through interactively selected control points,
Chebyshev polynomials with refinable coefficients, and the SNIP
algorithm \cite{Ryan1988snip}.

The interpretation of $\chi^2$ as a statistical goodness-of-fit
measure with an expected value near unity requires appropriate
inverse-variance weights, approximately independent observations,
and an adequate model. The basic weights generally do not meet
these requirements for area-detector lineouts. Each lineout point
averages contributions from multiple detector pixels, and the
recorded intensities may be expressed in analog-to-digital units
rather than photon counts. Interpolation can also introduce
correlations between samples. Consequently, the variance of a
lineout point is not generally equal to its intensity, and the
basic weighting scheme can yield $\chi^2 \ll 1$ without implying
an exceptionally accurate fit. For fixed preprocessing and
weighting, $R_{wp}$ and the background-adjusted $R_{wpb}$ defined
above remain useful descriptive measures of profile agreement.

An optional correction, enabled by default in \textit{HEXRDGUI},
multiplies the basic weight at each lineout point by the number of
contributing pixels, $N_i$, giving
$w_i = N_i/I_{\mathrm{obs},i}$. For an uncorrected arithmetic average
of $N_i$ independent Poisson counts, the estimated variance is
$\sigma_i^2 = I_{\mathrm{obs},i}/N_i$, so this correction gives the
appropriate inverse-variance weight. This correction remains
approximate for integrating detectors and processed images.
Detector gain, intensity corrections, and correlations must also be
considered when estimating uncertainties. In particular,
\textit{HEXRD} takes $N_i$ as the number of valid polar-view pixels,
and polar pixels finer than the detector pixels resample the same
detector pixels, so $N_i$ can exceed the number of independent
measurements; $\chi^2$ and $S$ therefore depend on the azimuthal bin
size and are comparable only between fits made with the same binning.
Scaling the weights by $N_i$ alone does not guarantee a statistically
calibrated $\chi^2$: for the Rietveld fit of the high-count data of
Fig.~\ref{fig:wppf}, with 1$^\circ$ azimuthal bins, $S = 4.3$, because the residual is
dominated by model error rather than counting noise.  In practice we
therefore recommend comparing fits by $R_{wp}$ and $R_{wpb}$ computed
with the same preprocessing and binning, and reading $S$ as a
diagnostic of model error rather than as a calibrated statistic.

Amorphous scattering from the sample or target assembly can be
modeled using split-Gaussian, split-pseudo-Voigt, or experimentally
measured profiles. During refinement, \textit{HEXRD} reports the
degree of crystallinity (DOC), defined here as the integrated
intensity assigned to the crystalline peaks divided by the sum
of the integrated crystalline and amorphous intensities over
the analyzed angular range. This quantity measures the relative
contributions of the fitted scattering components. Its
interpretation as a crystalline phase fraction requires
appropriate normalization of their relative scattering
intensities and accounting for other sources of diffuse scattering 
such as thermal diffuse scattering (TDS) and incoherent scattering, 
even when the crystalline and amorphous components have identical 
chemical compositions.

For Rietveld refinement, preferred orientation can be modeled
using either the March--Dollase model \cite{Dollase1986} or a
spherical-harmonic expansion of the general axis distribution
function \cite{VonDreele1997}. Thermal diffuse scattering (TDS)
can be included using the analytical Warren model
\cite{Warren1969} for monatomic FCC and BCC structures or an experimentally measured
diffuse-scattering profile. Models for more complex structures, such as that of \citeasnoun{sakuma1995}, are being implemented. Accounting for TDS is particularly important at the high temperatures reached during dynamic compression experiments \cite{Wark2025}.

Two tests validate the fitting module.  The first is a closure test
against the calibration of \S\ref{sec:validation}: a Le~Bail
refinement of the CeO$_2$ lattice parameter against the azimuthally
averaged lineout of the calibrated GE instrument, with the beam energy
fixed and the lattice parameter started 1200~ppm away from its true
value, recovers the certified value, at which the calibration held
it, to 0.3~ppm at $R_{wp} = 2.1\%$.  The calibration, the polar
mapping, and the peak-shape model are therefore mutually consistent.

The second test is a comparison with \textit{GSAS-II}
\cite{Toby2013gsas2}, version 5.8.4, in which both codes fitted the same
two patterns by the Le~Bail method.  The first pattern is synthetic and
noise-free, calculated with the symmetric Thompson--Cox--Hastings
profile \cite{Thompson1987} that \textit{HEXRD} fits, with a known
lattice parameter, which \textit{HEXRD} recovers to within 0.1~ppm.
The profile of \textit{GSAS-II} includes axial-divergence asymmetry
\cite{Finger1994}, with a minimum of $\mathrm{SH/L} = 0.002$, which
shifts the low-angle peaks of this pattern by the equivalent of a 23~ppm
change in wavelength.  With the wavelength adjusted by this amount, the
two lattice parameters differ by 0.6~ppm.  The Caglioti width coefficients are defined
differently in the two codes, so the peak widths were compared directly.
The full widths at half maximum of 12 isolated reflections in the fitted
patterns are within 0.02\% (\textit{HEXRD}) and 0.5\% (\textit{GSAS-II})
of those in the synthetic pattern.

The second pattern is the measured GE lineout used in the closure test.
Here the lattice parameters from the two codes differ by 62~ppm.  The
minimum asymmetry, calculated for the peak widths
of this pattern, accounts for 24~ppm of the difference.  The remaining
38~ppm reflects other differences between the two codes' profile
models, so a lattice
parameter from this pattern depends on the profile model at the level
of a few tens of ppm.  Both patterns and the fitting scripts are included with the
validation scripts.  Neither pattern tests the calculated intensities,
on which Rietveld refinement depends.  For these, the two codes were
given the same CeO$_2$ structure, and the relative intensities of all 18
reflections with $d \geq 0.8$~\AA\ agree to 0.4\% at both 12.5 and
80.7~keV.  The difference increases smoothly with scattering angle,
consistent with the different atomic scattering-factor
parameterizations used by the two codes.  Preferred orientation and
thermal diffuse scattering were not tested.

\section{Program overview}\label{sec:program}

This article describes \textit{HEXRD} version~0.10.7
\cite{hexrd_zenodo} and \textit{HEXRDGUI} version~1.2.6
\cite{hexrdgui_zenodo}; the archived releases corresponding to these
versions are deposited on Zenodo at the DOIs given in the reference
list, and each is also available as a tagged release on GitHub.

\textit{HEXRD} is implemented as a Python library (requiring Python~3.11
or later; versions through 3.14 are supported) with performance-critical routines
optimized using Numba \cite{Lam2015numba} just-in-time compilation and
core coordinate transform operations implemented in C/C++.
\textit{HEXRDGUI} provides an interactive
graphical interface built on the Qt6 framework (PySide6) that uses
the \textit{HEXRD} library as its computational backend.  This separation
of library and application enables both programmatic analysis via the
Python API for scripting and batch processing, and point-and-click
operation via the GUI for interactive exploration.

\begin{figure}[htbp]
\begin{center}
\includegraphics[width=\columnwidth]{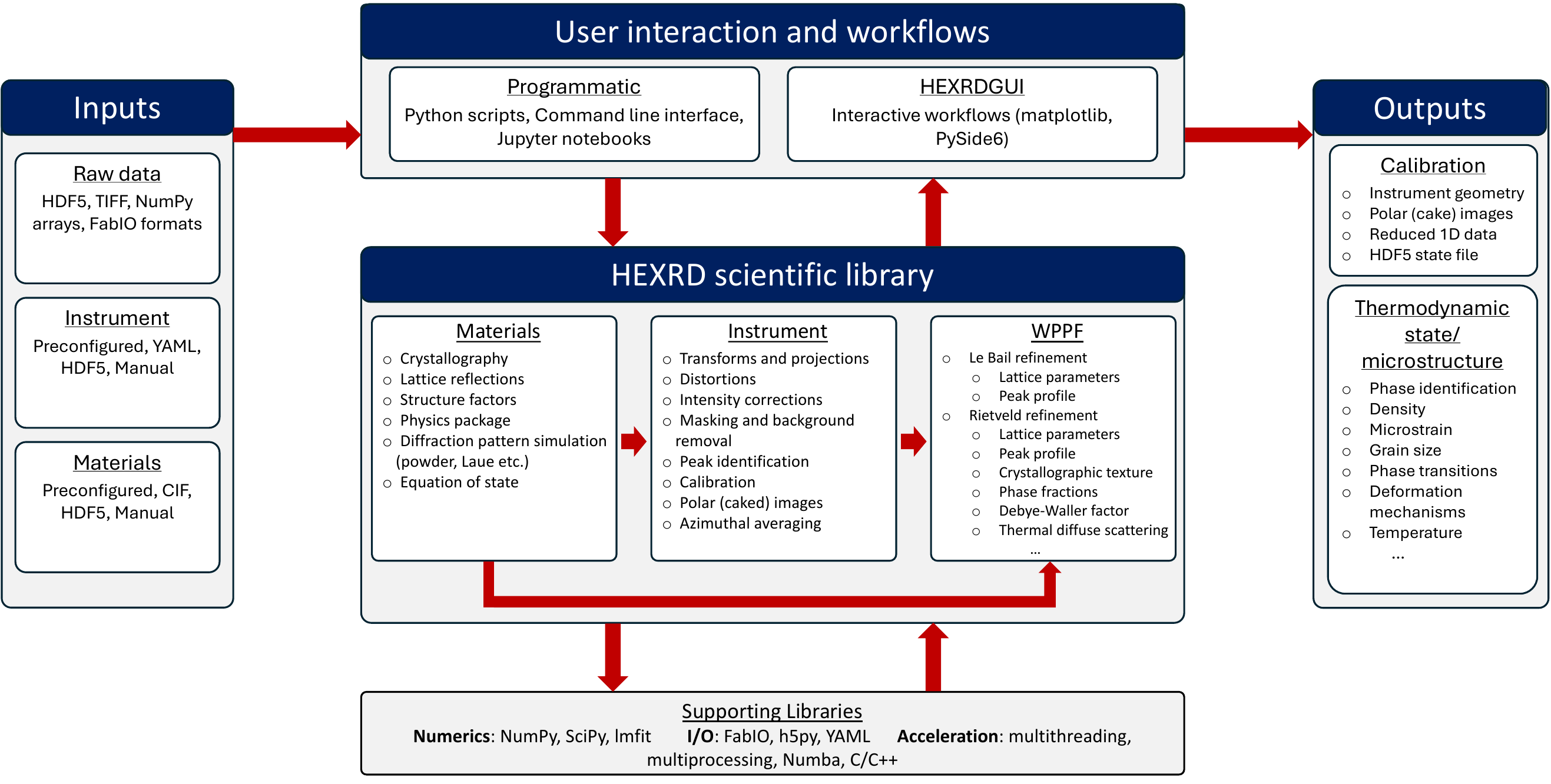}
\end{center}
\caption{Architecture of the \textit{HEXRD} toolkit.  Raw data,
instrument configurations, and material definitions (left) are consumed
either programmatically or through the \textit{HEXRDGUI} desktop
application, both of which drive the same \textit{HEXRD} scientific
library.  Within the library, the materials, instrument, and
whole-pattern fitting components build on shared supporting libraries
(NumPy, SciPy, \textit{lmfit}, FabIO) with Numba and C/C++ kernels for
performance-critical routines.  Outputs (right) comprise the calibrated
instrument geometry and reduced data, together with the derived
thermodynamic state and microstructure: phase identification and
transitions, density, microstrain and grain size, and, through the
refined texture, Debye--Waller factors and thermal diffuse scattering,
indicators of deformation mechanisms and temperature (the equation of
state, in contrast, takes the temperature as an input).  Following the focus of this article,
the diagram shows the capabilities relevant to dynamic compression
experiments; the ff-HEDM workflows that \textit{HEXRD} also provides
(see Introduction) are omitted.}
\label{fig:architecture}
\end{figure}

Figure~\ref{fig:architecture} shows how the two packages fit together.
All of the scientific functionality (the materials and
equation-of-state models of \S\ref{sec:materials}, the geometric model,
distortion and intensity corrections, and calibration engines of
\S\ref{sec:geometry}, \S\ref{sec:corrections} and
\S\ref{sec:calibration}, and the whole-pattern fitting module of
\S\ref{sec:wppf}) lives in the \textit{HEXRD} library and is reachable
both programmatically (Python scripts, Jupyter notebooks, and a command
line interface) and through \textit{HEXRDGUI}, which contributes the
interactive views, overlay and mask management, point picking, workflow
orchestration, and session state while calling into the same library
routines that a script would.  Analyses set up interactively can
therefore be reproduced programmatically: the instrument
configuration, materials, masks and fit parameters are all saved to
files that a script can read, so nothing depends on the interface.

\subsection{Software architecture and dependencies}

The \textit{HEXRD} library is organized into subpackages by technique:
\texttt{hexrd.core} holds the shared foundations described next, the
powder workflows and whole-pattern fitting live in
\texttt{hexrd.powder}, Laue simulation and fitting in
\texttt{hexrd.laue}, and the pinhole and physics-package corrections
in \texttt{hexrd.hed}, while the ff-HEDM workflows occupy their own
\texttt{hexrd.hedm} subpackage.
Within \texttt{hexrd.core}, the \texttt{transforms} module
implements the core geometric calculations, including the transfer
functions described in \S\ref{sec:geometry} and coordinate mappings
between laboratory, detector, and angular representations.  Vectorized
operations over arrays of diffraction vectors, including the transfer
functions and their inverse, are compiled C and C++ kernels; as a
representative figure, mapping every pixel of a 2048$\times$2048 panel
to diffraction angles takes about 0.2~s on one core of an Intel Core
Ultra~9 285K desktop processor.  Numba \cite{Lam2015numba} accelerates
other vectorized routines (rotations, distortion models, fitting).  The
\texttt{material} module provides crystallographic definitions, structure
factor calculations, and equation-of-state models.  The \texttt{fitting}
module implements the calibration algorithms and their parameter
handling; whole-pattern fitting lives in \texttt{hexrd.powder}.  The \texttt{instrument} module defines
detector models through an abstract base class, allowing arbitrary
panel shapes to be implemented by subclassing; planar and
cylindrical detectors are provided, and the framework accommodates
future extensions to other surfaces.

Core dependencies include NumPy and SciPy for numerical operations,
\textit{lmfit} \cite{Newville2014lmfit} for the least-squares
refinements of \S\ref{sec:calibration} and \S\ref{sec:wppf}, and the
FabIO library \cite{Knudsen2013fabio} for image formats.  \textit{HEXRDGUI}
adds PySide6 for the interface framework and Matplotlib for plotting.
A C/C++ compiler (gcc or Visual Studio) is required only when
installing from source (\textit{e.g.}\ via pip) to build the compiled
extensions; the binary conda packages and standalone bundles described
below ship prebuilt extensions.

\subsection{Distribution and platform support}

Binary packages for \textit{HEXRD} and \textit{HEXRDGUI} are distributed
via conda from the \texttt{hexrd} channel, and
as standalone application bundles for Windows (MSI installer and
portable .zip), macOS (disk images for both Apple Silicon and Intel
architectures), and Linux (tarball).  Installation via conda requires approximately 2~GB of
disk space for the full environment including dependencies.  Standalone
bundles include all dependencies and do not require a separate Python
installation.  The GUI is optimized for displays with resolution
$\geq$1920$\times$1080 pixels; smaller displays are usable but may
require scrolling within dialogs.

Both packages are under active development with stable releases tagged
and distributed under the \texttt{hexrd} channel's \texttt{main} label,
and pre-release versions
available from the \texttt{hexrd/\allowbreak label/\allowbreak prerelease} channel for early
testing.  The open-source development model allows community
contributions through GitHub pull requests; all contributions must use
the BSD 3-Clause license consistent with the upstream code
(LLNL-CODE-819716).  Documentation for \textit{HEXRDGUI} is hosted at
\mbox{\texttt{hexrdgui.readthedocs.io}}, including installation,
configuration and calibration guides and tutorial workflows.

\subsection{Availability}\label{sec:availability}

\textit{HEXRD} \cite{hexrd_zenodo} and \textit{HEXRDGUI}
\cite{hexrdgui_zenodo} are released under a BSD~3-Clause license, and
the source code for both is hosted in the HEXRD organization on
GitHub (\texttt{github.\allowbreak com/\allowbreak HEXRD}).

\section{Suite of examples}\label{sec:examples}

To illustrate the range of experimental configurations supported by
\textit{HEXRD}, we present examples from six different instrument
platforms spanning several major dynamic compression facilities.  Each
highlights a different part of the workflow: pink-beam profiles at DCS,
dual-source calibration with a cylindrical plate at TARDIS, liquid and
amorphous scattering on multi-panel detectors at the European XFEL and
MEC, the sample-layer and pinhole corrections at PXRDIP, and rigid-body
calibration of a five-camera array at FIDDLE.

\subsection{DCS at APS}

\begin{figure}[htbp]
\begin{center}
\includegraphics[width=0.6\columnwidth]{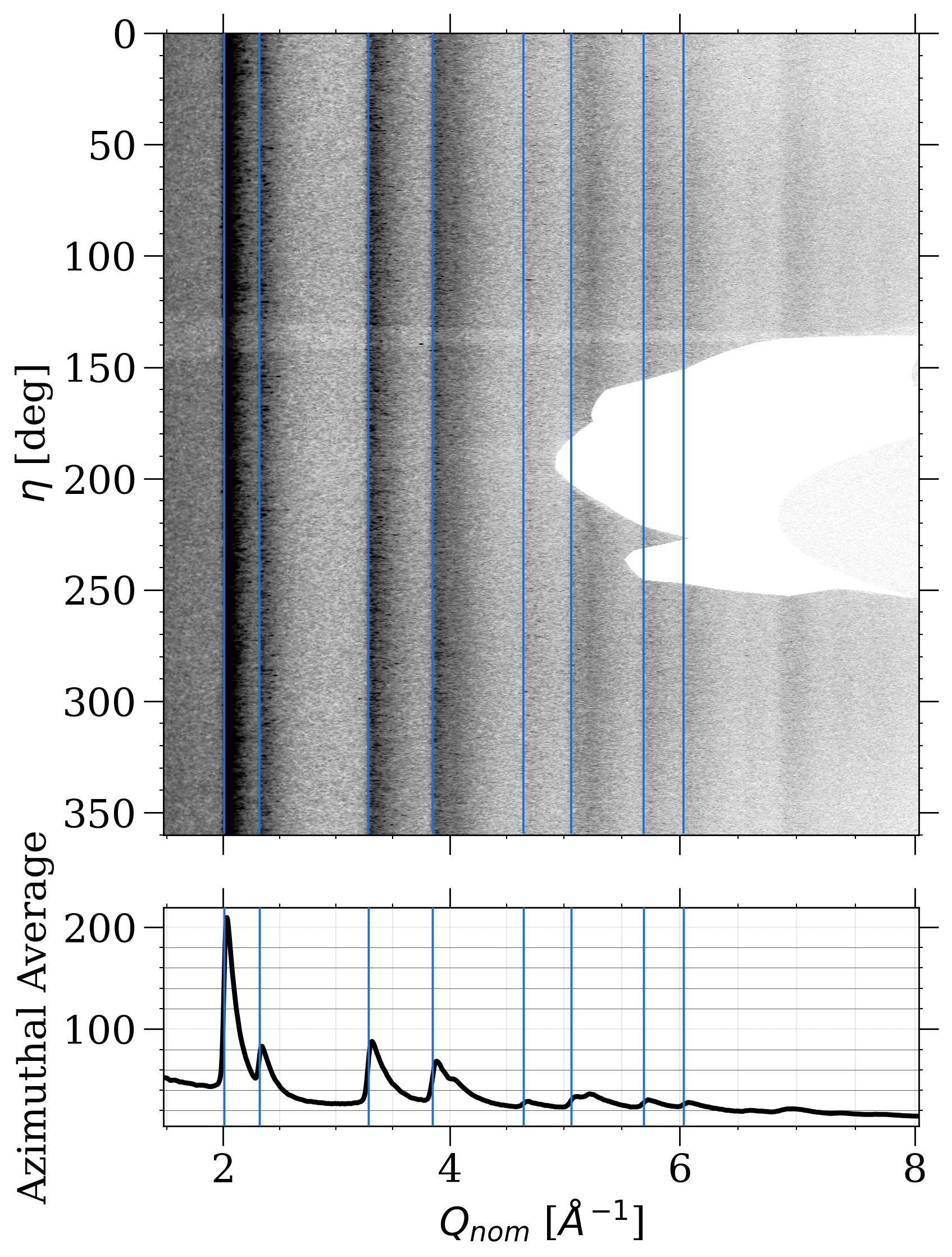}
\end{center}
\caption{DCS pink-beam CeO$_2$ calibration example with a Rayonix
detector, shown as a polar view (azimuth $\eta$ vs.\ nominal momentum
transfer $Q$, computed from the measured angles with the nominal beam
energy) with simulated ring positions (vertical blue lines) and
the azimuthal average below.  The asymmetric peak shapes characteristic
of the pink-beam spectrum (sharp onsets with extended tails toward
higher $Q$) are modeled by the pink-beam peak profile, used here for the
calibration peak fits and also available for whole-pattern fitting.  The
simulated lines mark the ring positions at
the nominal energy and therefore sit at the sharp low-$Q$ onset of
each peak, slightly below the apparent peak maximum, as expected for a
pink beam.}
\label{fig:example_dcs}
\end{figure}

The Dynamic Compression Sector beamline at the Advanced Photon Source
provides pink-beam or monochromatic \xray\ diffraction for gas-gun- and laser-driven experiments.
Figure~\ref{fig:example_dcs} shows a pink-beam CeO$_2$ calibration
dataset acquired with a Rayonix detector and calibrated with the powder
calibration workflow (Fig.~\ref{fig:fastpowder} shows the same dataset
during calibration).  For whole-pattern fitting of such data the WPPF
module provides the pink-beam peak-profile model of
\citeasnoun{VonDreele2021pink}, which was used for the structural
analysis of shock-compressed iron at DCS in \citeasnoun{Singh2023iron}.

\subsection{TARDIS at NIF}

\begin{figure}[htbp]
\begin{center}
\includegraphics[width=0.49\columnwidth]{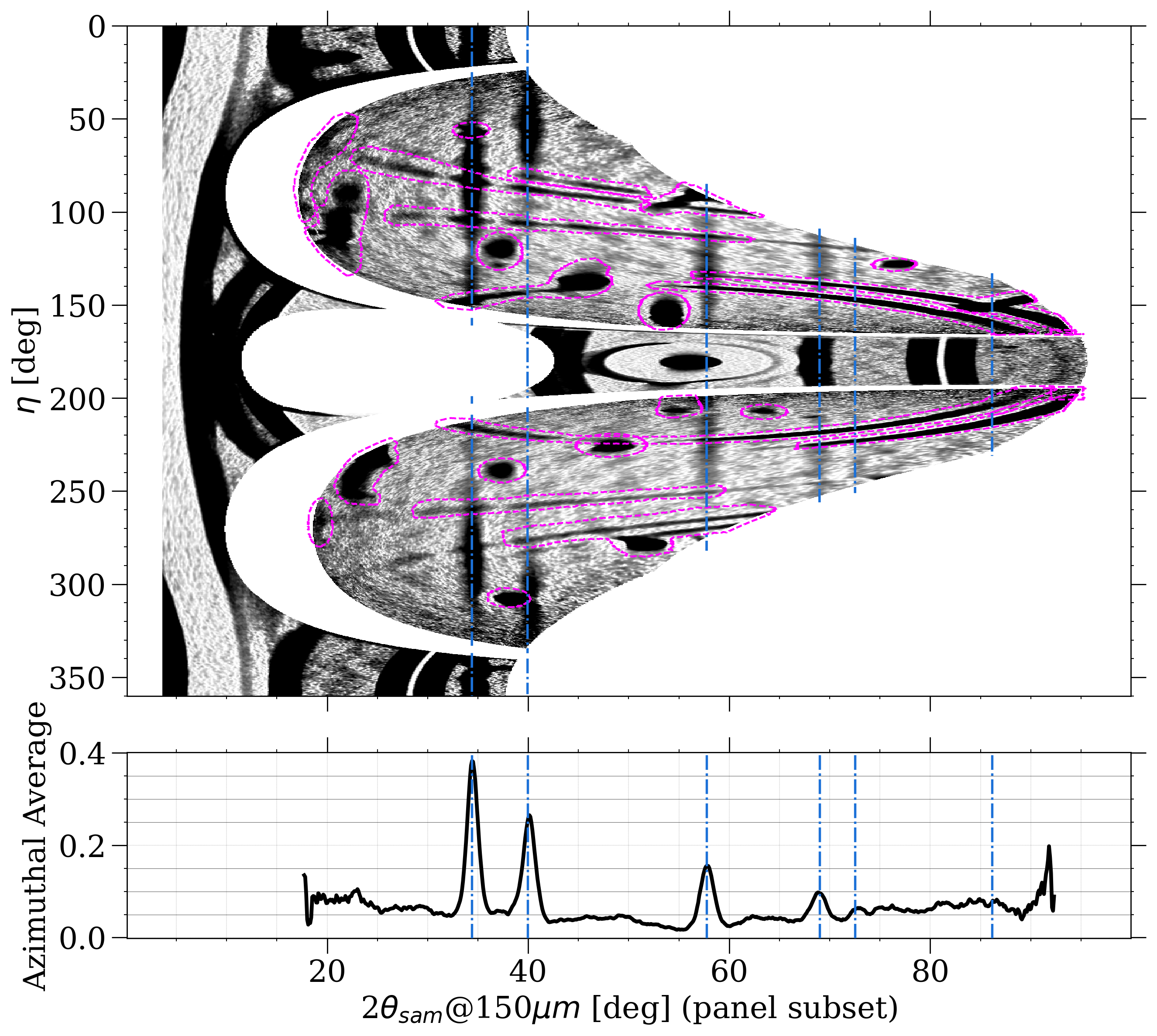}\hfill
\includegraphics[width=0.49\columnwidth]{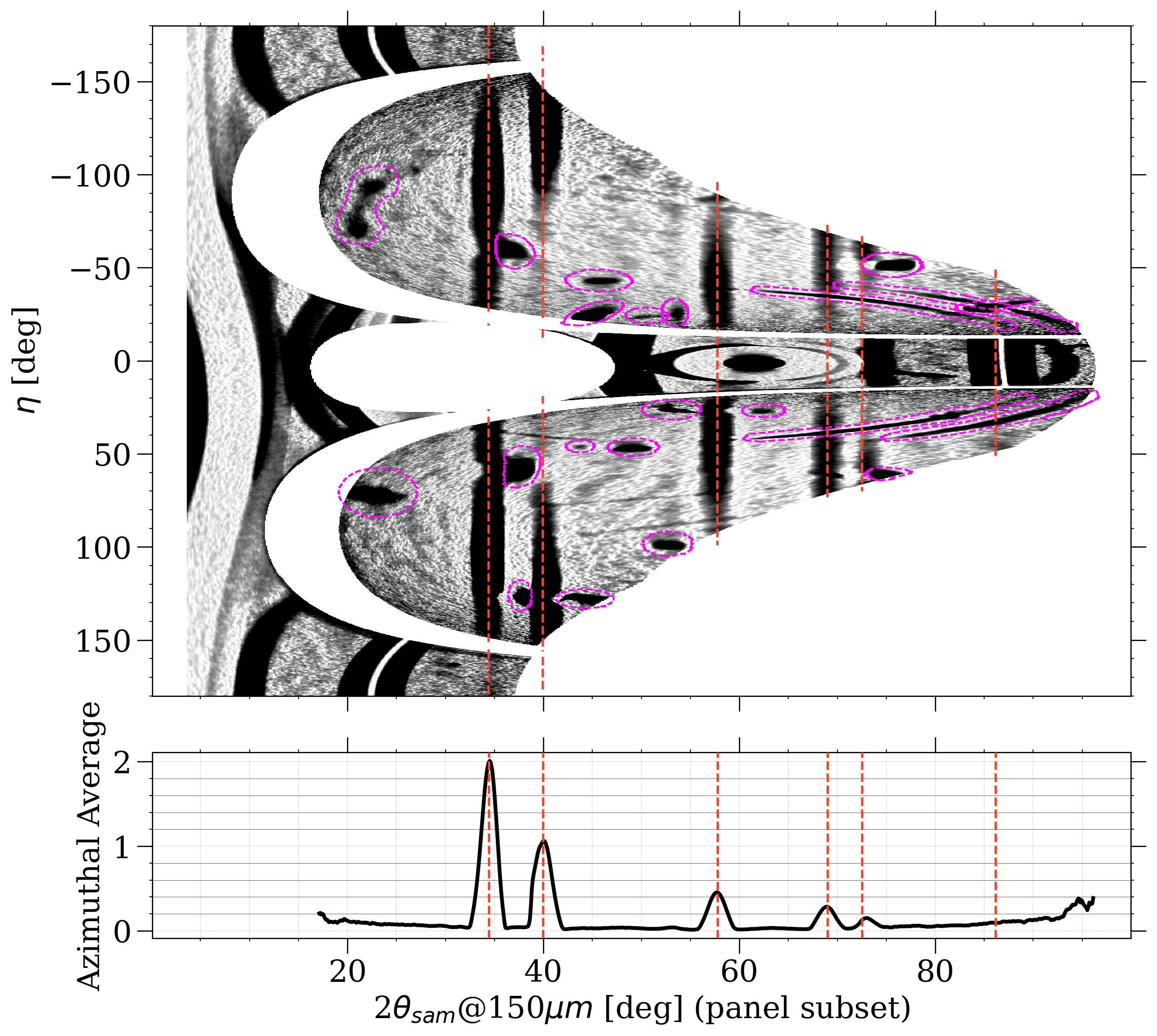}\\[6pt]
\includegraphics[width=0.55\columnwidth]{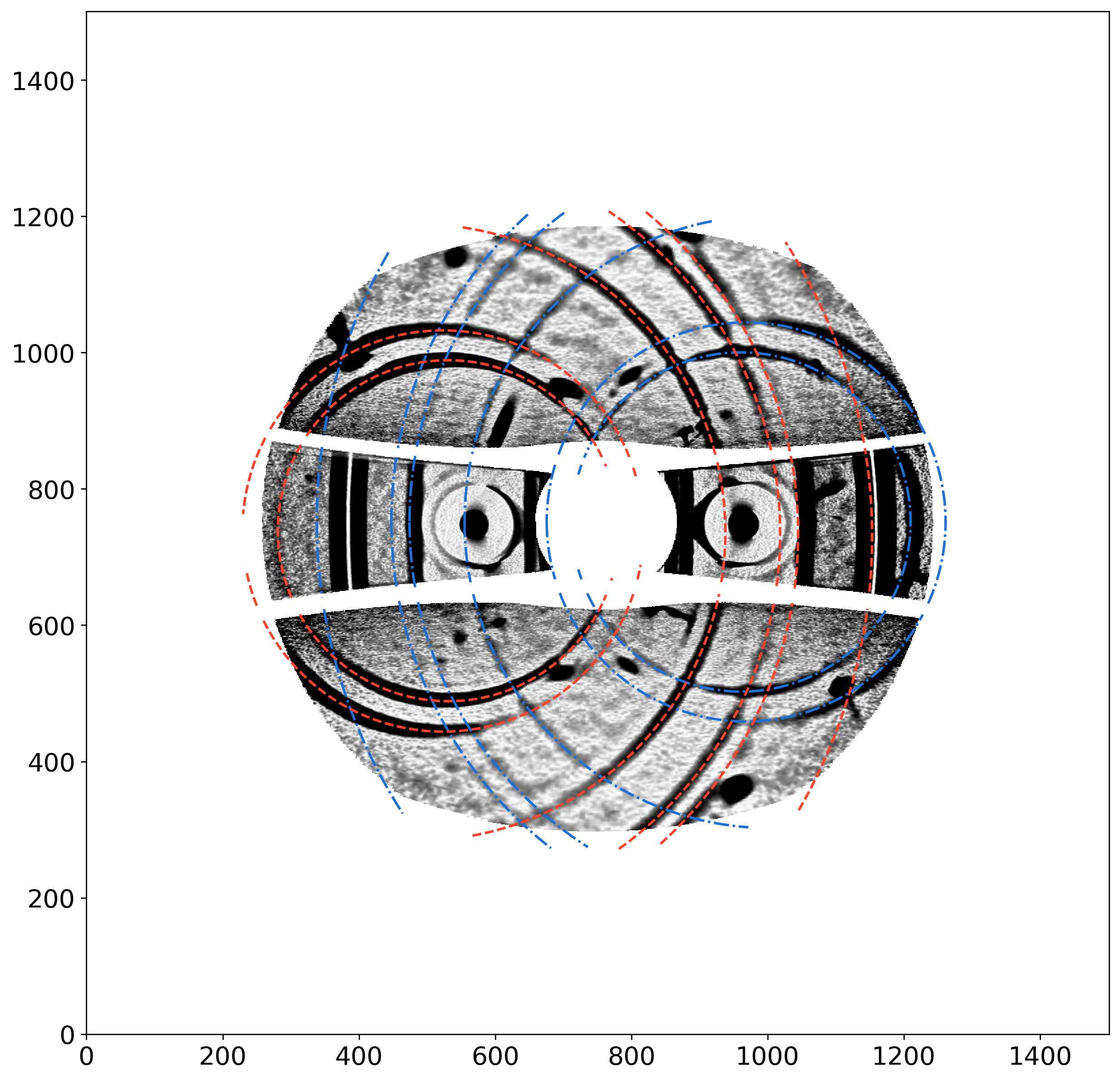}
\end{center}
\caption{TARDIS examples showing ramp-compressed Pt at NIF.  Blue and
red lines are simulated Pt diffraction patterns from XRS1 and XRS2,
respectively.  Top left: polar view centered at XRS1.  Top right: polar view
centered at XRS2.  Both polar views have a sample layer distortion
applied, with IMAGE-PLATE-3 excluded from the azimuthal lineout;
magenta dashed lines indicate manually masked regions.  Bottom:
stereographic view showing simulated patterns from both sources.}
\label{fig:example_tardis}
\end{figure}

The TARDIS platform at NIF \cite{Rygg2020} uses image plates arranged
around the target with a pinhole collimator and quasi-monochromatic
\xray\ sources generated by laser irradiation of metal foils.  At LLNL,
TARDIS and PXRDIP data have also been analyzed with two
platform-specific \textit{IGOR Pro} programs, \textit{AnalyzeTARDIS}
and \textit{AnalyzePXRDIP} \cite{Millot2026}, which share a common
origin but have since diverged; \textit{HEXRD} treats both platforms,
and the others in this section, with one geometric model.
Figure~\ref{fig:example_tardis} shows data from a Pt sample
ramp-compressed and probed with two \xray\ sources, shown in three views:
the polar view centered at \xray\ source 1 (XRS1),
the polar view centered at \xray\ source 2 (XRS2), and the stereographic view,
demonstrating
\textit{HEXRD}'s multi-beam calibration capability.

\subsection{European XFEL}

\begin{figure}[htbp]
\begin{center}
\includegraphics[width=\columnwidth]{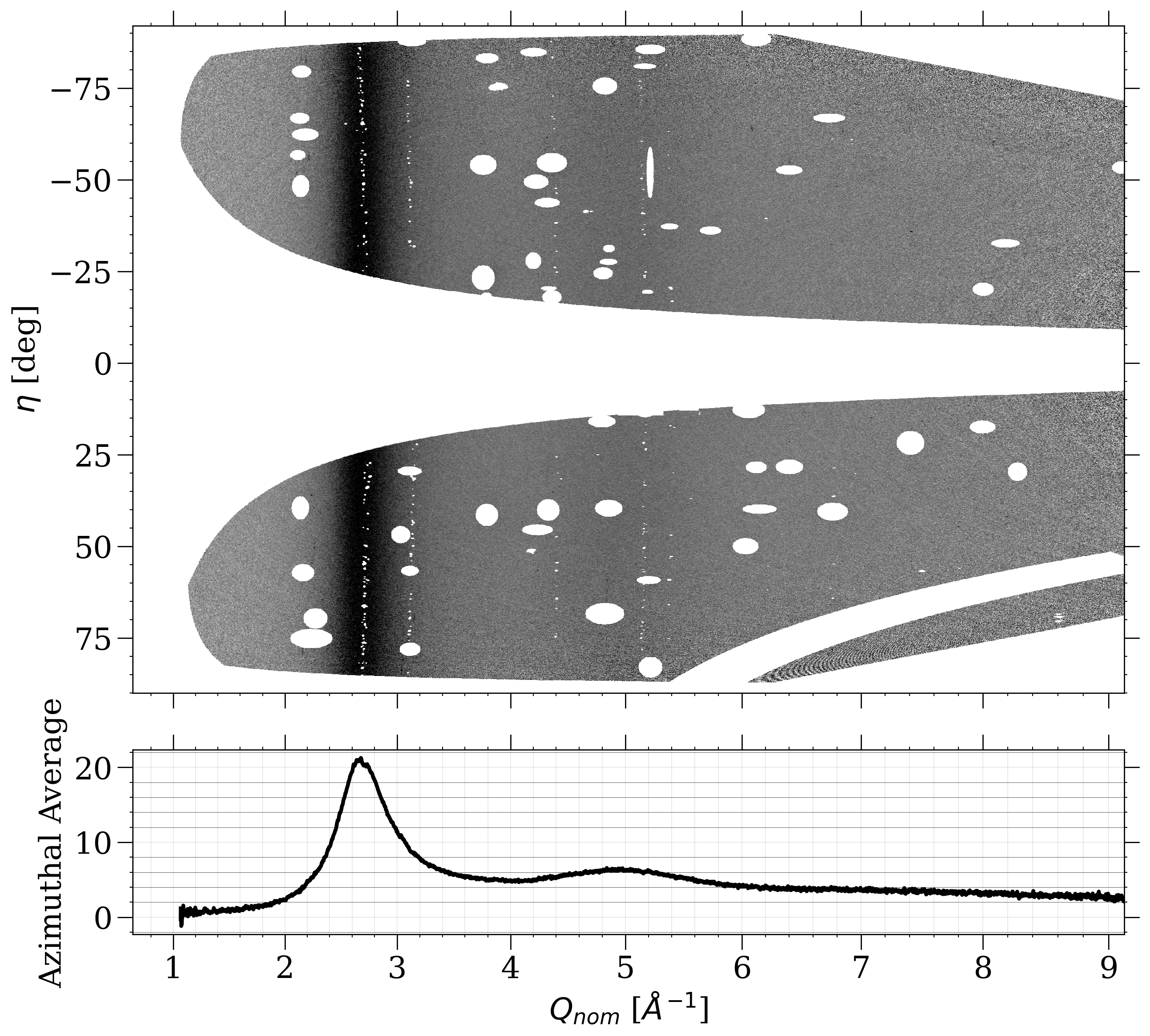}
\end{center}
\caption{Shock-compressed liquid Sn at the European XFEL using two
Varex detectors \protect\cite{Gorman2024xfel}, shown as a polar view
($\eta$ vs.\ nominal momentum transfer $Q$) with the azimuthal average
below.  Diffraction spots from solid Sn were masked (white ellipses)
in the images before import, so that the average shows the broad scattering of the
liquid.}
\label{fig:example_euxfel}
\end{figure}

At the High Energy Density (HED) instrument of the European XFEL,
shock-compressed liquid Sn was studied on two
Varex detectors using the DiPOLE 100-X laser
(Fig.~\ref{fig:example_euxfel}).  This example demonstrates
\textit{HEXRD}'s ability to handle free-electron-laser diffraction
data with minimal adaptation \cite{Gorman2024xfel}. Amorphous and liquid samples give broad halos rather than sharp rings; once the detectors are calibrated, the azimuthal average gives their scattering profile directly.

\subsection{MEC at LCLS}

\begin{figure}[htbp]
\begin{center}
\includegraphics[width=\columnwidth]{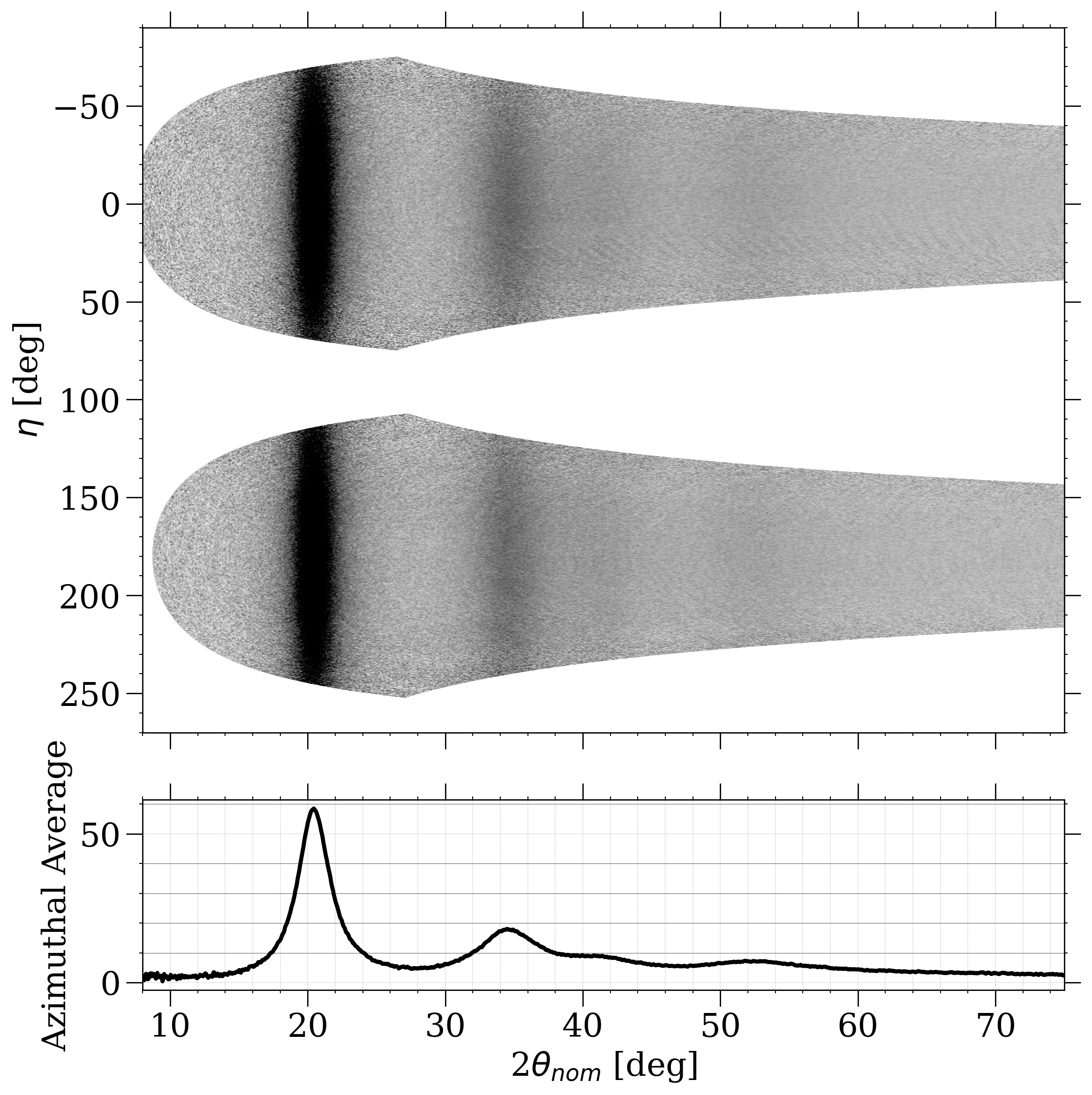}
\end{center}
\caption{Ambient metallic glass with single-shot XRD at MEC (LCLS)
using two Varex detectors, shown as a polar view with the azimuthal
average below, in which the broad peaks are the amorphous scattering of
the glass; unpublished data from LCLS proposal
L1013376, courtesy of C. McGuire and D. Khaghani.  A $7 \times 7$-pixel
median filter was applied to the detector images to reduce noise.}
\label{fig:example_mec}
\end{figure}

The Matter in Extreme Conditions end-station at LCLS uses femtosecond
\xray\ pulses to probe materials under extreme conditions.
Figure~\ref{fig:example_mec} shows an ambient metallic glass sample
recorded on two Varex detectors rather than the ePix10k detectors of the standard configuration (Fig.~\ref{fig:views}), which shows that \textit{HEXRD} adapts to non-standard configurations with minimal changes.

\subsection{PXRDIP at OMEGA EP}

\begin{figure}[htbp]
\begin{center}
\includegraphics[width=0.48\columnwidth]{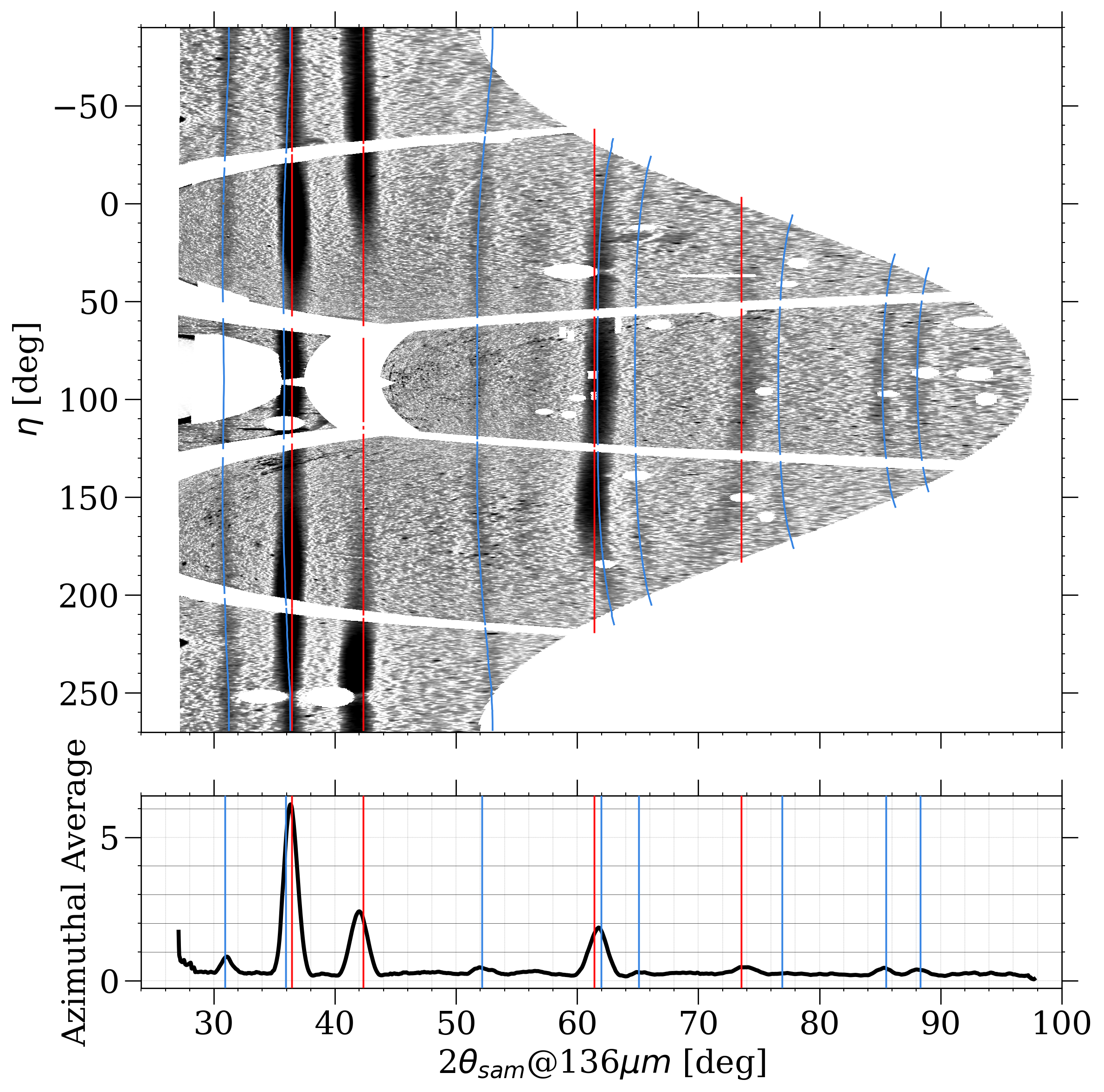}\hfill
\includegraphics[width=0.48\columnwidth]{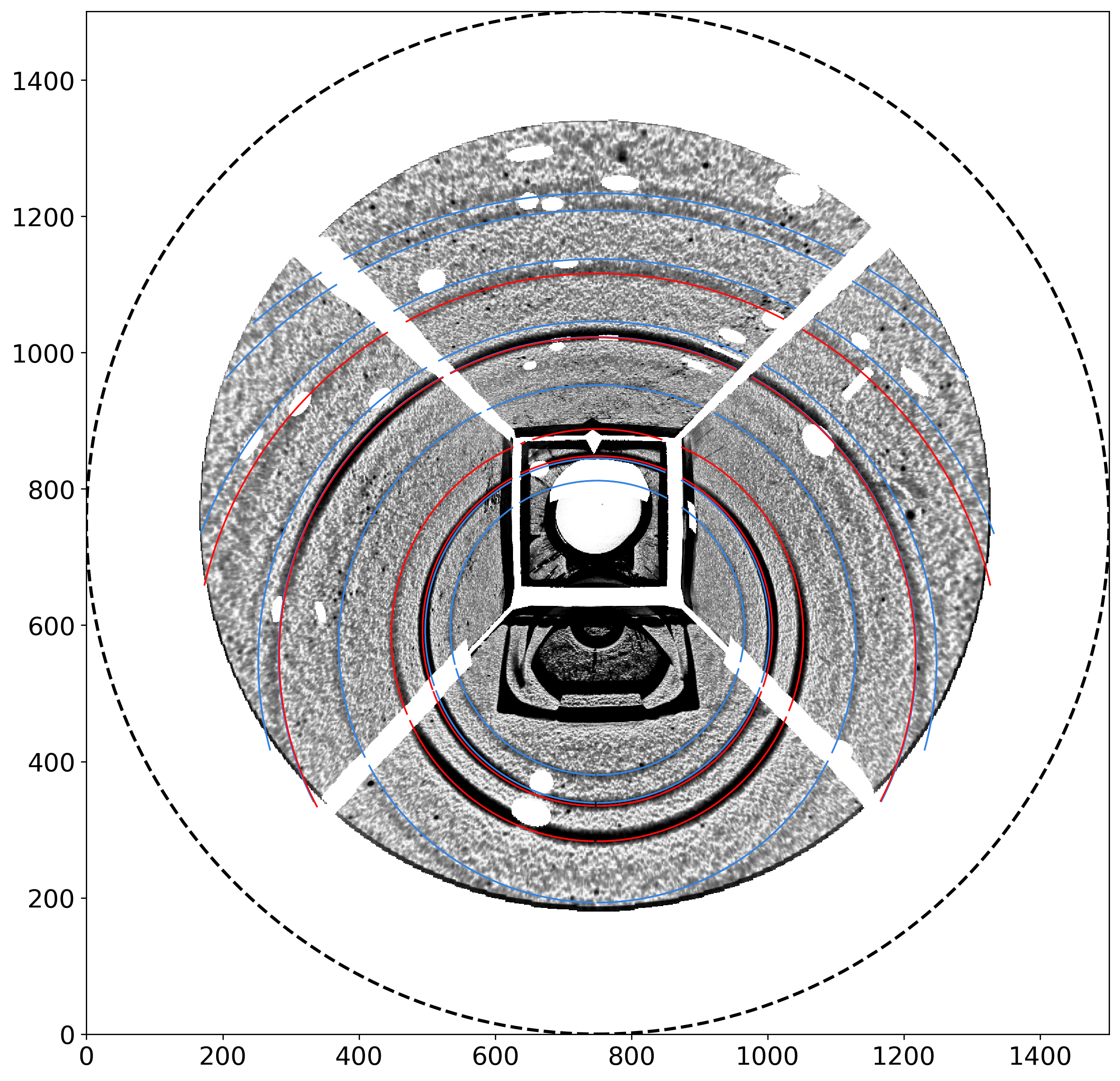}
\end{center}
\caption{Shock-ramp-compressed gold at OMEGA~EP (LLE) on the
PXRDIP platform. Red simulated pattern lines correspond to the high-pressure FCC Au sample and blue lines to the ambient Pt pinhole.  Left: polar view
with a sample distortion applied, which straightens the red sample lines
while curving the blue pinhole lines.  Right: stereographic view, with the
$\twotheta$ range taken low enough to include the low-angle plates, so that
the coverage extends into the center of the projection.  The dashed line
shows the edge of the stereographic circle.  The missing high-angle data in
the stereographic view are clipped due to the pinhole camera geometry.}
\label{fig:example_pxrdip}
\end{figure}

The PXRDIP platform at the OMEGA~EP laser facility captures powder
diffraction from shock-ramp-compressed samples on image plates, using
quasi-monochromatic \xray\ sources produced by laser irradiation of
metal foils.
Figure~\ref{fig:example_pxrdip} shows data from a gold sample under
shock-ramp compression \cite{Coleman2025}.

\subsection{FIDDLE at NIF}\label{sec:fiddle}

\begin{figure}[htbp]
\begin{center}
\includegraphics[width=\columnwidth]{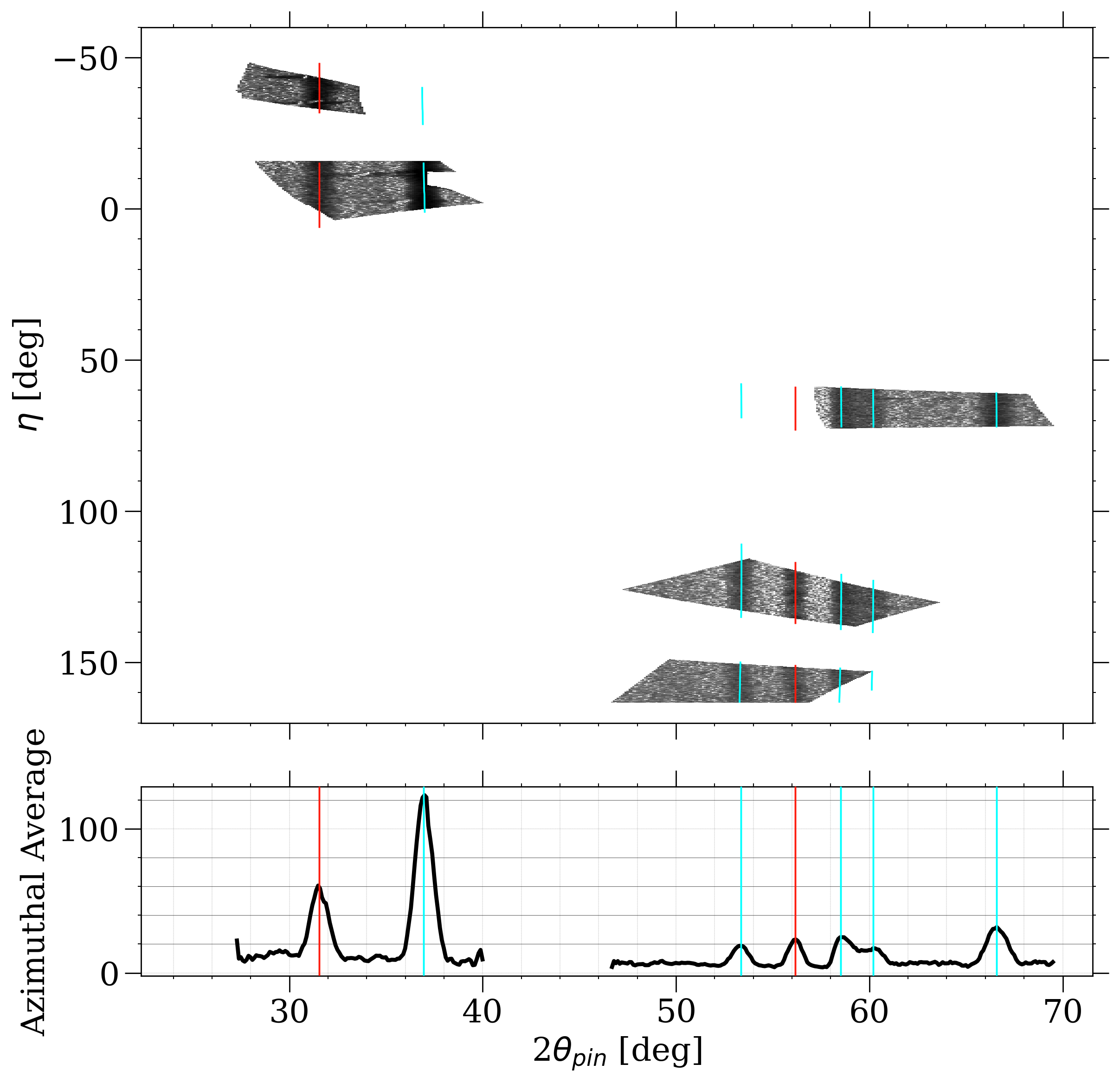}
\end{center}
\caption{Ambient CeO$_2$ dataset from the FIDDLE instrument at NIF,
showing multi-panel Icarus data.  Cyan simulated pattern lines correspond to the
CeO$_2$ sample and red lines to the Mo pinhole.  The polar view has the
pinhole correction applied ($\twotheta_{pin}$), and each set of simulated
lines carries its own target-assembly correction (sample layer for
CeO$_2$, pinhole for Mo).}
\label{fig:example_fiddle}
\end{figure}

The FIDDLE diagnostic captures time-resolved XRD using arrays of
Icarus detectors with 2~ns time resolution.
Figure~\ref{fig:example_fiddle} shows an ambient CeO$_2$ dataset
imaged using FIDDLE.
It was recorded separately from the dataset of
\citeasnoun{Vennari2024} and analyzed with the same workflow.  The positions and orientations of the five
cameras relative to one another were determined by the NIF Advanced
Tracking Laser Alignment System (ATLAS), and only the position and
orientation of the instrument as a whole were then refined, as one
rigid body, against hand-picked rings of both phases present.  With
those six parameters the simulated rings line up with the data on all
five cameras and with the peaks of the lineout
(Fig.~\ref{fig:example_fiddle}).  The distributed state file carries
the picks, each phase with its own target-assembly correction: 53 on the
CeO$_2$ powder (36~\textmu m of powder in epoxy against the source side
of a Mo pinhole 110~\textmu m thick and 380~\textmu m in diameter),
which takes the sample-layer correction, and 24 on the Mo pinhole
substrate itself, which takes the pinhole correction.  Holding that
calibration fixed and allowing only the CeO$_2$ lattice parameter to
vary, the composite calibrator returns
$a = 5.4131 \pm 0.0009$~\AA\ (standard error from the CeO$_2$ picks),
290~ppm or 1.7 standard errors above the value certified for NIST
SRM~674b, which is
less than one 25~\textmu m pixel (0.8~pixel for a ring at
$\twotheta = 58^{\circ}$) at these camera distances of 57--74~mm.
The placement of the instrument was refined against these same rings,
so this result tests the consistency of the calibration.  An offset of
this size could also come from the effective energy of the
laser-plasma source alone: 290~ppm corresponds to about 3~eV at the
10.25~keV Ge He-$\alpha$ line, much less than the energy spread of the
He-$\alpha$ line complex.  The picks fit with an
RMS residual of 75~mdeg for CeO$_2$ and 62~mdeg for Mo (the Mo rings
do not enter the lattice-parameter result).  A script distributed with the validation scripts of
\S\ref{sec:validation} repeats this from the state file.

\clearpage
\section{Conclusions}

Dynamic compression XRD relies on non-standard instrument configurations
(arbitrarily oriented multi-panel detectors, pinhole collimation,
curved image plates, polychromatic sources, and dual-beam geometries) that
no single general-purpose package covers in full.  \textit{HEXRD} and \textit{HEXRDGUI}
address this gap through a flexible geometric framework that supports a
broad range of detector arrangements and beam configurations.

The main advance of \textit{HEXRD} over existing software is that it
handles these instruments in one program, from calibration through whole-pattern
refinement.  The combination of three capabilities distinguishes
\textit{HEXRD}: (1) a generic geometric model supporting arbitrary
multi-detector configurations including cylindrical surfaces, (2)
integrated corrections for facility-specific geometries including
pinhole distortion and layered target assemblies, and (3) native
multi-beam calibration enabling simultaneous refinement against
diffraction from independent \xray\ sources.  Coupled with whole-pattern
peak fitting via Le~Bail and Rietveld methods, these features allow
experimentalists to carry out the analysis, from instrument
calibration to whole-pattern refinement, within a single tool
across multiple platforms.  Comparisons with independent codes and
synthetic ground-truth tests confirm the geometric model and the
multi-source calibration.  The software has been adopted
for routine analysis of data recorded at several major dynamic
compression platforms. Continued development focuses on automation of calibration workflows, support for emerging platforms, and expansion of the whole-pattern fitting module.

\begin{acknowledgements}
The authors thank Chris Harris, Marcus D. Hanwell and Donald E. Boyce
for their contributions to the code, Paul Shade (AFRL), the many
experimentalists who have provided feedback and testing, and the beamline 
and facility scientists at the facilities where the data shown here were 
recorded.  We thank Chris McGuire (LLNL) and
Dimitri Khaghani (SLAC) for providing the metallic glass data recorded
at MEC using Varex detectors under LCLS proposal L1013376; the
ePix10k CeO$_2$ data were recorded at MEC in LCLS experiment LV7318.
Use of the Linac Coherent Light Source (LCLS), SLAC National
Accelerator Laboratory, is supported by the U.S.\ Department of Energy
(DOE), Office of Science, Office of Basic Energy Sciences under Contract
No.\ DE-AC02-76SF00515.  This research used resources of the Advanced
Photon Source, a DOE Office of Science user facility operated for the
DOE Office of Science by Argonne National Laboratory under Contract
No.\ DE-AC02-06CH11357.  This publication is partially based upon work
performed at the Dynamic Compression Sector, which is operated by
Washington State University under the U.S.\ DOE/National Nuclear
Security Administration under Award No.\ DE-NA0003957.  
We acknowledge European XFEL in Schenefeld, Germany, 
for provision of X-ray free-electron laser beamtime at the HED instrument under 
community proposal 2720, and thank its staff for their assistance.  
We acknowledge the use of Anthropic's Claude in revising the text
during the preparation of this manuscript; the final contents
remain the sole responsibility of the authors.

\end{acknowledgements}

\begin{funding}
The development of \textit{HEXRD} and \textit{HEXRDGUI} has been
supported through subcontracts with Lawrence Livermore National
Laboratory and the Air Force Research Laboratory.  This work was
performed under the auspices of the US Department of Energy by
Lawrence Livermore National Laboratory under Contract No.
DE-AC52-07NA27344 (LLNL-JRNL-2024959).
\end{funding}

\ConflictsOfInterest{The authors declare no conflicts of interest.}

\DataAvailability{\textit{HEXRD} and \textit{HEXRDGUI} are freely
available as open-source software at \texttt{github.\allowbreak com/\allowbreak HEXRD}.
The \textit{HEXRDGUI} state files of every
dataset shown in this article (image data, calibration, materials,
masks and, where used, the calibration picks), the
validation scripts that reproduce the validation numbers of
\S\ref{sec:transfer}, \S\ref{sec:composite_cal}, \S\ref{sec:pinhole}, \S\ref{sec:validation}
and \S\ref{sec:wppf} and of the FIDDLE example, the converters between
\textit{HEXRD} instrument files and \textit{pyFAI} \texttt{poni} files,
and the \textit{Dioptas} calibration with its picked positions are
deposited on Zenodo \cite{paper_data_zenodo}.  The GE
CeO$_2$ dataset was recorded at beamline 1-ID-E of the Advanced Photon
Source under an Air Force Research Laboratory Partner User Proposal.}

\bibliography{references}

\end{document}